\documentclass[12pt]{iopart}
\usepackage{graphicx}
\usepackage{dcolumn}
\usepackage{bm}
\usepackage{iopams}
\usepackage{setstack}

\newcommand{\ket}[1]{\left| #1 \right \rangle }

\newcommand{\hc}{\hat{c}^{\phantom{\dagger}}}
\newcommand{\hcd}{\hat{c}^{\dagger}}
\newcommand{\hcdd}{\hat{c}^{(\dagger)}}
\newcommand{\hh}{\hat{h}^{\phantom{\dagger}}}
\newcommand{\hhd}{\hat{h}^{\dagger}}

\begin{document}

\title{The time-dependent Gutzwiller theory for multi-band Hubbard models} 

\author{E.~v.~Oelsen, G.~Seibold, and  J.~B\"unemann}

\address{Institut f\"ur Physik, BTU Cottbus, P.O. Box 101344, 03013 Cottbus, Germany}

\begin{abstract}%
We formulate a multi-band generalisation of the time-dependent 
 Gutzwiller theory.  This approach allows for the calculation of general 
 two-particle response functions, which are crucial for an understanding
 of various experiments in solid-state physics. As a first application, we 
study the momentum- and frequency-resolved magnetic susceptibility in a 
two-band Hubbard model. Like in the underlying ground-state approaches  we find 
significant differences between the results of our method and those 
from a time-dependent Hartree-Fock approximation.         
\end{abstract}
\pacs{71.10.-w,71.10.Fd,71.27.+a,75.10.-b,75.30.Ds}

 \submitto{\NJP}
        
\today
  
\maketitle 
             
\section{Introduction}
The study of materials with medium to strong Coulomb-interaction effects 
 has been a central subject for experimental and theoretical solid-state 
physics over many years. Despite enormous efforts and significant 
progress  in some fields, however,
 our theoretical toolbox is still far from satisfactory for such systems. 
 For quite some time, theoreticians in many-particle physics
have focused on relatively simple model 
systems, such as the Heisenberg or the single-band Hubbard models. 
Only in the past ten years, attention shifted towards the study
 of more realistic models, e.g., multi-band Hubbard models. 
A very important impulse in that direction 
came from the limit of infinite spatial dimensions 
($D\to \infty$). The exact solution of Hubbard models in this limit leads 
to the Dynamical Mean Field Theory (DMFT), in which the original lattice model 
is mapped onto an effective single-impurity system that has to be solved 
numerically \cite{metzner1989,vollhardt1989,vollhardt1993,georges1996,gebhard1997}. Although significant progress has been made in recent years 
in developing numerical techniques for the solution of the DMFT equations, 
it is still quite challenging  and can be carried out only with 
limited  accuracy. It is particularly difficult for the DMFT 
to study  multi-orbital Hubbard models when the full (local) Coulomb and 
exchange interaction is included.

An alternative method that also relies on infinite-$D$ techniques is the 
Gutzwiller 
 variational approach. It allows for the approximate study of ground-state
 properties and single-particle excitations with much less numerical
 effort than within DMFT and has been applied in a number of works in 
 recent years \cite{buenemann1997c,buenemann1998,buenemann2003,attaccalite2003,buenemann2003b,ferrero2005,julien2005,buenemann2007b,buenemann2007d,buenemann2008,lanata2008,ho2008,deng2009,zhouang2009,borghi2009,hofmann2009,wang2010,zhou2010,buenemann2011}. Another approach that leads to the
 same energy functional for multi-band models is the slave-boson mean field 
 theory 
\cite{lechermann2007,buenemann2007c,ferrero2009,isidori2009,lechermann2009,buenemann2010}.

The theoretical interpretation of a number of experiments requires the study of 
 two-particle response functions. For example, in magnetic neutron scattering
 the frequency- and momentum-resolved magnetic susceptibility is measured. 
The textbook method for the calculation of such  response functions
 is the random-phase approximation (RPA), which can be interpreted as a 
 time-dependent generalisation of the Hartree-Fock (HF) theory in the
small amplitude limit, i.e., where the perturbation is considered to be 
sufficiently small. For 
 electronic systems with medium or strong correlation effects, however, 
the ground-state description of a  HF theory is well known to be often 
inaccurate.  Therefore the RPA, as the time-dependent generalisation of the 
  HF theory,  is also 
 a questionable approach for such systems.

 A time-dependent Gutzwiller theory for the 
 calculation of two-particle response functions  was developed 
 for single-band Hubbard models by Seibold et al.\ 
\cite{seibold1998,seibold2001}. 
 In recent years this approach has been applied with astonishing success 
to quite a number of 
 such models and response functions \cite{seibold1998b,seibold2003,lorenzana2003,seibold2004,seibold2004b,lorenzana2005,seibold2005,seibold2006,seibold2007,seibold2008,seibold2008b}.
It is the main purpose of the present work to generalise the time-dependent 
 Gutzwiller theory for the investigation of 
  multi-band models. A brief introduction into our method has 
already been given in Ref. \cite{buenemann2011b}. All technical details, 
however,  will be first presented here.

Our presentation is organised as follows: In chapters~\ref{sec1} 
and \ref{ven} we summarise the main results of the Gutzwiller 
variational theory for multi-band Hubbard models. 
In chapter~\ref{chap7.3.2}  the reader will be reminded of the 
 derivation which introduces  the RPA as 
 a time-dependent generalisation of the HF theory. In a very 
 similar way  the  time-dependent Gutzwiller theory 
 (`Gutzwiller RPA') is introduced  in chapter~\ref{chap7.3.3}. 
 The general Gutzwiller RPA equations are 
used in  chapter~\ref{tpr} for the calculation  
of response functions for Hubbard-type lattice models. As a first application 
 we study the magnetic susceptibility in a two-band model in 
  chapter~\ref{ms}. A summary and conclusions close our presentation 
 in chapter~\ref{summary}. The more technical parts of our derivation
 are referred to four appendices.

\section{Multi-band Hubbard models and Gutzwiller wave functions }\label{sec1}
We study the general class of multi-band Hubbard models
\begin{equation}
\label{1.1}
\hat{H}=\sum_{i \ne j;\sigma,\sigma'}t_{i,j}^{\sigma,\sigma'}
\hat{c}_{i,\sigma}^{\dagger}\hat{c}_{j,\sigma'}^{\phantom{+}}
+\sum_i \hat{H}_{{\rm loc},i}=\hat{H}_0+\hat{H}_{\rm loc}\;.
\end{equation}
 Here, the first term describes the hopping of electrons between $N$ 
spin-orbital 
states $\sigma,\sigma'$ on~$L_{\rm s}$ lattice sites $i,j$, respectively. 
The Hamiltonian 
\begin{equation}\label{1.2}
\hat{H}_{{\rm loc},i}=\frac{1}{2}\sum_{\sigma_1,\sigma_2,\sigma_3,\sigma_4}
U_{i}^{\sigma_1,\sigma_2,\sigma_3,\sigma_4}\hat{c}^{\dagger}_{i,\sigma_1}
 \hat{c}^{\dagger}_{i,\sigma_2} 
\hat{c}_{i,\sigma_3} \hat{c}_{i,\sigma_4}+\sum_{\sigma_1,\sigma_2}\epsilon^{\sigma_1,\sigma_2}_i
\hat{c}^{\dagger}_{i,\sigma_1} 
\hat{c}_{i,\sigma_2}
\end{equation}
contains all local terms, i.e., the two-particle Coulomb 
interactions ($\sim U_{i}$) and the orbital onsite-energies 
($\sim \epsilon_{i}$).
We further introduce the eigenstates $|\Gamma \rangle_i $ 
of $\hat{H}_{{\rm loc},i}$ 
and the corresponding energies $E^{\rm loc}_{\Gamma,i} $, i.e.,
\begin{equation}\label{1.2b}
\hat{H}_{{\rm loc},i} |\Gamma\rangle_i  = E^{\rm loc}_{\Gamma,i}  
|\Gamma \rangle_i\;. 
\end{equation}

Within the Gutzwiller theory, the Hamiltonian~(\ref{1.1}) is investigated by 
means of the variational wave function 
\begin{equation}\label{1.3}
|\Psi_{\rm G}\rangle=\hat{P}_{\rm G}|\Psi_0\rangle
=\prod_{i}\hat{P}_{i}|\Psi_0\rangle\;,
\end{equation}
where $|\Psi_0\rangle$ is a normalised single-particle product state and the 
local Gutzwiller correlator is defined as 
\begin{equation}\label{1.4}
\hat{P}_{i}=\sum_{\Gamma,\Gamma^{\prime}}\lambda_{i;\Gamma,\Gamma^{\prime}}
|\Gamma \rangle_{i} {}_{i}\langle \Gamma^{\prime} |\;.
 \end{equation} 

For example, in case of the single-band Hubbard model
\begin{equation}\label{one_band}
\hat{H}_{\rm sb}=\sum_{i \ne j}\sum_{\sigma=\uparrow,\downarrow}t_{i,j}
\hat{c}_{i,\sigma}^{\dagger}\hat{c}_{j,\sigma}^{\phantom{+}}+\sum_{i}
U|d\rangle_i {}_{i}\langle d|\;,
\end{equation}
the local correlation operator reads 
\begin{eqnarray}\nonumber
 \hat{P}_{i}&=&\lambda_{i,d}|d\rangle_i {}_{i}\langle d|+
\lambda_{i,\uparrow}|\!\!\uparrow\rangle_i {}_{i}\langle\uparrow\! |+
\lambda_{i,\downarrow}\!|\!\downarrow\rangle_i {}_{i}\langle\downarrow\! |
+\lambda_{i,\downarrow,\uparrow}|\!\!\downarrow\rangle_i {}_{i}\langle\uparrow\! |
\\\label{pi_one}
&&+\lambda_{i;\uparrow,\downarrow}|\!\!\uparrow\rangle_i {}_{i}\langle\downarrow \!|
+
\lambda_{i,\emptyset}|\emptyset\rangle_i {}_{i}\langle\emptyset |\;.
\end{eqnarray}
Here, we introduced the atomic states $|\Gamma \rangle_i$ for 
doubly occupied sites $|d\rangle_i$, singly occupied sites 
$|\!\!\uparrow\rangle_i$ and $|\!\!\downarrow\rangle_i$,  
and for empty sites $|\emptyset\rangle_i$, as well as the 
  abbreviation $\lambda_{i,\Gamma}$ for the diagonal 
 variational parameters $\lambda_{i;\Gamma,\Gamma}$. 
In terms of the fermionic operators $\hat{c}^{(\dagger)}_{i,\sigma}$, 
the operator~(\ref{pi_one}) has the form
\begin{eqnarray}\label{pi_one2}
 \hat{P}_{i}&=&\lambda_{i,d}\hat{n}_{i,\uparrow}\hat{n}_{i,\downarrow}+
\lambda_{i,\uparrow}\hat{n}_{i,\uparrow}(1-\hat{n}_{i,\downarrow})+
\lambda_{i,\downarrow}\hat{n}_{i,\downarrow}(1-\hat{n}_{i,\uparrow})
\\\nonumber
&&+\lambda_{i,\downarrow,\uparrow}\hat{c}^{\dagger}_{i,\downarrow}
\hat{c}^{}_{i,\uparrow}
+\lambda_{i;\uparrow,\downarrow}\hat{c}^{\dagger}_{i,\uparrow}
\hat{c}^{}_{i,\downarrow}
+
\lambda_{i,\emptyset}(1-\hat{n}_{i,\uparrow})(1-\hat{n}_{i,\downarrow})
\end{eqnarray}
where $\hat{n}_{i,\sigma}\equiv \hat{c}^{\dagger}_{i,\sigma}
\hat{c}^{}_{i,\sigma}$.
The correlation 
 operator~(\ref{pi_one}) is the most general Ansatz for single-band models  
 without superconductivity. The latter would require additional 
 terms of the form $\sim |d\rangle_i {}_{i}\langle\emptyset |$ and
 $\sim|\emptyset\rangle_i {}_{i}\langle d |$; c.f., 
Ref.\ \cite{buenemann2005b}.

\section{Variational energy}\label{ven}
As shown in Refs.~\cite{buenemann1998,buenemann2005}, 
the expectation value of the Hamiltonian~(\ref{1.1}) with respect to 
the variational wave-function~(\ref{1.3}) can be evaluated in the limit of 
infinite spatial dimensions. 
We consider the expectation values of the local Hamiltonian 
$\hat{H}_{\rm loc}$,  and the one-particle Hamiltonian $\hat{H}_{1}$ 
 separately in sections~\ref{locen} and \ref{onepen}.   
 The additional constraints, which arise through the derivation in 
infinite dimensions are discussed in section~\ref{constraints}.  
In section~\ref{recovery}, we recall how the standard Gutzwiller 
energy functional for a single-band model is recovered from our general 
multi-band results.

\subsection{Local energy}\label{locen}
 The expectation value of the local Hamiltonian 
(\ref{1.2b}) in infinite 
dimensions reads \cite{note1}  
\begin{equation}\label{1.5}
\langle \hat{H}_{{\rm loc}} \rangle_{\Psi_{\rm G}}=\sum_{\Gamma}
E^{\rm loc}_{\Gamma}m_{\Gamma,\Gamma}\equiv E_{\rm loc}\;,
\end{equation}
where 
\begin{equation}\label{1.6}
m_{\Gamma,\Gamma'} \equiv\langle \left( |\Gamma \rangle   \langle  \Gamma' 
|\right)  \rangle_{\Psi_{\rm G}}
=\left\langle\left( \hat{P}^{\dagger} |\Gamma \rangle   \langle  \Gamma' 
|\hat{P} \right) \right  \rangle_{\Psi_{0}}=\sum_{\tilde{\Gamma},\tilde{\Gamma}'}
\lambda^{*}_{\Gamma,\tilde{\Gamma}}\lambda_{\Gamma',\tilde{\Gamma}'}^{}
m^0_{\tilde{\Gamma},\tilde{\Gamma}'}
\end{equation}
and
\begin{equation}\label{1.6c}
m^0_{\Gamma,\Gamma'}\equiv\langle \left( |\Gamma \rangle   \langle  \Gamma' |\right)  \rangle_{\Psi_{0}}\;.
\end{equation}
To further evaluate the expectation value~(\ref{1.6c}), we introduce the 
 basis of Fock states (i.e., `Slater determinants')
\begin{equation}\label{idef}
|I\rangle= \prod_{\sigma \in I}\hat{c}^{\dagger}_{\sigma}|0\rangle 
\end{equation}
in which certain spin-orbit states $\sigma \in I$ are occupied. 
Mathematically, the indices 
$I=(\sigma_1,\sigma_2,\ldots,\sigma_n)$ are considered as ordered sets 
of spin-orbit states $\sigma$. Therefore we can use all standard set 
operations, such as $I\cup\sigma$ or $I\backslash \sigma$. In addition, we 
define the number of orbitals in $I$ as $|I|$. 
The states $|I\rangle$ provide a basis of the local 
(atomic) Hilbert space. Hence, we can use them
 for an expansion of the eigenstates $|\Gamma\rangle$,
\begin{equation}\
|\Gamma\rangle=\sum_{I}T_{I,\Gamma}|I\rangle
 \end{equation}
and write the expectation value~(\ref{1.6c}) as
\begin{equation}\label{m0ggs}
m^0_{\Gamma,\Gamma'}
=\sum_{I,I'}T_{I,\Gamma}^{} T^{*}_{I',\Gamma'}m^{0}_{I,I'}\;.
\end{equation}
Finally, the uncorrelated expectation values of the transfer operators $|I \rangle   \langle I' |$,
\begin{equation}\label{m0iis}
m^{0}_{I,I'}\equiv 
\langle \left( |I \rangle   \langle I' |\right)
\rangle_{\Psi_0}\;,
\end{equation}
can be written as the determinant
\begin{equation}\label{miiprime}
m^{0}_{I,I'}=\left|
\begin{array}{cc}
\Omega^{I,I'}&-\Omega^{I,J}\\
\Omega^{J,I'}&\bar{\Omega}^{J,J}
\end{array}
\right|\;.
\end{equation}
Here, $\Omega_{I,I'}$ are the matrices
\begin{equation}
\Omega_{I,I'}=\left(
\begin{array}{cccc}
C^0_{\sigma_1,\sigma'_1}&C^0_{\sigma_1,\sigma'_2}&\ldots&C^0_{\sigma_1,\sigma'_{|I'|}}\\
C^0_{\sigma_2,\sigma'_1}&C^0_{\sigma_2,\sigma'_2}&\ldots&C^0_{\sigma_2,\sigma'_{|I'|}}\\
\ldots&\ldots&\ldots&\ldots\\
C^0_{\sigma_{|I|},\sigma'_1}&C^0_{\sigma_{|I|},\sigma'_2}&\ldots&C^0_{\sigma_{|I|},\sigma'_{|I'|}}
\end{array}
\right)\;,
\end{equation}
in which the entries are the elements of the 
uncorrelated local density matrix
\begin{equation}\label{cmat}
C^0_{\sigma,\sigma'}=\langle \hat{c}^{\dagger}_{i,\sigma}\hat{c}^{}_{i,\sigma'} 
\rangle_{\Psi_0}
\end{equation}
that belong to the configurations $I=(\sigma_1,\ldots,\sigma_{|I|})$ and
$I'=(\sigma'_1,\ldots,\sigma'_{|I'|})$. The matrix $\bar{\Omega}^{J,J}$ 
in~(\ref{miiprime}) is defined as 
\begin{equation}
\bar{\Omega}_{J,J}=\left(
\begin{array}{cccc}
1-C^0_{\sigma_1,\sigma_1}&-C^0_{\sigma_1,\sigma_2}&\ldots&-C^0_{\sigma_1,\sigma_{|J|}}\\
-C^0_{\sigma_2,\sigma_1}&1-C^0_{\sigma_2,\sigma_2}&\ldots&-C^0_{\sigma_2,\sigma_{|J|}}\\
\ldots&\ldots&\ldots&\ldots\\
-C^0_{\sigma_{|J|},\sigma_1}&-C^0_{\sigma_{|J|},\sigma_2}&\ldots&1-C^0_{\sigma_{|J|},\sigma_{|J|}}
\end{array}
\right)\;,
\end{equation}
with $\sigma_i\in J\equiv (1,\ldots,N)\backslash (I\cup I') $. 

 In applications of the Gutzwiller theory to multi-band systems 
 it would be quite cumbersome to evaluate the determinants~(\ref{miiprime}) 
if the local density matrix~(\ref{cmat})  is non-diagonal. 
 Fortunately, we are free to chose the local orbital basis in a way 
 that suits us best.  
 Therefore, we introduce an orbital basis, defined 
by (local) operators  $\hat{h}^{(\dagger)}_{\gamma}$, for which 
\begin{equation}\label{hmat}
C^0_{\gamma,\gamma'}=\bar{C}^0_{\gamma,\gamma'}\equiv\delta_{\gamma,\gamma'}
\langle \hat{h}^{\dagger}_{i,\gamma}\hat{h}^{}_{i,\gamma'} 
\rangle_{\Psi_0}\equiv n^0_{\gamma}\;.
\end{equation}
 With such a basis the expectation value~(\ref{miiprime}) has the 
 simple form 
\begin{equation}\label{rty}
m^{0}_{I,I'}=\delta_{I,I'}
\prod_{\gamma\in I} n^0_{\gamma} \prod_{\gamma \notin I}(1-n^0_{\gamma})\;.
\end{equation}
Note that, for simplicity, we always use the same variable $I$ for 
configuration states of the form~(\ref{idef}) irrespective of the 
underlying orbital basis (e.g., 
$\hat{c}^{(\dagger)}_{\sigma}$ in~(\ref{idef}) or
$\hat{h}^{(\dagger)}_{\gamma}$ in~(\ref{rty})). 

As will be shown in chapter~\ref{chap7.3.3}, the time-dependent Gutzwiller 
theory requires to calculate the first and second derivatives of the energy
 with respect to 
 all elements of the local density matrix, including the non-diagonal 
terms. Since the local density matrix 
 enters the energy functional solely through matrices of the 
form~(\ref{miiprime}) we only need to expand these matrices 
  with respect to small perturbations  
\begin{equation}
C^0_{\gamma,\gamma'}=\bar{C}^0_{\gamma,\gamma'}+\delta C^0_{\gamma,\gamma'}
\end{equation}
up to second order in $\delta C^0_{\gamma,\gamma'}$ around the diagonal 
 ground-state matrix~(\ref{hmat}). 
This expansion is explicitly carried out
 in \ref{app1}.

In our derivation of the ground-state energy all local onsite energies were 
 considered as part of the local Hamiltonian~(\ref{1.2}). 
 For later use, however, we also need an expression for the expectation
 value of a general local one-particle Hamiltonian 
  \begin{equation}
\hat{H}_{\rm onsite}=\sum_{\sigma,\sigma'}\epsilon^{\sigma,\sigma'}
\hat{c}^{\dagger}_{\sigma} 
\hat{c}_{\sigma'}\;.
 \end{equation}
This expectation value is given as 
\begin{equation}\label{0agh}
\langle \hat{H}_{\rm onsite}\rangle_{\Psi_{\rm G}}=\sum_{\sigma,\sigma'}\epsilon^{\sigma,\sigma'}C^{\rm c}_{\sigma,\sigma'}
\end{equation}
where
 \begin{equation}\label{cmatc}
C^{\rm c}_{\sigma,\sigma'}=
\sum_{\Gamma_1,\Gamma_2,\Gamma_3,\Gamma_4}
\lambda^{*}_{\Gamma_2,\Gamma_1}\lambda^{}_{\Gamma_3,\Gamma_4}
\langle \Gamma_2 | \hat{c}^{\dagger}_{\sigma} 
\hat{c}_{\sigma'} |\Gamma_3\rangle m^0_{\Gamma_1,\Gamma_4}
\end{equation}  
is the `correlated' local density matrix.
 
\subsection{Kinetic energy}\label{onepen}
The expectation value of a hopping term in $\hat{H}_0$, Eq.~(\ref{1.1}), 
 is given as
\begin{equation}\label{1.13a}
\langle  \hat{c}_{i,\sigma_1}^{\dagger}\hat{c}_{j,\sigma_2}^{\phantom{+}} \rangle_{\Psi_{\rm G}}=\sum_{\sigma'_1,\sigma'_2}q_{i,\sigma_1}^{\sigma'_1}\left( q_{j,\sigma_2}^{\sigma'_2}\right)^{*}\langle  
\hat{c}_{i,\sigma'_1}^{\dagger}\hat{c}_{j,\sigma'_2}^{\phantom{+}} \rangle_{\Psi_{0}}\;,
\end{equation}
where we have introduced the (local) renormalisation matrix \cite{note1}  
\begin{eqnarray}\label{qmat}
q_{\sigma}^{\sigma'}=\sum_{\Gamma_1,\ldots,\Gamma_4}\lambda^{*}_{\Gamma_2,\Gamma_1}
\lambda^{}_{\Gamma_3,\Gamma_4}\langle \Gamma_2|\hat{c}^{\dagger}_{\sigma}  
|\Gamma_3 \rangle \sum_{I_1,I_4}T_{I_1,\Gamma_1}T^{*}_{I_4,\Gamma_4}
H^{\sigma'}_{I_1,I_4}\;.
\end{eqnarray}
The matrix $H^{\sigma'}_{I_1,I_4}$ contains three different contributions 
depending on whether the index $\sigma'$ is an element 
 of $I_1\cap I_4$, 
$I_4\backslash (I_1\cap I_4)$, or $J=(1,\ldots,N)\backslash(I_1\cup I_4)$. 
With the abbreviation 
$f_{\sigma,I}\equiv\langle I |\hat{c}^{\dagger}_{\sigma}\hat{c}^{}_{\sigma} |I  \rangle$  
we can write $H^{\sigma'}_{I_1,I_4}$ as
\begin{eqnarray}\label{8sgdd}
H^{\sigma'}_{I_1,I_4}&\equiv&(1-f_{\sigma',I_1})\langle I_4  |\hat{c}^{}_{\sigma'} |I_4\cup \sigma'  \rangle
m^{0}_{I_1,I_4\cup \sigma'}\\\nonumber
&&\!\!\!\!\!\!\!\!\!\!\!\!\!\!\!\!\!\!\!\!\!\!\!\!\!\!
+\langle I_1 \backslash \sigma' |\hat{c}^{}_{\sigma'} |I_1  \rangle
\left(
f_{\sigma',I_4}m^{0}_{I_1\backslash \sigma',I_4}+
(1-f_{\sigma',I_4})m^{0;\sigma'}_{I_1\backslash \sigma',I_4}
\right)\;.
\end{eqnarray}
The expectation value $m^{0;\sigma'}_{I_1\backslash \sigma',I_4}$ in~(\ref{8sgdd})
 has the same 
form as the one in~(\ref{miiprime}), except that the index $J$ 
 has to be replaced by $J \backslash \sigma'$. 
In case of a diagonal local density matrix one finds 
\begin{equation}
H^{\sigma'}_{I_1,I_4}=\delta_{I_1\backslash \sigma',I_4}
\langle I_1 \backslash \sigma' |\hat{c}^{}_{\sigma'} |I_1  \rangle
\frac{m^0_{I_4,I_4}}{1-C^{0}_{\sigma',\sigma'}}
\end{equation}
in agreement with results derived earlier \cite{buenemann1998}. Note 
that, in general, the renormalisation matrix is {\sl not} Hermitian, 
i.e., it is
\begin{equation}
 q_{\sigma}^{\sigma'}\neq ( q_{\sigma'}^{\sigma})^*\;.
\end{equation}

Using~(\ref{1.13a}), the expectation value of the one particle 
Hamiltonian $\hat{H}_0$ can be written as
\begin{equation}\label{h0exp}
\langle \hat{H}_0 \rangle_{\Psi_{\rm G}}=L_s
\sum_{\sigma_1,\sigma_2,\sigma'_1,\sigma'_2}
q^{\sigma'_1}_{\sigma_1}\big(q^{\sigma'_2}_{\sigma_2}\big)^*
E_{\sigma_1,\sigma_2,\sigma'_1,\sigma'_2}
\end{equation}
where we introduced the tensor 
\begin{equation}\label{0etsy}
E_{\sigma_1,\sigma_2,\sigma'_1,\sigma'_2}\equiv\frac{1}{L_s} 
\sum_{i\neq j}t^{\sigma_1,\sigma_2}_{i,j}
\langle  \hat{c}^{\dagger}_{i,\sigma'_1}\hat{c}_{j,\sigma'_2}
\big \rangle_{\Psi_0}\;.
\end{equation}   

\subsection{Physical constraints}\label{constraints}
As it turns out through the evaluation of expectation values 
 in infinite dimensions, the variational parameters 
 $\lambda_{\Gamma,\Gamma'}$ need to obey certain local
 constraints.  These are
\begin{eqnarray}\label{1.10a}
\langle\hat{P}^{\dagger}\hat{P}^{}\rangle_{\Psi_0}&=&1\;,\\\label{1.10b}
\langle  \hat{c}^{\dagger}_{\sigma} \hat{P}^{\dagger}\hat{P}^{} 
\hat{c}^{}_{\sigma'} \rangle_{\Psi_0}&=&\langle
 \hat{c}^{\dagger}_{\sigma}\hat{c}^{}_{\sigma'}   \rangle_{\Psi_0}=
C^0_{\sigma,\sigma'}\;. 
\end{eqnarray}
Note  that moving the operator $\hat{P}^{\dagger}\hat{P}^{}$  relative to 
$\hat{c}^{\dagger}_{\sigma}$ or $\hat{c}^{}_{\sigma'}$ 
 in~(\ref{1.10b})  would not alter the whole 
set of constraints. With the explicit form~(\ref{1.4}) of the correlation 
 operator  $\hat{P}^{}$, the constraints read
\begin{eqnarray}\label{1.10c}
1&=&\sum_{\Gamma,\Gamma_1,\Gamma_2}
\lambda_{\Gamma,\Gamma_1}^{*}\lambda_{\Gamma,\Gamma_2}^{}
m^{0}_{\Gamma_1,\Gamma_2}\;,\\\label{1.10d}
C^0_{\sigma,\sigma'}&=&\sum_{\Gamma,\Gamma',\Gamma_1,\Gamma_2,\Gamma_3}
\lambda_{\Gamma_2,\Gamma_1}^{*}\lambda_{\Gamma_2,\Gamma_3}^{}
\langle\ \Gamma | \hat{c}^{\dagger}_{\sigma}|\Gamma_1   \rangle
\times\langle\ \Gamma_3 |  \hat{c}^{}_{\sigma'}     |\Gamma'   \rangle
m^{0}_{\Gamma,\Gamma'}\;.
\end{eqnarray}  

\subsection{Recovery of the `standard' single-band energy functional}\label{recovery}
In case of a  single-band model, the atomic eigenstates $|\Gamma \rangle$ 
coincide 
with the configuration states~$|I \rangle$. If we assume a translationally 
 invariant ground state and the most general form of a local density 
matrix \cite{note1}  
 \begin{equation}\label{cmat_one}
C^0=\left(
\begin{array}{cc}
\langle \hat{c}^{\dagger}_{\uparrow}\hat{c}^{}_{\uparrow}\rangle_{\Psi_0}&
\langle \hat{c}^{\dagger}_{\uparrow}\hat{c}^{}_{\downarrow}\rangle_{\Psi_0}\\
\langle \hat{c}^{\dagger}_{\downarrow}\hat{c}^{}_{\uparrow}\rangle_{\Psi_0}&
\langle \hat{c}^{\dagger}_{\downarrow}\hat{c}^{}_{\downarrow}\rangle_{\Psi_0}
\end{array}
 \right)=
\left(
\begin{array}{cc}
n^{0}_{\uparrow}&\Delta^0_{\uparrow,\downarrow}\\
\Delta^0_{\downarrow,\uparrow}&n^{0}_{\downarrow}
\end{array}
 \right)\;,
\end{equation}
where $(\Delta^0_{\downarrow,\uparrow})^{*}
=\Delta^0_{\uparrow,\downarrow}\equiv\Delta^0$, we find
 \begin{eqnarray}
m^0_{\emptyset,\emptyset}&=&(1-n^{0}_{\uparrow})(1-n^{0}_{\downarrow})-|\Delta^0|^2\;,\\\label{m0ss}
m^0_{\sigma,\sigma}&=&n^{0}_{\sigma}(1-n^{0}_{\bar{\sigma}})+|\Delta^0|^2\;,\\
m^0_{\sigma,\bar{\sigma}}&=&\Delta^0_{\sigma,\bar{\sigma}}\;,\\
m^0_{d,d}&=&n^{0}_{\uparrow}n^{0}_{\downarrow}-|\Delta^0|^2\;,
\end{eqnarray}
for those of the expectation values~(\ref{m0iis}) which are finite. 
Here we used 
the notation 
\begin{equation}
\bar{\uparrow}=\downarrow\;\;\; {\rm and}\;\;\;  \bar{\downarrow}=\uparrow\;.
\end{equation}
As a consequence, the expectation value of the local Coulomb interaction 
in~(\ref{one_band}) reads
 \begin{equation}
\sum_{i}
U\Big\langle |d\rangle_i {}_{i}\langle d| \Big\rangle_{\Psi_{\rm G}}=
L_{s} U|\lambda_{d}|^2m^0_{d,d}\;.
\end{equation}
 For the single-band model with  
the correlation operator~(\ref{pi_one}) 
and the local density matrix~(\ref{cmat_one}) the 
elements of the renormalisation matrix have the form
\begin{eqnarray}
q_{\sigma}^{\sigma}&=&\lambda^*_{\sigma} \lambda_{\emptyset}
(1-n^0_{\bar{\sigma}})+\lambda^*_{d}\lambda_{\bar{\sigma}}
n^0_{\bar{\sigma}}+(\lambda^*_{d}\lambda_{\bar{\sigma},\sigma}
+\lambda^*_{\sigma,\bar{\sigma}}\lambda_{\emptyset})
\Delta^0_{\bar{\sigma},\sigma}\;,\\
q_{\sigma}^{\bar{\sigma}}&=&\Delta^0_{\sigma,\bar{\sigma}}
(\lambda^*_{\sigma}\lambda_{\emptyset}-\lambda^*_{d}\lambda_{\bar{\sigma}})
-\lambda^*_{d}\lambda_{\bar{\sigma},\sigma}n^0_{\sigma}
+\lambda^*_{\sigma,\bar{\sigma}}\lambda_{\emptyset}(1-n^0_{\sigma})\;.
\end{eqnarray} 
Finally, the constraints in this case are given as 
\begin{eqnarray}\nonumber
1&=&|\lambda_{\emptyset}|^2m^0_{\emptyset,\emptyset}+|\lambda_{d}|^2m^0_{d,d}
+(|\lambda_{\uparrow}|^2+|\lambda_{\downarrow,\uparrow}|^2)m^0_{\uparrow,\uparrow}+(|\lambda_{\downarrow}|^2+|\lambda_{\uparrow,\downarrow}|^2)m^0_{\downarrow,\downarrow}\\\label{con1a}
&& +(\lambda^*_{\downarrow,\uparrow} \lambda_{\downarrow}
+ \lambda^*_{\uparrow}\lambda_{\uparrow,\downarrow}  )\Delta^0_{\uparrow,\downarrow}
+(\lambda^*_{\uparrow,\downarrow}\lambda_{\uparrow}+
\lambda^*_{\downarrow}\lambda_{\downarrow,\uparrow}
 )\Delta^0_{\downarrow,\uparrow}\;,\\\label{con1b}
n^0_{\sigma}&=&(|\lambda_{\bar{\sigma}}|^2
+|\lambda_{\sigma,\bar{\sigma}}|^2)
m^0_{d,d}
+|\lambda_{\emptyset}|^2m^0_{\sigma,\sigma}\;,\\\label{con1c}
\Delta^0_{\sigma,\bar{\sigma}}&=&
-(\lambda^*_{\sigma,\bar{\sigma}}\lambda_{\sigma}
+\lambda^*_{\bar{\sigma}}\lambda_{\bar{\sigma},\sigma})m^0_{d,d}+
|\lambda_{\emptyset}|^2\Delta^0_{\sigma,\bar{\sigma}}
\;.
\end{eqnarray}  
As mentioned before, it is possible to overcome the complications 
 that arise from a non-diagonal local density matrix by a simple transformation
 \begin{equation}
\hat{h}^{\dagger}_{\gamma}=\sum_{\sigma}u_{\sigma,\gamma}\hat{c}^{\dagger}_{\sigma}
\end{equation}
 to a new orbital basis
 for which the local density matrix is diagonal by definition,
\begin{equation}
\left(
\begin{array}{cc}
\langle \hat{h}^{\dagger}_{1}\hat{h}^{}_{1}\rangle_{\Psi_0}&
\langle \hat{h}^{\dagger}_{1}\hat{h}^{}_{2}\rangle_{\Psi_0}\\
\langle \hat{h}^{\dagger}_{2}\hat{h}^{}_{1}\rangle_{\Psi_0}&
\langle \hat{h}^{\dagger}_{2}\hat{h}^{}_{2}\rangle_{\Psi_0}
\end{array}
 \right)=
\left(
\begin{array}{cc}
\tilde{n}^{0}_{1}&0\\
0&\tilde{n}^{0}_{2}
\end{array}
 \right)\;.
\end{equation}
In this new basis the constraints have the rather simple form 
\begin{eqnarray}\label{conh}
1&=&\tilde{\lambda}^2_{\emptyset}m^0_{\emptyset,\emptyset}+
\tilde{\lambda}^2_{1}m^0_{1,1}+\tilde{\lambda}^2_{2}m^0_{2,2}+
\tilde{\lambda}^2_{d}m^0_{d,d}\;,
\\\label{conh2}
\tilde{n}^{0}_{\gamma}&=&\tilde{\lambda}^2_{\gamma}m^0_{\gamma,\gamma}+\tilde{\lambda}^2_{d}m^0_{d,d}\;.
\end{eqnarray} 
where the non-diagonal constraints are automatically fulfilled 
 by working with a diagonal correlation operator 
($\tilde{\lambda}_{1,2}=0$). Equations~(\ref{conh})-(\ref{conh2}) 
can be readily solved by introducing the expectation values 
 \begin{eqnarray}
\tilde{m}_{d}&\equiv& \tilde{\lambda}^2_{d}m^0_{d,d}\;,\\
\tilde{m}_{\emptyset}&\equiv& \tilde{\lambda}^2_{\emptyset}m^0_{\emptyset,\emptyset}
= 1-\tilde{n}^{0}_{1}-\tilde{n}^{0}_{2}+\tilde{m}_{d}\;,\\
\tilde{m}_{\gamma}&\equiv&\tilde{\lambda}^2_{\gamma}m^0_{\gamma,\gamma}=
\tilde{n}^{0}_{\gamma}-\tilde{m}_{d}\;,
\end{eqnarray}  
which leaves us with only one variational parameter, the expectation value
 $\tilde{m}_{d}$ for a double occupancy. The resulting renormalisation 
 matrix is then diagonal and its elements have the well known form
 \cite{gutzwiller1963,gutzwiller1964,gutzwiller1965}
\begin{equation}
 q_{\gamma}\equiv q_{\gamma}^{\gamma}=\tilde{\lambda}_{\emptyset}\tilde{\lambda}_{\gamma}
(1-\tilde{n}^0_{\bar{\gamma}})+
\tilde{\lambda}_{d}\tilde{\lambda}_{\bar{\gamma}}
\tilde{n}^0_{\bar{\gamma}}\frac{1}{\sqrt{\tilde{n}^0_{\gamma}(1-\tilde{n}^0_{\gamma})}}
\left(\sqrt{\tilde{m}_{\emptyset}\tilde{m}_{\gamma}}
+ \sqrt{\tilde{m}_{d}\tilde{m}_{\bar{\gamma}}}
\right)\;.
\end{equation}
Hence, the single-particle energy~(\ref{h0exp}) is given as
\begin{equation}
 E_0=\sum_{\gamma}
q_{\gamma}^2    
\sum_{i\neq j} t_{i,j} \langle \hat{h}^{\dagger}_{i,\gamma}\hat{h}^{}_{j,\gamma}
\rangle_{\Psi_0}
\end{equation}
where we used the orthonormality relation 
 \begin{equation} \label{orth1}
\sum_{\sigma}u_{\sigma,\gamma}u^{*}_{\sigma,\gamma'}=\delta_{\gamma,\gamma'}\;.
 \end{equation}

In order to formally show the equivalence of both approaches we 
 write the parameters $\lambda_{\Gamma,\Gamma'}$ in~(\ref{pi_one})
as 
\begin{eqnarray}\label{para12a} 
\lambda_{\emptyset}&=&\tilde{\lambda}_{\emptyset}\;,\\
\lambda_{d}&=&\tilde{\lambda}_{d}\;,\\\label{para12c} 
\lambda_{\sigma,\sigma'}&=&\sum_{\gamma}u_{\sigma,\gamma}
u^{*}_{\sigma',\gamma}\tilde{\lambda}_{\gamma}\;.
\end{eqnarray}  
 With these relations it is easy to show that the 
constraints~(\ref{con1a})-(\ref{con1c}) are indeed fulfilled. 
For example, the first 
constraint~(\ref{con1a}) can be written as 
  \begin{equation}\label{con67} 
 1=\lambda^2_{\emptyset}m^0_{\emptyset,\emptyset}+
\lambda^2_{d}m^0_{d,d}
+\sum_{\sigma,\sigma',\sigma^{\prime\prime}}
\lambda^*_{\sigma^{\prime\prime},\sigma}
\lambda_{\sigma^{\prime\prime},\sigma'}m^{0}_{\sigma,\sigma'}\;.
 \end{equation}
If we use 
 \begin{equation} 
m^{0}_{\sigma,\sigma'}=\sum_{\gamma}u^{*}_{\sigma,\gamma}
u_{\sigma',\gamma}m^{0}_{\gamma}
 \end{equation}
and the orthonormality relation~(\ref{orth1})
we readily find that~(\ref{con67}) is indeed solved by the 
 parameters~(\ref{para12a})-(\ref{para12c}). In the same way one can show 
 the equivalence of the ground-state energy functionals.

\section{The Time-Dependent Hartree-Fock Approximation}\label{chap7.3.2}
The approximation most frequently applied to two-particle Green's functions 
 is the `random-phase approximation' (RPA). This approach can be derived 
in various ways, e.g., by an equation of motion technique or in diagrammatic
 perturbation theory \cite{buenemann2001b}. 
In this section, we use a different derivation which introduces the RPA as
  a time-dependent generalisation of the Hartree-Fock theory; see, e.g., 
Refs.\ \cite{ring1980,blaizot1986}. If derived 
 in this way, the approach can be generalised quite naturally in order 
 to formulate a time-dependent Gutzwiller theory. This will be the subject
 of chapter~\ref{chap7.3.3}.

\subsection{The Hartree-Fock approximation}\label{hfa}
In the Hartree-Fock approximation a single-particle product wave 
function $|\Psi_0\rangle$ is used in order to investigate the 
ground-state properties of a many-particle 
 system. Note that such wave functions are included in the Gutzwiller 
 variational space by setting $\lambda_{i;\Gamma,\Gamma'}=\delta_{\Gamma,\Gamma'}$.
The   expectation value of a many-particle Hamiltonian with respect to a 
Hartree-Fock wave function is a function of the single-particle
 density matrix. For example, for the Hamiltonian~(\ref{1.1}) it reads
 \begin{eqnarray}\label{56781} 
E^{\rm HF}(\tilde{\rho})&\equiv&\langle \hat{H}  \rangle_{\Psi_0}\\\nonumber
&=&\sum_{i \ne j;\sigma,\sigma'}t_{i,j}^{\sigma,\sigma'}
\rho_{(j\sigma'),(i\sigma)}
+\sum_{i;\sigma_1,\sigma_2}\epsilon^{\sigma_1,\sigma_2}_i
\rho_{(i\sigma_2),(i\sigma_1)}+\sum_i E^{\rm HF}_{\rm loc,i}(\tilde{\rho})
 \end{eqnarray}
where
\begin{equation} \label{10.780}
\rho_{(j\sigma'),(i\sigma)}\equiv 
\langle \hat{c}_{i,\sigma}^{\dagger}
\hat{c}_{j,\sigma'}^{\phantom{+}}   \rangle_{\Psi_0}
\end{equation} 
are the elements of the single-particle density matrix $\tilde{\rho}$
 and 
\begin{equation}
 E^{\rm HF}_{\rm loc,i}(\tilde{\rho})=\frac{1}{2}\sum_{\sigma_1,\sigma_2,\atop\sigma_3,\sigma_4}
U_{i}^{\sigma_1,\sigma_2,\sigma_3,\sigma_4}\Big [\rho_{(i\sigma_4),(i\sigma_1)}\rho_{(i\sigma_3),(i\sigma_2)}-\rho_{(i\sigma_3),(i\sigma_1)}\rho_{(i\sigma_4),(i\sigma_2)}\Big] 
\end{equation} 
is the expectation value of the Coulomb interaction in the Hamiltonian 
(\ref{1.2}). Note that it will be more convenient both in the
 time-dependent Hartree-Fock and Gutzwiller theory, 
to use a different order 
 of subscripts in the definition~(\ref{10.780}) of density matrices 
than, e.g., in section~\ref{ven} or in previous work on the Gutzwiller theory. 

To keep notations simple, we use the abbreviations  
 $\upsilon\equiv (i,\sigma)$ for local single-particle states and 
 $Y=(\upsilon,\upsilon')$ for pairs of these indices. For example,
 the elements of $\tilde{\rho}$ can then be written as
\begin{equation}
\rho_{Y}=\rho_{\upsilon_1,\upsilon_2}=\rho_{(i_1,\sigma_1),(i_2,\sigma_2)}\;.
\end{equation}  
 With these new notations, the
Hartree-Fock energy~(\ref{56781}) reads 
\begin{eqnarray}\nonumber
 E^{\rm HF}(\tilde{\rho})&=&\sum_{\upsilon_1,\upsilon_2}
\varepsilon_{\upsilon_1,\upsilon_2}\rho_{\upsilon_2,\upsilon_1}+
\frac{1}{2}\sum_{\upsilon_1,\upsilon_2\atop{\upsilon_3,\upsilon_4}}
\rho_{\upsilon_4,\upsilon_1} W_{(\upsilon_1,\upsilon_4),(\upsilon_3,\upsilon_2)}
\rho_{\upsilon_3,\upsilon_2}\\\label{5923}
&=&\sum_{Y}\varepsilon_{Y}\rho_{\bar{Y}}
+\frac{1}{2}\sum_{Y,Y'}\rho_{\bar{Y}}W_{Y,Y'}\rho_{Y'}
\end{eqnarray} 
 where 
\begin{equation}
\varepsilon_{(i\sigma_1),(j\sigma_2)}\equiv t_{i,j}^{\sigma_1,\sigma_2}
+\delta_{i,j}\epsilon^{\sigma_1,\sigma_2}_i
\end{equation} 
and 
\begin{equation}
W_{(\upsilon_1,\upsilon_4),(\upsilon_3,\upsilon_2)}\equiv
U_i^{\sigma_1,\sigma_2,\sigma_3,\sigma_4}-U_i^{\sigma_1,\sigma_2,\sigma_4,\sigma_3}
\end{equation} 
for indices $\upsilon_{k}=(i,\sigma_k)$ that belong to the same lattice site 
$i$. Further, we introduced  the `inverse' index 
$\bar{Y}\equiv(\upsilon',\upsilon)$ for
  $Y=(\upsilon,\upsilon')$.
Note the symmetries 
\begin{equation}\label{sym}
W_{(\upsilon_1,\upsilon_4),(\upsilon_3,\upsilon_2)}
=W_{(\upsilon_2,\upsilon_3),(\upsilon_4,\upsilon_1)}
=-W_{(\upsilon_1,\upsilon_3),(\upsilon_4,\upsilon_2)}\;,
\end{equation}
which will be employed in the following section.

The energy functional~(\ref{5923}) has to be minimised with respect to all 
 density matrices which belong to a single-particle product state. Such 
matrices are idempotent, i.e., they obey the matrix equation
\begin{equation}\label{rt56}
\tilde{\rho}^2=\tilde{\rho}\;.
\end{equation}
If one imposes this constraint via a Lagrange parameter matrix 
 $\tilde{\eta}$ with elements
$\eta_{\upsilon,\upsilon'}$, the following equation has to be solved
\begin{eqnarray}\label{pw56}
\frac{\partial}{\partial \rho_{\upsilon,\upsilon'}}
\left[ 
 E^{\rm HF}(\tilde{\rho})-{\rm tr} \Big( 
 \tilde{\eta}(  \tilde{\rho}^2-\tilde{\rho}    )   \Big)
 \right]=0\:.
\end{eqnarray}
This condition leads to
\begin{equation}\label{we3}
\tilde{h}(\tilde{\rho})+\tilde{\eta}-\tilde{\eta}\tilde{\rho}
-\tilde{\rho}\tilde{\eta}=0
\end{equation} 
where we introduced the matrix $\tilde{h}(\tilde{\rho})$ with the elements
\begin{equation}\label{we32}
h_{Y}(\tilde{\rho})=\frac{\partial}{\partial \rho_{\bar{Y} }}
E^{\rm HF}(\tilde{\rho})=\varepsilon_{Y}+
\sum_{Y'}W_{Y,Y'}\rho_{Y'}\;.
\end{equation} 
Equation~(\ref{we3}) is solved if $\tilde{\rho}$ satisfies both 
(\ref{rt56}) and
\begin{equation}\label{ui6}
[ \tilde{h}(\tilde{\rho}), \tilde{\rho} ]=0\;.
\end{equation} 
Starting with a certain density matrix $\tilde{\rho}$  we can
 introduce the `Hartree-Fock' basis 
\begin{equation}\label{ui7}
\ket{\alpha}=\sum_{\upsilon} u_{\upsilon,\alpha}\ket{\upsilon}
\end{equation}
of  states which diagonalise the 
Hamilton matrix $\tilde{h}(\tilde{\rho})$, i.e.,
\begin{equation}\label{10.900} 
\sum_{\upsilon'} h_{\upsilon,\upsilon'}(\tilde{\rho})u_{\upsilon',\alpha}
=E_{\alpha} u_{\upsilon,\alpha}\;.
 \end{equation} 
Equation~(\ref{ui6}) is then solved by setting
\begin{equation}\label{10.890} 
\rho_{\alpha,\alpha'}=\delta_{\alpha,\alpha'}
\Theta(E_{\rm F}-E_{\alpha})
 \end{equation} 
where the Fermi energy $E_{\rm F}$ is determined by the total number
 of particles
\begin{equation}
N=\sum_{\alpha}\Theta(E_{\rm F}-E_{\alpha})\;.
\end{equation} 
The density matrix~(\ref{10.890}) has to be reinserted 
into~(\ref{5923}),(\ref{we32}) until
 self-consistency is reached. We denote the solution of these equations
 as $\tilde{\rho}^0$ and introduce the corresponding 
Hamilton matrix 
\begin{equation}
\tilde{h}^0\equiv \tilde{h}(\tilde{\rho}^0)\;.
\end{equation} 
\subsection{Equation of Motion for the Density Matrix}
We consider two-particle Green's functions of the form
\begin{eqnarray}\label{10.700}   
&&G_{(\upsilon_2,\upsilon_1),(\upsilon_3,\upsilon_4)}(t-t')
\equiv \langle \langle   \hcd_{\upsilon_1}(t)\hc_{\upsilon_2}(t); 
\hcd_{\upsilon_3}(t')\hc_{\upsilon_4}(t')  \rangle \rangle\\
\nonumber
&&\;\;\;\;\;\;\;\;\equiv-\rmi \Theta(t-t')
\langle\Phi_0|[\hcd_{\upsilon_1}(t)\hc_{\upsilon_2}(t),
\hcd_{\upsilon_3}(t')\hc_{\upsilon_4}(t')]|\Phi_0\rangle\;,
\end{eqnarray}
where $\ket{\Phi_0}$ is the exact ground state of our 
 multi-band Hubbard Hamiltonian~(\ref{1.1}), and 
$\hcdd_{\upsilon}(t)$ is the 
 Heisenberg representation of the operators 
$\hcdd_{\upsilon}$ with respect to $\hat{H}$. As shown in most textbooks 
 on many-particle physics,  the Green's 
functions~(\ref{10.700}) naturally 
 arise in `linear-response theory' because they describe the 
 time-dependent changes
 \begin{eqnarray}   \nonumber
\delta\langle \hcd_{\upsilon_1}\hc_{\upsilon_2}  \rangle_{t}&\equiv& 
\langle \hcd_{\upsilon_1}\hc_{\upsilon_2}  \rangle_{t}-
\langle \hcd_{\upsilon_1}\hc_{\upsilon_2}  \rangle_{-\infty}
\equiv\delta \rho_{\upsilon_2,\upsilon_1}(t)\\\label{10.710}
&=&\sum_{\upsilon_3,\upsilon_4}
\int_{-\infty}^{\infty}
 {\rm d}t'  G_{(\upsilon_2,\upsilon_1),(\upsilon_3,\upsilon_4)}(t-t') 
f_{\upsilon_3,\upsilon_4}(t')
\end{eqnarray}
of the density matrix $\tilde{\rho}$  
in the presence of a small time-dependent perturbation 
 \begin{equation}\label{10.720}   
\hat{V}_f(t)=\sum_{\upsilon,\upsilon'}f_{\upsilon,\upsilon'}(t)
\hcd_{\upsilon}\hc_{\upsilon'}
\end{equation}
added to $\hat{H}$ \cite{kubo1957,kubo1959,mahan2005}.
After a Fourier transformation and using again the abbreviation 
$Y=(\upsilon,\upsilon')$, Eq.\  (\ref{10.710})
 reads 
 \begin{equation}\label{10.730}   
\delta\rho_{Y}(\omega)
=\sum_{Y'}
G_{Y,Y'}(\omega)
f_{Y'}(\omega)
\end{equation}
with 
\begin{equation}\label{10.740}   
G_{Y,Y'}(\omega)\equiv 
\int^{\infty}_{-\infty} {\rm d}\tau \,
G_{Y,Y'}(\tau)
e^{\rmi \omega \tau}\;,
\end{equation}
and $f_{Y}(\omega)$ and 
$\delta\rho_{Y}(\omega)$ defined
 accordingly.

Ideally, we would like to calculate the time dependence of the 
density matrix 
  \begin{equation}   \label{10.780b}     
\rho_{\upsilon',\upsilon}(t)\equiv 
\langle \Psi(t)|
  \hcd_{\upsilon}\hc_{\upsilon'} |
  \Psi(t)\rangle\;,
 \end{equation} 
where $\ket{\Psi(t)}$ is the exact solution of the time-dependent 
Schr\"odinger equation for the Hamiltonian
\begin{equation} \label{10.785}     
\hat{H}(t)=\hat{H}+\hat{V}_f(t)\;.
  \end{equation} 
The expectation value~(\ref{10.780b}) obeys the Heisenberg equation
 \begin{equation} \label{10.790}     
- \rmi  \dot{\rho}_{\upsilon',\upsilon}(t)=
 \langle \Psi(t)|[\hat{H},\hcd_{\upsilon}\hc_{\upsilon'}] |
  \Psi(t)\rangle\;,
 \end{equation} 
which contains the  commutator 
  \begin{eqnarray}  \label{10.800} 
 &&[\hat{H}(t),\hcd_{\upsilon}\hc_{\upsilon'}]=\sum_{\upsilon_1}
\big(\varepsilon_{\upsilon_1,\upsilon}+f_{\upsilon_1,\upsilon}(t)\big)
\hcd_{\upsilon_1}\hc_{\upsilon'}-\sum_{\upsilon_1}\big(\varepsilon_{\upsilon',\upsilon_1}+f_{\upsilon',\upsilon_1}(t)\big)
\hcd_{\upsilon}\hc_{\upsilon_1}\\ \nonumber 
&&+\frac{1}{2}\sum_{\upsilon_1,\upsilon_2,\upsilon_3}\left(
W_{(\upsilon_1,\upsilon_3),(\upsilon,\upsilon_2)}
\hcd_{\upsilon_1}\hcd_{\upsilon_2}\hc_{\upsilon'}\hc_{\upsilon_3}
+W_{(\upsilon_1,\upsilon_2),(\upsilon_3,\upsilon')}
\hcd_{\upsilon_1}\hcd_{\upsilon}\hc_{\upsilon_2}\hc_{\upsilon_3}\right)
\;.
\end{eqnarray} 

In the time-dependent Hartree-Fock approximation, it is assumed 
 that the solution $\ket{\Psi(t)}$ of the Schr\"odinger 
equation at any time $t$ is approximately given by a 
single-particle product wave function. 
 In this case, the expectation value of the commutator 
(\ref{10.800}) can be evaluated by means of Wick's theorem. This leads to
the equation of motion 
\begin{equation}\label{10.810}    
 \rmi  \dot{\tilde{\rho}}(t)=
[\tilde{h}(\tilde{\rho}(t))+\tilde{f}(t),\tilde{\rho}(t)]
\end{equation} 
for $\tilde{\rho}(t)$, where the matrix  $\tilde{h}(\tilde{\rho})$ 
 has been introduced in~(\ref{we32}). 
Equations~(\ref{we32}) and~(\ref{10.810}) will be crucial also for 
our formulation of a 
time-dependent Gutzwiller theory in chapter~\ref{chap7.3.3}.

\subsection{Expansion for Weak Perturbations}
We are only interested in cases  where 
\begin{equation}
\hat{V}_f(t)\rightarrow \delta\hat{V}_f(t)
=\sum_{\upsilon,\upsilon'}\delta f_{\upsilon,\upsilon'}(t)
\hcd_{\upsilon}\hc_{\upsilon'}
 \end{equation} 
 is  a weak perturbation 
 to the time-independent 
Hamiltonian $\hat{H}$. In this case, the density matrix 
$\tilde{\rho}(t)$ and the  Hamilton matrix $\tilde{h}(t)$  
 are given as  
\begin{eqnarray}\label{10.860a}  
  \tilde{\rho}(t)&\approx& \tilde{\rho}^0+\delta  \tilde{\rho}(t)\;,\\\label{10.860b}  
\tilde{h}(t)&\approx&\tilde{h}^0+\delta  \tilde{h}(t)\;,
  \end{eqnarray}  
where $\delta  \tilde{\rho}(t)$ describes a `small' time-dependent 
 perturbation around the ground-state density matrix $\tilde{\rho}^0$, and 
\begin{eqnarray}
h^{0}_{Y}&=&\varepsilon_{Y}+
\sum_{Y'}W_{Y,Y'}
\rho^0_{Y'}\;,\\
\delta h_{Y}(t)&=&\sum_{Y'}W_{Y,Y'}
\delta \rho_{Y'}(t)\;.
\end{eqnarray}   
With the expansion~(\ref{10.860a})-(\ref{10.860b}), the equation of
 motion~(\ref{10.810}) 
becomes
\begin{eqnarray}\label{10.870a}  
 0&=&[\tilde{h}^0,\tilde{\rho}^0]\;,\\\label{10.870b}  
\rmi  \delta\dot{\tilde{\rho}}(t) &=&[\tilde{h}^0,\delta\tilde{\rho}(t)]
+[\delta \tilde{h}(t)+\delta\tilde{f}(t),\tilde{\rho}^0]\;.
\end{eqnarray}
These equations have to be solved for density matrices 
$\tilde{\rho}(t)$
that obey the matrix equation~(\ref{rt56}).
After applying the expansion~(\ref{10.860a}), 
Eq.\  (\ref{rt56})
reads (to leading order in $\delta \tilde{\rho}(t)$) 
\begin{eqnarray}\label{10.885a}  
\tilde{\rho}^0&=&\left(\tilde{\rho}^0\right)^2\;,\\\label{10.885b} 
\delta \tilde{\rho}(t)&=&\tilde{\rho}^0 \delta \tilde{\rho}(t)+
\delta \tilde{\rho}(t)\tilde{\rho}^0\;.
\end{eqnarray}  
Note that Eqs.\  (\ref{10.870a}),(\ref{10.885a})  
just recover the time-independent 
Hartree-Fock equations derived in section~\ref{hfa}.  
\subsection{Random -phase approximation (RPA) equations}
Mathematically, the density matrix 
is a projector onto `hole'-states, $\tilde{\rho}_{\rm h}\equiv\tilde{\rho}^0$. 
In addition, we define the  projector onto `particle'-states as
\begin{equation}\label{10.910} 
\tilde{\rho}_{\rm p}\equiv 1-\tilde{\rho}^0\;.
\end{equation} 
With these two operators,  
we can decompose all matrices into their four components 
\begin{eqnarray}\label{10.920a} 
\delta  \tilde{\rho}^{ vw}(t)&\equiv& \tilde{\rho}_{ v} 
\delta  \tilde{\rho}(t)
\tilde{\rho}_{w}\;,\\\label{10.920b} 
\delta \tilde{f}^{ vw}(t)&\equiv&\tilde{\rho}_{ v} \delta \tilde{f}(t)\tilde{\rho}_{ w}\;,\\\label{10.920c} 
 \tilde{h}^{0;vw}&\equiv&\tilde{\rho}_{ v}  \tilde{h}^0\tilde{\rho}_{ w}\;,
\end{eqnarray} 
where $v,w\in \{{\rm p,h } \}$. Note that $ \tilde{h}^{0;vw}$ has the elements 
\begin{equation}
h^{0;vw}_{\alpha,\alpha'}=\delta_{v,w}\delta_{\alpha,\alpha'}E_{\alpha}\;.
\end{equation} 
An evaluation of the condition~(\ref{10.885b}) for the components 
$\delta  \tilde{\rho}^{vw}(t)$ yields
\begin{equation}\label{10.930} 
\delta  \tilde{\rho}^{ vw}(t)=\delta  \tilde{\rho}^{ vw}(t)
+\delta\tilde{\rho}^{ vv}(t)\delta \tilde{\rho}^{ vw}(t)+
\delta\tilde{\rho}^{ vw}(t) \delta\tilde{\rho}^{ ww}(t)\;,
\end{equation}
 and
\begin{equation} \label{10.940} 
\delta  \tilde{\rho}^{ww}(t)=0
 \end{equation}  
where $v \neq w$. Hence  the components 
  $\delta  \tilde{\rho}^{\rm pp}(t)$ and $\delta  \tilde{\rho}^{\rm hh}(t)$
  can be neglected in~the following compared to the leading 
fluctuations $\delta  \tilde{\rho}^{\rm hp}(t)$
 and  $\delta  \tilde{\rho}^{\rm ph}(t)$.  

 We express the time-dependent quantities $\delta\tilde{\rho}^{vw}(t)$ and  
$\delta\tilde{f}^{vw}(t)$ by their respective Fourier transforms 
$\delta\tilde{\rho}^{vw}(\omega)$  and $\delta\tilde{f}^{vw}(\omega)$.
The equation of motion~(\ref{10.870b})  then leads to
\begin{equation}\label{oitde}
+\omega \delta \rho_{\alpha_1,\alpha_2}^{vw}(\omega)
=(E_{\alpha_1}-E_{\alpha_2})
 \delta \rho_{\alpha_1,\alpha_2}^{vw}(\omega)\pm 
(\delta h_{\alpha_1,\alpha_2}^{vw}(\omega)
+ \delta f^{vw}_{\alpha_1,\alpha_2}(\omega))
\end{equation}
 where the plus and minus signs correspond to $vw={\rm ph}$ and 
$vw={\rm hp}$, respectively. With the abbreviation 
$A=(\alpha_1,\alpha_2)$ for pairs of indices $\alpha$ we find 
\begin{equation}\label{oitde2}
\delta h_{A}^{vw}(\omega)=-\sum_{A'}U_{A,A'}(\delta \rho_{A'}^{vw}(\omega)+
\delta \rho_{A'}^{wv}(\omega))\;.
\end{equation}
Here, the elements of the matrix $\tilde{U}$  are given as 
\begin{equation}  \label{10.960} 
U_{A,A'}=U_{(\alpha_1,\alpha_2),(\alpha'_1,\alpha'_2)}
\equiv-\sum_{\upsilon_1,\upsilon_2,\atop \upsilon'_1,\upsilon'_2}
u^{*}_{\upsilon_1,\alpha_1}u^{}_{\upsilon_2,\alpha_2}
W_{(\upsilon_1,\upsilon_2),(\upsilon'_1,\upsilon'_2)}
u^{}_{\upsilon'_1,\alpha'_1}u^{*}_{\upsilon'_2,\alpha'_2}\;.
\end{equation}
 The coefficients $u_{\upsilon,\alpha}$ in~(\ref{10.960})
  have been introduced in Eq.~(\ref{ui7}) and
 determine the solutions $\ket{\alpha}$ of the Hartree-Fock equations. 
Equations~(\ref{oitde}) and~(\ref{oitde2}) then yield
\begin{equation}  \label{10.950} 
\left[
(\omega - \tilde{E})
\left(
\begin{array}{cc}
1&0\\
0&-1
\end{array}
\right)
+\tilde{U}
\right]
\left(
\begin{array}{c}
\delta \tilde{\rho}^{\rm ph}(\omega)\\
\delta \tilde{\rho}^{\rm hp}(\omega)
\end{array}
\right)=
\left( 
\begin{array}{c}
\delta \tilde{f}^{\rm ph}(\omega)\\
\delta \tilde{f}^{\rm hp}(\omega)
\end{array}
\right)\;.
\end{equation}
with a matrix $\tilde{E}$ defined as 
\begin{equation}  \label{10.970} 
E_{A,A'}=E_{(\alpha_1,\alpha_2),(\alpha'_1,\alpha'_2)}
=\delta_{\alpha_1,\alpha'_1}
\delta_{\alpha_2,\alpha'_2}
(E_{\alpha_1}-E_{\alpha_2})\;.
 \end{equation}
 By comparing Eqs.\  (\ref{10.950}) and~(\ref{10.730}) we find 
\begin{equation}\label{10.980} 
\tilde{G}^{-1}(\omega)=\left[
(\omega+\rmi \delta  - \tilde{E})
\left(
\begin{array}{cc}
1&0\\
0&-1
\end{array}
\right)
+\tilde{U}
\right]
 \end{equation}
for the inverse of the two-particle Green's function
\begin{eqnarray}\nonumber
G_{A,A'}(\omega)&=&
G_{(\alpha_1,\alpha_2),(\alpha'_1,\alpha'_2)}(\omega)\\
&=&\sum_{\upsilon_1,\upsilon_2,\upsilon'_1,\upsilon'_2}
u^{}_{\upsilon_1,\alpha_1}u^{*}_{\upsilon_2,\alpha_2}
G_{(\upsilon_1,\upsilon_2),(\upsilon'_1,\upsilon'_2)}(\omega)
u^{*}_{\upsilon'_1,\alpha'_1}u^{}_{\upsilon'_2,\alpha'_2}\;.
\end{eqnarray}
 Here we have added an increment $\rmi \delta $ with 
$\delta=0^+$  in order to ensure the correct boundary conditions of
 a retarded Green's function.
 For $\tilde{U}=0$, the inverse Green's function~(\ref{10.980}) reads
\begin{equation}\label{10.990} 
\tilde{\Gamma}^{-1}(\omega)\equiv\pm (\omega+\rmi \delta  -\tilde{E})
 \end{equation}
which leads to
\begin{equation}\label{10.1000} 
\Gamma_{A,A'}(\omega)
=\Gamma_{(\alpha_1,\alpha_2),(\alpha'_1,\alpha'_2)}(\omega)
=\delta_{\alpha_1,\alpha'_1}\delta_{\alpha_2,\alpha'_2}
\frac{\rho^0_{\alpha_2,\alpha_2}-\rho^0_{\alpha_1,\alpha_1} }
{\omega -(E_{\alpha_1}-E_{\alpha_2})+\rmi \delta}\;.
 \end{equation}
Note that $\tilde{\Gamma}$ is not the exact Green's function for
 the single-particle Hamiltonian $\hat{H}_{0}$ since 
 we just set $\tilde{U}=0$ in~(\ref{10.980}),
  but kept finite the `Hartree-Fock self-energy' contributions 
\begin{equation}
\Sigma_{A}\equiv\sum_{A'}
W_{A,A'}\rho^0_{A'}
 \end{equation}
which usually change the `eigenvalues' $E_{\alpha}$ 
in~(\ref{10.1000}); c.f. Eqs.\ (\ref{we32}) and~(\ref{10.900}). 

With the Green's function~(\ref{10.1000})  we can write~(\ref{10.980})  as
\begin{eqnarray}\label{10.1010a} 
\tilde{G}(\omega)&=&\tilde{\Gamma}(\omega)[1+\tilde{U}\tilde{\Gamma}(\omega)]^{-1}\\\label{10.1010b} 
&=&\tilde{\Gamma}(\omega)+\tilde{\Gamma}(\omega)\tilde{U}\tilde{G}(\omega)
 \end{eqnarray} 
where, in the second line, we expanded 
$[1+\tilde{U}\tilde{\Gamma}(\omega)]^{-1}$ into a power series with respect to 
 $\tilde{U}\tilde{\Gamma}$. Both Eqs.\  (\ref{10.1010a}),(\ref{10.1010b})  are 
familiar expressions for the two-particle Green's function in 
the random-phase approximation. 

\section{Time-Dependent Gutzwiller Theory}\label{chap7.3.3}
The time-dependent Gutzwiller approximation has been first introduced 
 for single-band Hubbard models by Seibold et al. 
\cite{seibold1998,seibold2001}. 
In this section, we generalise this approach for the investigation of 
  multi-band models.  To this end, we set up an effective 
 energy functional of the density matrix in section~\ref{jkio1},
 which is used in sections~\ref{jkio4}-\ref{lpe} to derive 
the Gutzwiller RPA equations. 

\subsection{Effective Energy Functional}\label{jkio1}
As summarised in chapter~\ref{ven}, the expectation value
 of the multi-band Hamiltonian~(\ref{1.1})  
in the Gutzwiller theory  is a function of the variational 
 parameters $\lambda_{\Gamma,\Gamma'}$ and of the one-particle
 wave function $|\Psi_0\rangle$. 
 Like in the Hartree-Fock theory, the single-particle wave function 
$|\Psi_0\rangle$ enters the  energy functional solely through the 
elements~(\ref{10.780}) of the non-interacting density matrix $\tilde{\rho}$.
 It is therefore possible to consider the energy 
 \begin{equation}\label{10.1030} 
E=E(\vec{\lambda},\tilde{\rho})
 \end{equation}
  as a function of the density matrix $\tilde{\rho}$ and of the  
 `vector' 
\begin{equation}\label{10.1030b} 
\vec{\lambda}=(\{\lambda^*_{\Gamma,\Gamma'} \},\{\lambda_{\Gamma,\Gamma'} \} )
=(\lambda_1,\ldots,\lambda_{n_{\rm p}})
\end{equation} 
of $n_{\rm p}$ variational parameters $\lambda_{\Gamma,\Gamma'}$ 
(and $\lambda^*_{\Gamma,\Gamma'}$ for $\Gamma\neq \Gamma'$). 
The density matrix in the energy functional~(\ref{10.1030}) must be 
 derived from a single-particle wave function and, therefore, it has to obey 
the condition~(\ref{rt56}).
 Note that, in the following considerations,  the 
  density matrix will either be considered as a matrix (with respect
 to its two indices $(i,\sigma)$ and $(j,\sigma')$) or as a vector 
  (with respect to its single index $Y$). To distinguish both cases,
  we will denote the density matrix $\tilde{\rho}$ in some equations  
 as $\vec{\rho}$
 in order to indicate its vector interpretation.        

The constraints~(\ref{1.10c})-(\ref{1.10d}) are also 
 functions of $\vec{\lambda}$ and $\vec{\rho}$ and will be denoted as 
\begin{equation}\label{const}
g_{n}(\vec{\lambda},\vec{\rho})=0\;\;\;,\;\;\;1\leq n \leq n_{\rm c}\;.
\end{equation}
Here, $n_{\rm c}$ is the (maximum) number of independent 
constraints, which, due to symmetries, is usually smaller than its 
maximum value $N_{\rm so}^2+1$, where~$N_{\rm so}$ is the number of 
spin-orbital states per lattice site. We assume that the 
 functions (\ref{const}) are real, i.e., in case of complex
 equations (\ref{1.10a})-(\ref{1.10b}) 
 their real and imaginary parts are treated separately.

By solving Eqs.\  (\ref{const}) we can, at least in principle, 
 express $n_{\rm c}$ of the variational parameters ($\equiv\lambda^{\rm d}_{X}$) 
 through the density matrix $\rho_Y$ and the remaining `independent'
 parameters ($\equiv \lambda^{\rm i}_{Z}$),
\begin{equation}\label{2eys}
\lambda^{\rm d}_{X}=\lambda^{\rm d}_{X}(\vec{\lambda}^{\rm i},\vec{\rho})\;.
\end{equation} 
 In this way, we obtain an energy 
functional 
\begin{equation}\label{1uyr}
E^{\rm GA}(\vec{\lambda}^{\rm i},\vec{\rho})\equiv
E(\vec{\lambda}^{\rm d}(\vec{\lambda}^{\rm i},\vec{\rho}),\vec{\lambda}^{\rm i},\vec{\rho})\;.
\end{equation}
which has to be minimised without constraints apart from Eq.\  (\ref{rt56}) 
and the condition that the total particle number
\begin{equation}
N=\sum_\upsilon \rho_{\upsilon,\upsilon}
\end{equation}
is conserved. 

For a fixed density matrix $\tilde{\rho}$, the minimisation of 
(\ref{1uyr}) with respect to the parameters 
$ \lambda^{\rm i}_{Z}$,
\begin{equation}\label{9retww}
\frac{\partial}{\partial \lambda^{\rm i}_{Z}}
E^{\rm GA}(\vec{\lambda}^{\rm i},\vec{\rho})
=0\;,
\end{equation}
determines these parameters
\begin{equation} \label{9yeu}
\vec{\lambda}^{\rm i}=\vec{\lambda}^{\rm i}(\vec{\rho})
\end{equation}
as a function of $\vec{\rho}$. This allows us to define the `effective'
 energy functional 
\begin{equation}\label{e3445}
 E^{\rm eff}(\vec{\rho})=E^{\rm GA}(\vec{\lambda}^{\rm i}
(\vec{\rho}),\vec{\rho})
\end{equation}
which, for a fixed density matrix  $\vec{\rho}$, is given as the 
 minimum of  $E^{\rm GA}$ with respect to $\vec{\lambda}^{\rm i}$. 
With this  effective functional we will formulate 
 the time-dependent Gutzwiller theory in the following 
 section.

 Using a Lagrange-parameter matrix $\tilde{\eta}$ as in chapter \ref{hfa},
 we find 
\begin{equation}\label{pw56b}
\left.\frac{\partial}{\partial \rho_{\upsilon,\upsilon'}}
\left[ 
E^{\rm eff}(\tilde{\rho})
-{\rm tr} \Big( 
 \tilde{\eta}(\tilde{\rho}^2-\tilde{\rho})\Big)
 \right]\right|_{\tilde{\rho}=\tilde{\rho}^0}=0
\end{equation}  
 which leads to 
\begin{equation}\label{jjj1}
0=[\tilde{h}(\tilde{\rho}),\tilde{\rho}]\;.
\end{equation}  
Here we introduced the matrix $\tilde{h}(\tilde{\rho})$ with the elements
\begin{equation}\label{hmat2}
h_{Y}(\tilde{\rho})=\frac{\partial E^{\rm eff}(\tilde{\rho})}{\partial \rho_{\bar{Y}}}
\;.
\end{equation}
and used again the notation $\bar{Y}\equiv(j,\sigma';i,\sigma)$ for 
  $Y=(i,\sigma;j,\sigma')$. The self-consistent solution of Eqs.\ 
 (\ref{jjj1})-(\ref{hmat2}) then yields the ground-state density matrix 
$\tilde{\rho}^0$, the matrix $\tilde{h}^0\equiv \tilde{h}(\tilde{\rho}^0)$,
 and the corresponding single-particle 
`Gutzwiller-Hamiltonian'
\begin{equation}\label{hop}
\hat{h}^0\equiv\sum_{i,j;\sigma,\sigma'}h^0_{i,\sigma;j,\sigma'}
\hat{c}_{i,\sigma}^{\dagger}\hat{c}_{j,\sigma'}^{\phantom{+}}\;.
\end{equation} 
  
\subsection{Gutzwiller RPA Equations}\label{jkio4}
The derivation of RPA-type equations within the time-dependent Gutzwiller
 theory goes along the same lines as discussed in chapter~\ref{chap7.3.2} for
 the time-dependent Hartree-Fock theory. We add a small time-dependent field 
\begin{equation}\label{10.1090} 
\delta \hat{V}_f(t)=\sum_{i,j;\sigma,\sigma'}\delta f^0_{i,\sigma;j,\sigma'}(t)
\hat{c}_{i,\sigma}^{\dagger}\hat{c}_{j,\sigma'}^{\phantom{+}}+{\rm h.c.}
\end{equation}
to our multi-band Hamiltonian~(\ref{1.1}).
With the particular time dependence 
\begin{equation}
\delta f^0_{i,\sigma;j,\sigma'}(t)=\delta \tilde{f}^0_{i,\sigma;j,\sigma'}(\omega)
e^{-{\rm i}\omega t}
\end{equation}
the expectation value of $\delta \hat{V}(t)$ reads 
\begin{equation}\label{10.1100} 
E_f(\tilde{\rho})=\sum_{i,j;\sigma,\sigma'}\delta \tilde{f}_{i,\sigma;j,\sigma'}(\omega)
e^{-{\rm i}\omega t}\rho_{j,\sigma';i,\sigma}+{\rm c.c.}\;,
\end{equation}
where 
\begin{eqnarray}\label{10.1110a} 
\delta \tilde{f}_{i,\sigma_1;j,\sigma_2}(\omega)&=&
\delta_{i,j}\delta \tilde{f}^0_{i,\sigma_1;i,\sigma_2}(\omega)
\frac{C^{\rm c}_{i,\sigma_1;i,\sigma_2}}
{\rho_{i,\sigma_2;i,\sigma_1}}\\\nonumber
&&+(1-\delta_{i,j})\sum_{\sigma'_1,\sigma'_2}
\delta f^0_{i,\sigma'_1;j,\sigma'_2}(\omega)
q_{\sigma'_1}^{\sigma_1}
\Big( q_{\sigma'_2}^{\sigma_2}\Big)^{*}\;.
\end{eqnarray}
The renormalisation matrix $\tilde{q}$ and the (correlated) local
 density matrix $\tilde{C}^{\rm c}$ are defined in equations 
(\ref{qmat}) and~(\ref{cmatc}), respectively.  
With Eq.\ (\ref{9yeu}) they can both be considered as functions of 
$\tilde{\rho}$. 

The time-dependent field induces small fluctuations of the density matrix, 
\begin{equation}\label{10.1120} 
\rho_Y=\rho_Y^{0}+\delta \rho_Y(t)\;.
\end{equation}  
Our main assumption is now that $\delta \rho_Y(t)$ obeys the same 
equation of motion,
\begin{equation}\label{10.1230} 
\rmi  \delta\dot{\tilde{\rho}}(t) =[\tilde{h}^0,\delta\tilde{\rho}(t)]
+[\delta \tilde{h}(t)+\delta\tilde{f}(t),\tilde{\rho}^0]\;,
\end{equation}
as the density matrix in the  time-dependent Hartree-Fock theory; see 
Eq.\  (\ref{10.870b}). Here, however, the 
Hamilton matrix 
\begin{equation}
\tilde{h}(t)\approx \tilde{h}^0(t)+\delta \tilde{h}(t) 
\end{equation}
is  not derived from the  Hartree-Fock functional~(\ref{5923}), but 
from the effective energy functional~(\ref{e3445}),
 \begin{equation}\label{10.1220} 
h_{Y}(t)=\frac{\partial }{\partial \rho_{\bar{Y}}}E^{\rm eff}(\tilde{\rho})
\approx h^0_{Y}+\sum_{Y'}K_{Y,Y'}\delta \rho_{Y'}(t)
\equiv h^0_{Y}+\delta h_{Y}(t)
\;,
\end{equation}
where the matrix $\tilde{K}$ is given as
 \begin{equation}\label{9dgf}
\tilde{K}_{Y,Y'}\equiv \left .\frac{\partial^2 E^{\rm eff}}
{\partial \rho_{\bar{Y}} \partial \rho_{Y'}}\right|_{\tilde{\rho}=\tilde{\rho}^0}\;.
\end{equation}

 The diagonalisation of $\tilde{h}^0$ 
(or equivalently of the Gutzwiller Hamiltonian $\hat{h}^0$)
yields  a basis $\ket{\alpha}$ with
\begin{equation} 
h^0_{\alpha,\alpha'}=h^0_{A}=\delta_{\alpha,\alpha'}E_{\alpha}
 \end{equation}
and a ground-state density matrix that is given as
\begin{equation}\label{10.1240}  
\rho^0_{\alpha,\alpha'}=\rho^0_{A}=\delta_{\alpha,\alpha'}
\Theta(E_{\rm F}-E_{\alpha})\;.
 \end{equation} 
With the projectors $\tilde{\rho}_{\rm h}\equiv\tilde{\rho}^0$ and
  $\tilde{\rho}_{\rm p}\equiv 1-\tilde{\rho}^0$, we define the particle 
 and hole components  of all matrices, as we did in Eqs.\  (\ref{10.920a})-(\ref{10.920c}). 
  The components $\delta\tilde{\rho}^{vw}(t)$ of the density-matrix 
 fluctuations obey Eqs.\  (\ref{10.930})-(\ref{10.940}), i.e., to leading order we can 
neglect $\delta\tilde{\rho}^{\rm hh}(t)$ and $\delta\tilde{\rho}^{pp}(t)$. 
Hence, after a Fourier transformation we end up with the same form 
of RPA equations,
\begin{equation} \label{10.1250} 
\left[
(\omega - \tilde{E})
\left(
\begin{array}{cc}
1&0\\
0&-1
\end{array}
\right)
+\tilde{K}
\right]
\left(
\begin{array}{c}
\delta\tilde{\rho}^{\rm ph}(\omega)\\
\delta\tilde{\rho}^{\rm hp}(\omega)
\end{array}
\right)=
 \left( 
\begin{array}{c}
\delta \tilde{f}^{\rm ph}(\omega)\\
\delta \tilde{f}^{\rm hp}(\omega)
\end{array}
\right)
\end{equation}
as in Eq.\  (\ref{10.950}). Here, however, the bare matrix of Coulomb 
parameters $\tilde{U}$ is replaced by the matrix $\tilde{K}$, 
defined in~(\ref{9dgf}),
 and the energies $E_\alpha$ in the matrix $\tilde{E}$, Eq.\  (\ref{10.970}), are the  eigenvalues of the 
Gutzwiller Hamiltonian~(\ref{hop}). 
The comparison with~(\ref{10.730}) leads to the final result 
\begin{equation}\label{10.1260} 
\tilde{G}(\omega)\equiv\left[
(\omega+\rmi \delta  - \tilde{E})
\left(
\begin{array}{cc}
1&0\\
0&-1
\end{array}
\right)
+\tilde{K}
\right]^{-1}\;.
 \end{equation}
for the two-particle Green's function matrix within the time-dependent 
 Gutzwiller approximation. 

One should keep in mind
that the external `fields'  $\delta \tilde{f}(\omega)$ in 
(\ref{10.1250}) are `renormalised', i.e., they are not the bare fields 
as they appear in~(\ref{10.730}), see Eq.\  (\ref{10.1110a}).
 On the other hand, on the l.h.s.\ of 
(\ref{10.730}) appears the `correlated' expectation value of the
 density matrix, while in~(\ref{10.1250}) we work with the fluctuations 
 of the uncorrelated density matrix. Therefore, the `true' Green's function  
 seen in experiments may, in fact, be given as   
\begin{equation}\label{10.1270} 
\underline{G}_{Y,Y'}(\omega)=c_{Y,Y'}G_{Y,Y'}(\omega)\;,
\end{equation}
with certain  frequency independent factors $c_{Y,Y'}$. 
These factors, however, are of minor importance since they only affect
 the overall  spectral weight and not  the frequency dependence  of the Green's 
function matrix  $ \tilde{G}(\omega)$. 
We can calculate  them with the assumption that correlated and uncorrelated 
   density-matrix fluctuations are related through the same renormalisation 
factors as the corresponding ground-state density matrices. 
For the one-band model it has been checked that this prescription
is in fact the correct procedure for which the correlation
functions fulfil the standard sum rules \cite{seibold2003,seibold2004b,seibold2008b}.

\subsection{Second-Order Expansion of the Energy Functional}\label{jkio2}
For an evaluation of the Gutzwiller RPA equations~(\ref{10.1250}), 
we need to determine the  matrix $\tilde{K}$ which is given by the second 
derivatives~(\ref{9dgf}) of the effective energy functional~(\ref{e3445}).
 To this end, we expand $E^{\rm GA}$ up to second order around the 
 ground state values ${\vec{\rho}\,}^0$ and 
$\vec{\lambda}^{\rm i;0}\equiv\vec{\lambda}^{\rm i}({\vec{\rho}\,}^0)$,
 \begin{eqnarray}\nonumber
E^{\rm GA}(\vec{\lambda}^{\rm i},\tilde{\rho})&=&E_0+{\rm tr}
(\tilde{h}^{0}\delta\tilde{\rho}) +
\frac{1}{2}\bigg[\sum_{Y,Y'}\delta \rho_Y M^{\rho\rho}_{Y,Y'}
\delta \rho_{Y'}+\sum_{Z,Z'}\delta \lambda^{\rm i}_Z M^{\lambda\lambda}_{Z,Z'}\delta \lambda^{\rm i}_{Z'}\\ \nonumber
&&+\sum_{Z,Y}\left(\delta\lambda^{\rm i}_Z 
M^{\lambda\rho}_{Z,Y}\delta \rho_Y
+\delta \rho_YM^{\rho\lambda}_{Y,Z}\delta\lambda^{\rm i}_Z \right)\bigg]\\\label{eee1}
&\equiv&E_0+{\rm tr}
(\tilde{h}^{0}\delta\tilde{\rho})+\delta E^{(2)}\;.
\end{eqnarray}
Here, we introduced the matrices $\tilde{M}^{\rho\rho}$, $\tilde{M}^{\lambda\rho}$, 
$\tilde{M}^{\rho\lambda}$, $\tilde{M}^{\lambda\lambda}$ with the elements 
\begin{eqnarray}\label{mat1}
M^{\rho\rho}_{Y,Y'}&=&\frac{\partial^2 E^{\rm GA}}
{\partial\rho_Y  \partial\rho_{Y'}} \;, \\\label{mat2}
M^{\lambda\rho}_{Z,Y}&=&\frac{\partial^2E^{\rm GA}}{\partial 
\lambda^{\rm i}_Z \partial\rho_Y}=M^{\rho\lambda}_{Y,Z}\;,\\\label{mat4}
M^{\lambda\lambda}_{Z,Z'}&=&\frac{\partial^2E^{\rm GA}}
{\partial \lambda^{\rm i}_Z \partial \lambda^{\rm i}_{Z'}}\;,
\end{eqnarray}  
where the second derivatives on the r.h.s.\ are evaluated for 
$\tilde{\rho}=\tilde{\rho}^{0}$ 
 and $\vec{\lambda}^{\rm i}=\vec{\lambda}^{{\rm i};0}$. Note that there is 
no linear term 
$\sim \lambda^{\rm i}_Z$ in~(\ref{eee1}) because 
of the minimisation condition~(\ref{9retww}).
For our further evaluation, it is useful to write the second order
 terms in Eq.\  (\ref{eee1}) in 
a more compact form by means of matrix-vector products, 
\begin{equation}\label{qqq1}
\delta E^{(2)}=\frac{1}{2}\Big[
 (\delta \vec{\rho})^{T} \tilde{M}^{\rho\rho}\delta \vec{\rho}
+2(\delta\vec{\lambda}^{\rm i})^{T}  \tilde{M}^{\lambda\rho}\delta \vec{\rho}
+(\delta \vec{\lambda}^{\rm i})^{T} \tilde{M}^{\lambda\lambda}\delta 
\vec{\lambda}^{\rm i}\Big]\;.
\end{equation}
Here we used the symmetry
\begin{eqnarray}
 \tilde{M}^{\lambda\rho}
&=&\big[\tilde{M}^{\rho\lambda}\big]^{T}\;.
\end{eqnarray}

In the effective energy functional~(\ref{e3445}) the parameters 
$\vec{\lambda}^{\rm i}$ are determined by the minimisation condition 
(\ref{9retww}). Applied to our second-order expansion~(\ref{qqq1})  this 
 condition yields 
\begin{equation} \label{10.1160} 
\frac{\partial}{\partial \delta \lambda^{\rm i}_{Z}}
\delta E^{(2)}(\delta\vec{\lambda}^{\rm i},\delta \vec{\rho})=0\;,
\end{equation}
which gives us the multiplet-amplitudes
\begin{equation} \label{10.1170} 
\delta \vec{\lambda}^{\rm i}=-\left[\tilde{M}^{\lambda\lambda}\right]^{-1}\tilde{M}^{\lambda\rho}
\delta\vec{ \rho}
\end{equation}
as a linear function of the densities $\delta \vec{\rho}$.
This result leads to the quadratic expansion 
 \begin{eqnarray}\label{enexp}
E^{\rm eff}({\vec{\rho}\,}^0+\delta\vec{\rho})&=&E_0+{\rm tr}(\tilde{h}^{0}\delta\tilde{\rho})+\frac{1}{2} (\delta\vec{\rho})^{T}
\tilde{K}\delta\vec{\rho}\;,\\\label{enexpb}
\tilde{K}&\equiv&\tilde{M}^{\rho\rho}-\tilde{M}^{\rho\lambda}
\left[\tilde{M}^{\lambda\lambda}\right]^{-1}
\tilde{M}^{\lambda\rho}\;,
\end{eqnarray}  
of the effective energy as a function of the density fluctuations 
$\delta\vec{\rho}$. In earlier work on the time-dependent Gutzwiller
 theory, Eqs.\ (\ref{10.1160}) and~(\ref{10.1170}) have been 
 denoted as the `antiadiabaticity assumption'. In fact, these equations 
 have the physical meaning that the local multiplet dynamics, described by 
 fluctuations $\delta \lambda^{\rm i}_{Z}(t)$, are fast compared to those 
 of the density-matrix fluctuations $\delta \rho_{Y}(t)$. We will
 use the phrase `antiadiabaticity assumption'  in this work too although, 
strictly speaking, in our derivation it does not constitute an
 additional approximation. 
   
With the functional~(\ref{enexp}), we could now proceed with our evaluation 
 of the Gutzwiller RPA Eqs.\ (\ref{10.1250}).
 For practical applications, however, it is more convenient to determine 
 the `interaction kernel'~(\ref{enexpb}) in a way that avoids the 
 explicit solution of the constraint equations~(\ref{const}). This alternative
 procedure is the subject of the following section. 
\subsection{Lagrange-functional expansion}\label{lpe}
In the second-order expansion, described in section~\ref{jkio2}, we 
implemented the constraints~(\ref{const}) by explicitly 
eliminating a  certain set of $n_{\rm c}$ variational parameters. Although 
such a procedure can, at least in principle, always be applied,  for the 
numerical implementation it is more convenient to impose the constraints 
by means of Lagrange parameters. 
 To this end, we define the `Lagrange functional'
\begin{equation}\label{lag1}
L(\vec{\lambda},\vec{\rho},\vec{\Lambda}) 
\equiv E(\vec{\lambda},\vec{\rho}) 
+\sum_{n=1}^{n_{\rm c}} \Lambda_n g_{n}(\vec{\lambda},\vec{\rho})
\end{equation}
 which depends on {\sl all} variational parameters  $\vec{\lambda}$, the 
 density matrix $\tilde{\rho}(\hat{=}\vec{\rho})$ and the $n_{\rm c}$
 Lagrange parameters  $\Lambda_n$. The optimum variational parameters  
$\lambda_{Z}^0$, density-matrix elements $\rho^0_Y$, and 
  Lagrange parameters  $\Lambda^0_n$
are then determined by the equations  
\begin{equation}\label{2eiusy}
\left. \frac{\partial L}{\partial \lambda_Z}
\right|_{\vec{\lambda}=\vec{\lambda}^0,\vec{\Lambda}=\vec{\Lambda}^0,
\vec{\rho}=\vec{\rho}^0}=\left. \frac{\partial L}{\partial \Lambda_n}
\right|_{\ldots}=  
\left. \frac{\partial L}{\partial \rho_Y}
\right|_{\ldots}=0\;.
\end{equation}
which have to be solved simultaneously. 

We expand the Lagrange functional to leading order with respect to 
parameter ($\delta \lambda_Z$, $\delta\Lambda_n$) and density fluctuations
 ($\delta \rho_Y$). The second-order contribution has the form
\begin{eqnarray}\nonumber
 \delta L^{(2)}&=&
\frac{1}{2}\sum_{Y,Y'}\delta \rho_Y L^{\rho\rho}_{Y,Y'}\delta \rho_{Y'}
+\sum_{Z,Y}\delta\lambda_Z 
L^{\lambda\rho}_{Z,Y}\delta \rho_Y+\frac{1}{2}\sum_{Z,Z'}\delta 
\lambda_Z L^{\lambda\lambda}_{Z,Z'}\delta \lambda_{Z'}\\\label{eq:lagrange}
&&+\sum_n\delta\Lambda_n\left\lbrace\sum_Z \frac{\partial g_n}{\partial\lambda_Z}\delta\lambda_Z + \sum_Y\frac{\partial g_n}{\partial\rho_Y}\delta\rho_Y\right\rbrace
\end{eqnarray}
 with matrices $\tilde{L}^{\rho\rho},\tilde{L}^{\lambda\rho}
,\tilde{L}^{\lambda\lambda}$ defined as in Eqs.\ (\ref{mat1})-(\ref{mat4}) 
only with 
 $E^{\rm GA}$ replaced by $L$.  The antiadiabaticity conditions  
\begin{eqnarray}
\frac{\partial }{\partial \delta \lambda_Z} \delta L^{(2)}  &=&0\;,\\
\frac{\partial }{\partial\delta \Lambda_n}\delta L^{(2)}&=&0\;,
\end{eqnarray}
 yield the $n_{\rm c}$ equations
\begin{equation}\label{lag2}
\sum_{Z}\frac{\partial g_n}{\partial\lambda_Z}\delta\lambda_Z 
+ \sum_Y \frac{\partial g_n}{\partial\rho_Y}\delta\rho_Y = 0\;,
\end{equation}
and the $n_{\rm p}$ equations
\begin{equation}\label{lag3}
\sum_{Z'}L^{\lambda\lambda}_{Z,Z'} \delta\lambda_{Z'}
+\sum_{Y}L^{\lambda\rho}_{Z,Y} \delta\rho_{Y}
+\sum_n \frac{\partial g_n}{\partial\lambda_Z}\delta\Lambda_n=0 \;.
\end{equation}
Together these equations allow us to express the 
$n_{\rm p}+n_{\rm c}$ parameter fluctuations
$\delta\Lambda_n$, $\delta\lambda_Z$ in terms of the density fluctuations
$\delta\rho_Y$.  
These can be reinserted into~(\ref{eq:lagrange}) to obtain
the desired quadratic functional solely of the density fluctuations,
 \begin{equation}\label{hj74}
 \delta L^{(2)}=
\frac{1}{2}\sum_{Y,Y'}\delta \rho_Y \bar{K}_{Y,Y'}\delta \rho_{Y'}\;.
\end{equation}
In  \ref{equi}, we prove that the interaction matrix 
$\bar{K}_{Y,Y'}$ in~(\ref{hj74}) is, in fact, 
 identical to $K_{Y,Y'}$ in Eqs.\  (\ref{enexp})-(\ref{enexpb}).

\section{Two particle response functions for lattice models} \label{tpr}
In the previous chapter we have developed the general formalism of the 
 time-dependent Gutzwiller theory for the calculation of  
 two-particle Green's functions. We will be  more specific  in this section 
and explain in detail how the response 
 functions which are of interest in solid-state physics
 can be calculated within our approach.    
\subsection{Two-particle response functions}
In solid-state physics one is usually  not interested in the full 
two-particle Greens\-function~$\tilde{G}$  
as it has been defined in~(\ref{10.700}).
 The properties, relevant for experiments, are certain linear combinations
 of elements of  $\tilde{G}$. For our translationally invariant model 
Hamiltonians~(\ref{1.1}) these are in particular the two-particle 
response functions 
\begin{equation}\label{2ysdui}
G_{(\sigma_2,\sigma_1),(\sigma_3,\sigma_4)}(\vec{R}_i-\vec{R}_j,t-t')\equiv
\langle \langle   \hcd_{i,\sigma_1}(t)\hc_{i,\sigma_2}(t); 
\hcd_{j,\sigma_3}(t')\hc_{j,\sigma_4}(t')  \rangle \rangle
\end{equation}
 or, more importantly, their Fourier transforms
\begin{eqnarray}\nonumber
G_{(\sigma_2,\sigma_1),(\sigma_3,\sigma_4)}(\vec{q},\omega)
&=&\frac{1}{L_{\rm s}}\int\limits_{-\infty}^{\infty}d\tau e^{{\rm i}\omega \tau} \sum_{i,j}
e^{{\rm i}(\vec{R}_i-\vec{R}_j)\cdot\vec{q}}
G_{(\sigma_2,\sigma_1),(\sigma_3,\sigma_4)}
(\vec{R}_i-\vec{R}_j,\tau)\\\label{4ytut}
&=&\frac{1}{L_{\rm s}}\sum_{k,k'}   \langle \langle
\hcd_{k,\sigma_1}\hc_{k+q,\sigma_2}; 
\hcd_{k'+q,\sigma_3}\hc_{k',\sigma_4}
\rangle \rangle_\omega\;.
\end{eqnarray}  
Here, we introduced the fermionic operators
\begin{equation}
\hat{c}_{k,\sigma}^{(\dagger)}=\frac{1}{\sqrt{L_{\rm s}}}\sum_i
e^{\mp{\rm i}\vec{R}_i\cdot\vec{k}}
\hat{c}_{i,\sigma}^{(\dagger)}
\end{equation}
and the usual notation 
\begin{equation}
\langle \langle   
\hat{O}; 
\hat{O}'
\rangle \rangle_\omega
=\int_{-\infty}^{\infty}d\tau 
\langle \langle
\hat{O}(\tau); 
\hat{O}'(0)
\rangle \rangle e^{{\rm i}\omega \tau}
\end{equation}
for the Fourier transform of a Green's function with arbitrary operators
 $\hat{O}$,$\hat{O}'$. With the abbreviation $v=(\sigma,\sigma')$ for 
spin-orbit indices and the operators 
\begin{equation}\label{4vbyt}
\hat{A}^{q}_{v}\equiv\hat{A}^{q}_{\sigma_2,\sigma_1}
\equiv\frac{1}{\sqrt{L_s}} \sum_{k}\hcd_{k,\sigma_1}\hc_{k+q,\sigma_2}
\end{equation}
we can write~(\ref{4ytut}) as
\begin{equation}\label{4bgd}
G_{v,v'}(\vec{q},\omega)= \langle\langle
\hat{A}^{q}_{v};(\hat{A}^{q}_{v'})^{\dagger}\rangle\rangle_\omega\;.
\end{equation}

The Green's functions~(\ref{4ytut}) are still quite general since they include 
 all possible channels of local coupling $\sigma_1\leftrightarrow \sigma_2$,
 $\sigma_3\leftrightarrow \sigma_4$. In experiments one usually measures
 response functions which are certain linear combinations,
\begin{equation}\label{2odu}
G_{\rm e}(\vec{q},\omega)=\sum_{v,v'}\kappa_{v}
G_{v,v'}(\vec{q},\omega)
\kappa_{v'}
\end{equation}
of some of the  Green's functions~(\ref{4ytut}), defined by the matrix 
 $\kappa_v=\kappa_{\sigma,\sigma'}$.
 For example, the transversal 
 spin-susceptibility $\chi(\vec{q},\omega)$ is given as
\begin{equation}\label{2odub}
\chi(\vec{q},\omega)=\frac{1}{L_{\rm s}}
\langle \langle   
\hat{S}^{+}_{q}; 
\hat{S}^{-}_{-q}
\rangle \rangle_\omega
\end{equation} 
where 
\begin{eqnarray}
 \hat{S}^{+}_{q}&=&\sum_i e^{-{\rm i}\vec{R}_i\cdot\vec{q}} \hat{S}^{+}_{i}
=\sum_k\sum_b\hcd_{k,(b\uparrow)}\hc_{k+q,(b\downarrow)}\;, \\
\hat{S}^{-}_{-q}&=&\sum_i e^{{\rm i}\vec{R}_i\cdot\vec{q}} \hat{S}^{-}_{i}
=\sum_k\sum_b\hcd_{k+q,(b\downarrow)}\hc_{k,(b\uparrow)}
\equiv  \Big( \hat{S}^{+}_{q} \Big)^{\dagger}\;, \\
\hat{S}^{+}_{i}&=&\sum_{b}\hcd_{i,(b\uparrow)}\hc_{i,(b\downarrow)}\;,\;
\hat{S}^{-}_{i}=\sum_{b}\hcd_{i,(b\downarrow)}\hc_{i,(b\uparrow)}\;,
\end{eqnarray}  
 are the usual spin-flip operators and $b$ is an index for the orbitals
 at each lattice site $i$. 
 The spin susceptibility 
 of a  two-band Hubbard model will be investigated  in chapter~\ref{ms}.  
\subsection{Response functions in the time-dependent Gutzwiller approximation}
In order to apply the time-dependent Gutzwiller approximation, 
as developed in chapter~\ref{chap7.3.3}, we have to expand the Lagrange
 functional~(\ref{lag1}) up to second order with respect to density-matrix
 ($\delta \tilde{\rho}$) and variational-parameter fluctuations  
($\delta \lambda_{\Gamma,\Gamma'}$). This means that we need an expansion 
 of the constraints~(\ref{1.10c})-(\ref{1.10d}), of the local energies  
(\ref{1.5})-(\ref{1.6c}) and~(\ref{0agh}), 
and of the kinetic energy~(\ref{h0exp})-(\ref{0etsy}). 
The second-order expansion of the kinetic energy is more involved than that of 
 the local energies and of the constraints. In the latter there are only 
 contributions from fluctuations at {\sl same} lattice sites while in the 
 kinetic energy local and non-local fluctuations 
(such as $\delta\langle  
\hat{c}_{i,\sigma}^{\dagger}\hat{c}_{j,\sigma'}^{\phantom{+}}
\rangle_{\Psi_{0}}$)  couple. Nevertheless, the calculation of the 
second-order Lagrange functional is  tedious but otherwise 
 straightforward. 
 We therefore refer to 
 \ref{seexp} where the details of this derivation are presented. 
 As shown in that Appendix, it is useful to introduce the operators 
\begin{eqnarray}\label{9yh}
\hat{B}^{q}_{w}&\equiv& \hat{B}^{q}_{\sigma_1,\sigma_2,\sigma'_1,\sigma'_2}
\equiv\frac{1}{\sqrt{L_s}} \sum_k\epsilon^{\sigma_2,\sigma_1}_{k}
\hcd_{k,\sigma'_2}\hc_{k+q,\sigma'_1}\;,\\\label{9yhb}
\hat{\bar{B}}^{q}_{w}&\equiv&
\hat{\bar{B}}^{q}_{\sigma_1,\sigma_2,\sigma'_1,\sigma'_2}
\equiv \frac{1}{\sqrt{L_s}}\sum_{k}\epsilon^{\sigma_2,\sigma_1}_{k+q}
\hcd_{k,\sigma'_2}\hc_{k+q,\sigma'_1}\;,
\end{eqnarray}  
and to define the 
auxiliary   Green's function matrix   
$\tilde{\Pi}(\vec{q},\omega)$   with the elements 
\begin{eqnarray}\nonumber
\Pi_{\underset{(w)}{v},\underset{(w')}{v'}}(\vec{q},\omega)
\equiv\left(
\begin{array}{ccc}
\langle\langle
\hat{A}^{q}_{v};\big(\hat{A}^{q}_{v'}\big)^{\dagger}\rangle\rangle_\omega& \langle\langle
\hat{A}^{q}_{v};
\big(\hat{B}^{q}_{w'}\big)^{\dagger}
\rangle\rangle_\omega&
\langle\langle
\hat{A}^{q}_{v};
\big(\hat{\bar{B}}^{q}_{w'}\big)^{\dagger}
\rangle\rangle_\omega \\
\langle\langle
\hat{B}^{q}_{w};\big(\hat{A}^{q}_{v'}\big)^{\dagger}
\rangle\rangle_\omega&
\langle\langle
\hat{B}^{q}_{w};
\big(\hat{B}^{q}_{w'}\big)^{\dagger}
\rangle\rangle_\omega
&\langle\langle
\hat{B}^{q}_{w};
\big(\hat{\bar{B}}^{q}_{w'}\big)^{\dagger}
\rangle\rangle_\omega
\\
\langle\langle
\hat{\bar{B}}^{q}_{w};\big(\hat{A}^{q}_{v'}\big)^{\dagger}
\rangle\rangle_\omega&
\langle\langle
\hat{\bar{B}}^{q}_{w};\big(\hat{B}^{q}_{w'}\big)^{\dagger}
\rangle\rangle_\omega
&
\langle\langle
\hat{\bar{B}}^{q}_{w};\big(\hat{\bar{B}}^{q}_{w'}\big)^{\dagger}
\rangle\rangle_\omega
\end{array}
\right)\;.\\\label{8dgdd}
 \end{eqnarray}
We are actually interested  only in the first `element' of this
 matrix, i.e., the Green's functions~(\ref{4bgd}) since they allow us to 
determine any 
response function of the form~(\ref{2odu}). 
As shown in \ref{app56}, however, 
the time-dependent Gutzwiller approximation leads to the following
equation for the entire matrix~(\ref{8dgdd}) from which~(\ref{4bgd}) 
can be extracted,
\begin{equation}\label{0897}
\tilde{\Pi}(\vec{q},\omega)=
\big(1+\tilde{\Pi}^0(\vec{q},\omega)\tilde{V}^{q}\big)^{-1}
\tilde{\Pi}^0(\vec{q},\omega)
\;.
\end{equation}
Here, $\tilde{V}^{q}$ is the effective second-order interaction matrix, 
introduced in~(\ref{vq}), and   $\tilde{\Pi}^0(\vec{q},\omega)$ is 
the Green's function matrix~(\ref{8dgdd}) evaluated for 
the single-particle Gutzwiller Hamiltonian~(\ref{hop}). As shown in Refs.\
 \cite{buenemann2003b,buenemann2005}, this Gutzwiller Hamiltonian 
$\hat{h}^0\equiv \hat{H}^{\rm eff}_0 $ for our 
 lattice Hamiltonian~(\ref{1.1}) has the form
\begin{equation}\label{9stys}
 \hat{H}^{\rm eff}_0=\sum_k\sum_{\sigma_1,\sigma_2} \big
(\bar{\epsilon}^{\sigma_1,\sigma_2}_{k}+
\eta_{\sigma_1,\sigma_2}\big)\hcd_{k,\sigma_1}\hc_{k,\sigma_2}
\equiv\sum_k\sum_{\alpha}E_{k,\alpha}\hhd_{k,\alpha}\hh_{k,\alpha}
\end{equation}
where  the Lagrange parameters $\eta_{\sigma_1,\sigma_2}$ are 
determined by the  minimisation of the variational ground-state energy and 
$\bar{\epsilon}^{\sigma_1,\sigma_2}_{k}$
is defined as
\begin{equation}
\bar{\epsilon}^{\sigma_1,\sigma_2}_{k}\equiv
\sum_{\sigma'_1,\sigma'_2}
q^{\sigma_1}_{\sigma'_1}\big(q^{\sigma_2}_{\sigma'_2}\big)^*
\epsilon^{\sigma'_1,\sigma'_2}_{k}\;.
\end{equation}
The creation and annihilation operators $\hat{h}^{(\dagger)}_{k,\alpha}$ of the
 effective single-particle Hamiltonian~(\ref{9stys})
 can be written as
\begin{eqnarray}\label{9oshss}
\hhd_{k,\alpha}&=&\sum_{\sigma}u^k_{\sigma,\alpha}\hcd_{k,\sigma}\;,\\\label{9oshssb}
\hh_{k,\alpha}&=&\sum_{\sigma}(u^k_{\sigma,\alpha})^*\hc_{k,\sigma}\;,
\end{eqnarray}
where the coefficients $u_{\sigma,\alpha}$ are determined by a diagonalisation 
 of~(\ref{9stys}). With these eigenstates the calculation of 
$\tilde{\Pi}^0(\vec{q},\omega)$ is now a simple task. For example, the 
 first element $\langle\langle
\hat{A}^{q}_{v};\big(\hat{A}^{q}_{v'}\big)^{\dagger}\rangle\rangle^0_\omega$ is given as
\begin{eqnarray}\label{9ghd}
&&\langle\langle\hat{A}^{q}_{\sigma_1,\sigma_2};\big(\hat{A}^{q}_{\sigma'_1,\sigma'_2}\big)^{\dagger}
\rangle\rangle^0_\omega\\\nonumber
&&=\frac{1}{L_s}\sum_{k,k'}
\sum_{\alpha_1,\alpha_2\atop \alpha'_1,\alpha'_2} 
\langle\langle \hhd_{k,\alpha_2} \hh_{k+q,\alpha_1};
\hhd_{k'+q,\alpha'_1} \hh_{k',\alpha'_2} \rangle\rangle^0_\omega\, \big (u^k_{\sigma_2,\alpha_2}\big)^*u^{k+q}_{\sigma_1,\alpha_1}
\big(u^{k'+q}_{\sigma'_1,\alpha'_1}\big)^*u^{k'}_{\sigma'_2,\alpha'_2}\\\nonumber
&&=\frac{1}{L_s}\sum_k\sum_{\alpha_1,\alpha_2}
\frac{\big (u^k_{\sigma_2,\alpha_2}\big)^*u^{k+q}_{\sigma_1,\alpha_1}
\big(u^{k+q}_{\sigma'_1,\alpha_1}\big)^*u^{k}_{\sigma'_2,\alpha_2}}{
\omega-(E_{k+q,\alpha_1}-E_{k,\alpha_2})+{\rm i} \delta}
\big(n^0_{k,\alpha_2}-n^0_{k+q,\alpha_1}\big)
\end{eqnarray}
where 
\begin{equation}
n^0_{k,\alpha}=\Theta(E_{\rm F}-E_{k,\alpha})
\end{equation}
is the ground-state distribution function~(\ref{10.1240}). 
In the same way, we can calculate all other elements of 
$\tilde{\Pi}^0(\vec{q},\omega)$ . The 
 result is always the same as in~(\ref{9ghd}) only with 
 additional factors $\sim \epsilon^{\sigma,\sigma'}_{k}$ or 
$\sim \epsilon^{\sigma,\sigma'}_{k+q}$ due to the definition of the 
operators~(\ref{9yh})-(\ref{9yhb}). For example, the second element in~(\ref{8dgdd}) 
leads to
\begin{eqnarray}\label{9ghdd}
\langle\langle\hat{A}^{q}_{\sigma_1,\sigma_2};
(\hat{B}^{q}_{\sigma_3,\sigma_4\sigma'_3,\sigma'_4})^{\dagger}\rangle\rangle^0_\omega\\\nonumber
=\frac{1}{L_s}
\sum_k\sum_{\alpha_1,\alpha_2}
\frac{\big(u^{k}_{\sigma_2,\alpha_2}\big)^*u^{k+q}_{\sigma_1,\alpha_1}
\big( u^{k+q}_{\sigma'_3,\alpha_1}\big)^*u^k_{\sigma'_4,\alpha_2}}
{\omega-(E_{k+q,\alpha_1}-E_{k,\alpha_2})+{\rm i}\delta}
\epsilon^{\sigma_3,\sigma_4}_{k}
\big(n^0_{k,\alpha_2}-n^0_{k+q,\alpha_1}\big)\;.
\end{eqnarray}

To summarise, with Eqs.\ (\ref{0897}),~(\ref{9ghd}),~(\ref{9ghdd}), and 
 the interaction matrix~(\ref{vq}) we are now in the position to 
 investigate any two-particle response function for our general class
 of multi-band models~(\ref{1.1}). As a first example, we  study 
 the magnetic susceptibility for a two-band model in the following section. 
 
 \section{Magnetic susceptibility of a two-band Hubbard model}\label{ms}
In this chapter we investigate the magnetic susceptibility of a two-band 
Hubbard model in three spatial dimensions. The model Hamiltonian 
and the Gutzwiller wave functions which we use for its investigation 
 are introduced in section~\ref{ms1}. In section~\ref{ms2} we discuss
 the Green's function matrices that we need to study in order to calculate
 the magnetic susceptibilities within the RPA and the Gutzwiller-RPA schemes. 
 The numerical results for the two-band model are presented in 
section~\ref{ms3}. 
\subsection{Model and  variational ground state}\label{ms1}
We investigate a Hubbard model with two degenerate $e_{\rm g}$ orbitals 
 per site on a  cubic lattice. The local 
 Hamiltonian~(\ref{1.2}) for this system can be written as 
\begin{eqnarray}\label{2.160a}
\hat{H}^{2{\rm b}}_{\rm I} &=&
U \sum_{e}\hat{n}_{e,\uparrow}\hat{n}_{e,\downarrow}
+U'\sum_{s,s'}\hat{n}_{1,s}\hat{n}_{2,s'}
-J\sum_{s}\hat{n}_{1,s}\hat{n}_{2,s}
\label{twoorbhamiltonian} \\[3pt]
&& +J\sum_{s}\hat{c}_{1,s}^{\dagger}
\hat{c}_{2,\bar{s}}^{\dagger}
\hat{c}_{1,\bar{s}}^{\vphantom{+}}
\hat{c}_{2,s}^{\vphantom{+}}
+J \Bigl(
\hat{c}_{1,\uparrow}^{\dagger}\hat{c}_{1,\downarrow}^{\dagger}
\hat{c}_{2,\downarrow}^{\vphantom{+}}\hat{c}_{2,\uparrow}^{\vphantom{+}}
+ 
\hat{c}_{2,\uparrow}^{\dagger}\hat{c}_{2,\downarrow}^{\dagger}
\hat{c}_{1,\downarrow}^{\vphantom{+}}\hat{c}_{1,\uparrow}^{\vphantom{+}}
\Bigr)\;.  \nonumber
\end{eqnarray}
Here, $e=1,2$ labels the $e_{\rm g}$ orbitals, $s=\uparrow,\downarrow$ 
is the spin index and we use the convention $\bar{\uparrow}\equiv \downarrow$, 
$\bar{\downarrow}\equiv\uparrow$.
Due to the cubic symmetry the Coulomb parameters $U$, $U'$ and the exchange 
parameter $J$ are related  to each other through 
\begin{equation}
U'=U-2J\;.
\end{equation}
 Hence, only
 two of these three parameters can be chosen independently. 

  There are four spin-orbital states $\sigma=(e,s)$  per atom, 
leading to a $2^4=16$-dimensional atomic Hilbert space. 
All eigenstates $\ket{\Gamma}$ of $\hat{H}^{2{\rm b}}_{\rm I}$ 
with particle numbers $N\neq 2$ 
 are simple Slater determinants of spin-orbital states $\ket{\sigma}$
  and their energies are
\begin{equation}
\begin{array}{ll}
E_{\Gamma}=0 & (N=0,1)\;,\\
E_{\Gamma}=U+2U'-J & (N=3)\;,\\
E_{\Gamma}=2U+4U'-2J & (N=4)\;.
\end{array}
\end{equation}

\begin{table}[t]
\label{tableone}
\centering
\begin{tabular}{c|c|c|c}
\# & Atomic eigenstate $|\Gamma \rangle $ & Symmetry  & energy $E_{\Gamma} $ \\
\hline 
1 & $\ket{\uparrow ,\uparrow } $ & ${}^3A_2$  & $U'-J$  \\ 
2 & $(\ket{\uparrow ,\downarrow } +\ket{\downarrow ,\uparrow } )/\sqrt{2}
$ & ${}^3A_2$  & $U'-J$  \\ 
3 & $\ket{\downarrow ,\downarrow } $ & ${}^3A_2$ 
& $U'-J$  \\ 
4 & $(\ket{\uparrow ,\downarrow} -\ket{\downarrow ,\uparrow } )/\sqrt{2}
$ & ${}^1E$ &  $U'+J$  \\ 
5 & $(\ket{\uparrow \downarrow ,0} -\ket{0,\uparrow \downarrow } )/\sqrt{2}$
 & ${}^1E$ &  $U-J$\\ 
6 & $(\ket{\uparrow \downarrow ,0} +\ket{0,\uparrow \downarrow } )/\sqrt{2}$
 & ${}^1A_1$  & $U+J$  \\ 
\end{tabular}
\caption{Two-particle eigenstates with symmetry specifications and energies.}
\end{table}

The two-particle eigenstates are slightly more 
 complicated because some of them are 
 linear combinations of Slater determinants. 
We introduce the basis
 \begin{eqnarray}\label{2.299a}
\ket{s,s'}&\equiv&\hcd_{1,s}\hcd_{2,s'}\ket{0}\;,\\
\ket{\uparrow\downarrow,0}&\equiv&\hcd_{1,\uparrow}
\hcd_{1,\downarrow}\ket{0}\;,\\
\ket{0,\uparrow\downarrow}&\equiv&\hcd_{2,\uparrow}
\hcd_{2,\downarrow}\ket{0}\;,
\end{eqnarray}
of two-particle states, which are used to set up 
 the eigenstates of $\hat{H}_{{\rm loc};i}$, see table~\ref{tableone}.
The states of lowest energy are the three triplet states 
with spin $S=1$, which belong to the 
 representation $A_2$ of the cubic point-symmetry group. Finding 
 a high-spin ground state is a simple consequence of 
Hund's first rule. Higher in energy
 are the two degenerate singlet states of symmetry $E$ 
and the non-degenerate singlet state  of symmetry $A_1$.  

For the variational ground state we can work with 
 a wave function~(\ref{1.3})  that contains only diagonal parameters
 $\lambda_{\Gamma,\Gamma}$. Non-diagonal parameters could only arise if we
 break the cubic symmetry or want to study states with magnetic orders
  not collinear to the chosen spin-quantisation axis. Note, however, that 
 for the study of spin excitations we {\sl must} allow for non-diagonal 
 variational parameters, see below. 

\begin{figure}
{\centering
\includegraphics[width=0.45\textwidth]{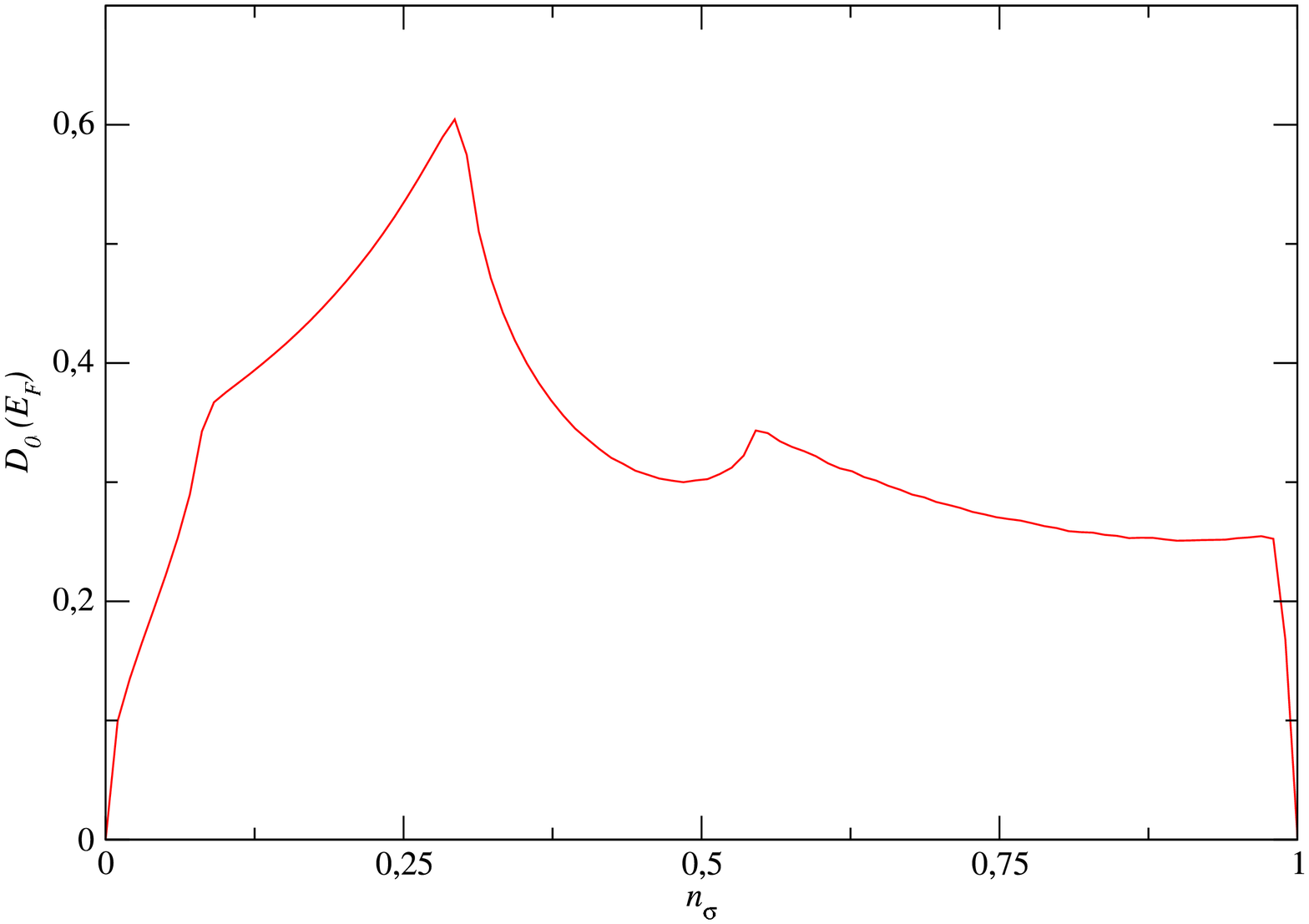}\hfill
\includegraphics[width=0.45\textwidth]{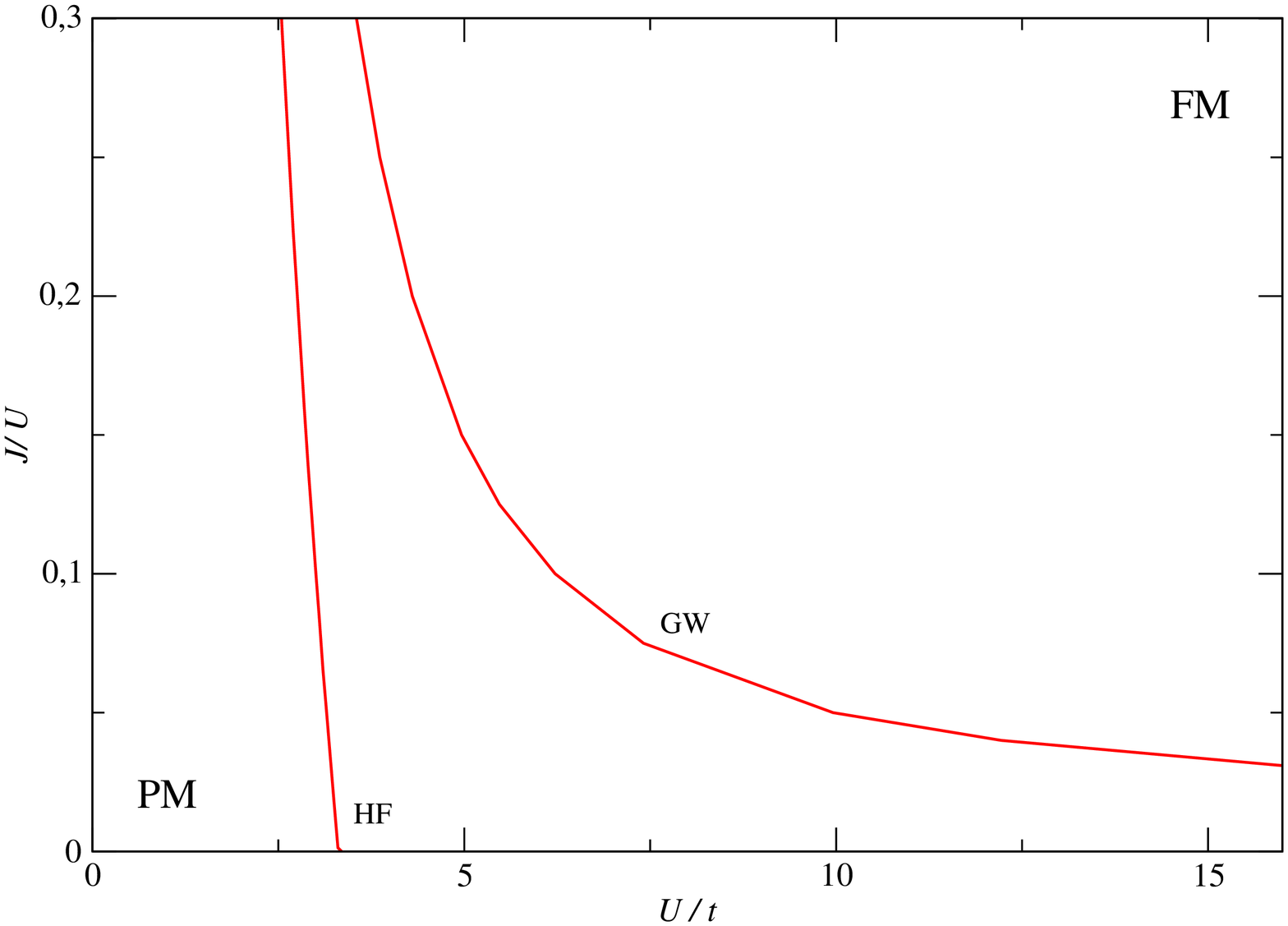}
\caption{Left: Model density of states at the Fermi energy as a function of
orbital filling $n_{\sigma}$. Right: Ground-state phase diagram for both HF and GW. The lines mark the instability for a transition from the paramagnetic (PM) to the ferromagnetic (FM) state. The orbital filling is $n_{\sigma}\approx0.29$ and $t\equiv |t_{dd\sigma}^{(1)}|$ 
 (c.f., Ref.~\cite{slater1954}).}\label{fig10.8}}
\end{figure}

In our numerical analysis of this two-band model we will consider 
a tight-binding Hamiltonian $\hat{H}_0$ with generic hopping parameters
 which were already used in previous works and lead to the density
 of states at the Fermi energy  shown in Fig.\ \ref{fig10.8} (left). 
Due to the maximum in the density of states at
 approx.\ $n_{\sigma}=0.29$, in that range
 of band fillings there is the strongest tendency for a ferromagnetic state 
 to be lower in energy than the paramagnet. This has already been demonstrated
 in Ref.\ \cite{buenemann1998}. Another important finding in that work is the 
huge importance of the exchange interaction~$J$ for the appearance of ferromagnetic order.
  This can be seen from the Gutzwiller phase
 diagram  for our model in Fig.\ \ref{fig10.8} (right). 
In contrast, the HF phase diagrams 
 shows almost no dependence on the size of~$J$; see also  
Ref.\ \cite{buenemann2011b} where similar results have been reported for a 
 two-band model in infinite dimensions.

\subsection{The magnetic susceptibility}\label{ms2}
For the calculation of the spin susceptibility~(\ref{2odub}), we need to 
 determine a Green's function matrix of the form~(\ref{8dgdd}) in which 
 the operators $\hat{A}^{q}_v$,  $\hat{B}^{q}_w$, $\hat{\bar{B}}^{q}_w$
 are given as 
 \begin{eqnarray}\label{qhr1}
 \hat{A}^{q}_{b_1,b_2}\equiv\frac{1}{\sqrt{L_s}} \sum_{k}\hcd_{k,(b_2\uparrow)}
\hc_{k+q,(b_1\downarrow)}\;,\\\label{qhr2}
\hat{B}^{q}_{b_1,b_2,b'_1,b'_2}\equiv\frac{1}{\sqrt{L_s}} \sum_{k}
\epsilon_{k}^{b_2,b_1}\hcd_{k,(b'_2\uparrow)}
\hc_{k+q,(b'_1\downarrow)}\;,\\\label{qhr3}
\hat{\bar{B}}^{q}_{b_1,b_2,b'_1,b'_2}\equiv\frac{1}{\sqrt{L_s}} \sum_{k}
\epsilon_{k+q}^{b_2,b_1}\hcd_{k,(b'_2\uparrow)}
\hc_{k+q,(b'_1\downarrow)}\;.
\end{eqnarray}
The matrix~(\ref{8dgdd}), which results from these operators 
 is $4+16+16=36$ dimensional. Due to symmetries, this 
dimension can be reduced to $20$ for a general wave-vector~$\vec{q}$. Along
 symmetry lines the symmetry reduction could even go  further. In our numerical
 calculations, however, we did not exploit such symmetry considerations 
since the numerical efforts for a two-band model
 are still moderate, even in three dimensions.

Note that there is a difference between Hartree-Fock and Gutzwiller RPA 
 calculations concerning the elements of $\tilde{\Pi}^0(\vec{q},\omega)$
 which have to be taken into account in our calculation of the 
 susceptibility 
 \begin{equation}\label{7azst}
\chi(\vec{q},\omega)=
\sum_{b,b'}\langle \langle 
\hat{A}^{q}_{b,b};(\hat{A}^{q}_{b',b'})^{\dagger}
\rangle \rangle_{\omega}\;.
 \end{equation}
 In Hartree-Fock RPA, due to the locality of the interaction
 terms in Hubbard models and the symmetries of $e_{\rm g}$ orbitals, 
the only elements of   $\tilde{\Pi}^0(\vec{q},\omega)$ which contribute are
 $\langle \langle 
\hat{A}^{q}_{b,b};(\hat{A}^{q}_{b',b'})^{\dagger}
\rangle \rangle^0_{\omega}$, i.e., those which are diagonal with respect
 to the local orbital indices. This is different from
 the Gutzwiller RPA equations in which all Green's functions defined by
 the operators (\ref{qhr1})-(\ref{qhr3})  have to be taken into account. 
In particular, 
 Green's functions $\langle \langle 
\hat{A}^{q}_{b_1,b_2};(\hat{A}^{q}_{b_3,b_4})^{\dagger}
\rangle \rangle^0_{\omega}$  with 
 $b_1\neq b_2$ or $b_3\neq b_4$ cannot be discarded. The reason for this
 difference  is the non-locality  of the interaction matrix  $\tilde{V}^q$
 in the time-dependent Gutzwiller theory.

\subsection{Results} \label{ms3}
\begin{figure}
  \includegraphics[width=0.8381\textwidth]{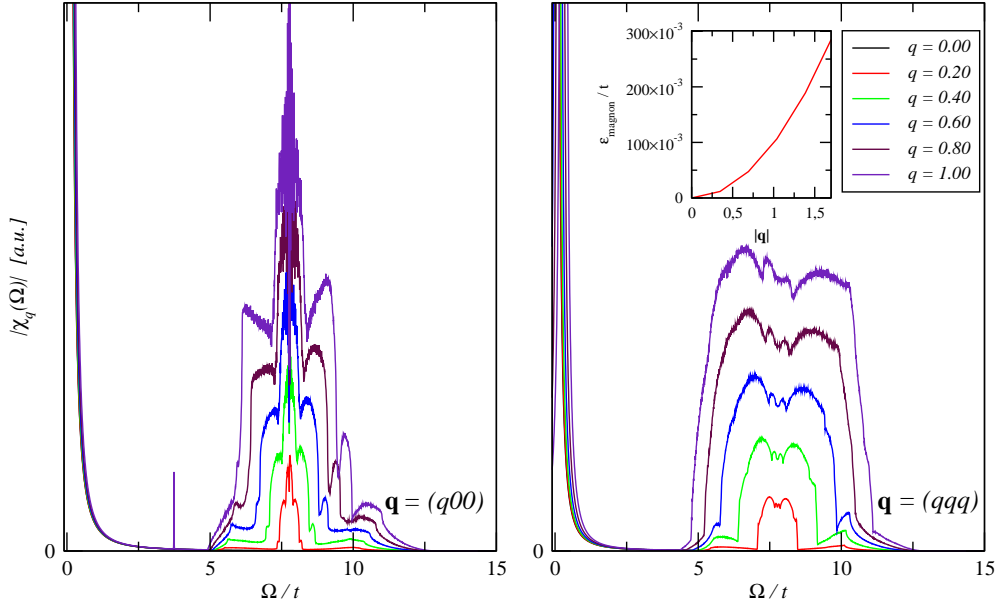}
  \caption{HF excitations for $U=10.0 t$, $J/U=0.30$ resulting in $m=0.5975$ (fully polarised) and $\lambda^{-}=E_{F}^{\uparrow}-E_{F}^{\downarrow}\approx 7.77 t$.
 The magnon dispersion is fitted by 
$\varepsilon_{magnon}(\vec{q})/t=D|\vec{q}|^{2}(1+\beta|\vec{q}|^{2})$ with $D^{100}\approx D^{111}=100\times10^{-3}$. Scaling: $t\equiv |t_{dd\sigma}^{(1)}|$ 
 (c.f., Ref.\ \cite{slater1954}) and $q_{x,y,z}\in (-\pi,\pi)$.}\label{fig1}
\end{figure}

\begin{figure}
  \includegraphics[width=0.8381\textwidth]{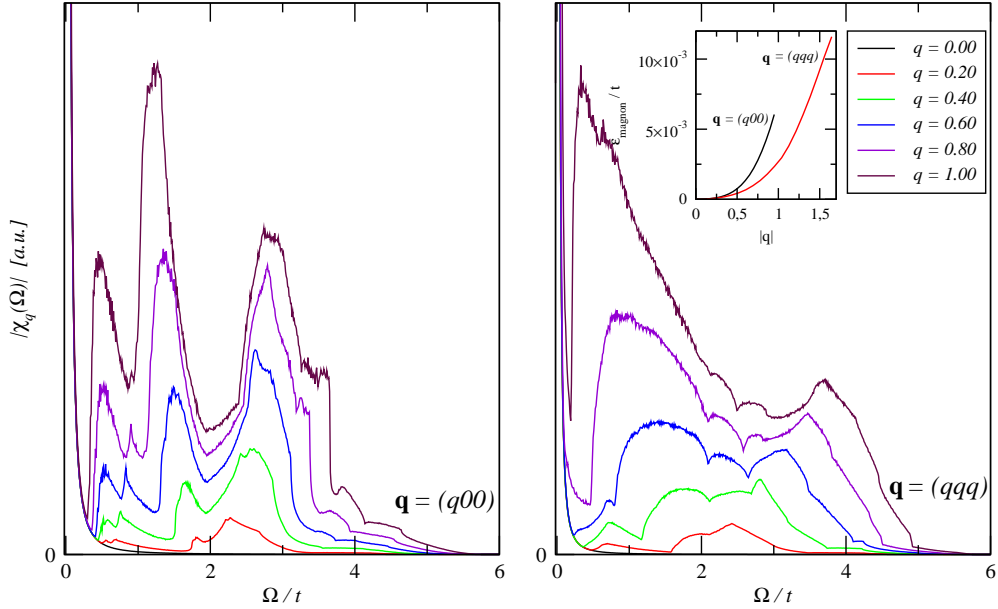}
  \caption{GW excitations for $U=10.0 t$, $J/U=0.30$ resulting in $m=0.5728$ (almost fully polarized) and $\lambda^{-}=E_{F}^{\uparrow}-E_{F}^{\downarrow}\approx2.128 t$. The 
Magnon dispersion is fitted by $\varepsilon_{magnon}(\vec{q})/t=D|\vec{q}|^{2}(1+\beta|\vec{q}|^{2})$ with $D^{100}=1.34\times10^{-3}$ and $D^{111}=1.30\times10^{-3}$ for small wave vectors. Scaling: $t\equiv |t_{dd\sigma}^{(1)}|$ 
 (c.f., Ref.\ \cite{slater1954}) and $q_{x,y,z}\in (-\pi,\pi)$.}\label{fig2}
\end{figure}

We prepare a ferromagnetic ground state in both HF and Gutzwiller 
approximation at band filling
$n_{\sigma} \approx 0.2987$ 
in order to be close to the maximum of the DOS. In general both schemes
 will give different magnetisations for the same set of interaction
 parameters. Therefore, one
could either perform the comparison for fixed parameters or fixed
magnetisation (cf.\ also Ref.\ \cite{guenther2010}).
To avoid this inconsistency, we present results for interaction
 parameters which lead to a fully polarised
ferromagnetic ground state in both approximations,  i.e. 
$m=2 n_{\sigma}$.
Note that due to numerical reasons we have to stay slightly below this
value in case of the Gutzwiller approximation. The corresponding interaction
parameters are specified in the captions to Figs.\ \ref{fig1}, \ref{fig2}
which display the magnetic excitations obtained within both approximations.

These spectra are composed of a low-energy magnon part due
to the breaking of spin-rotational invariance and
a high energy Stoner continuum which reflects the particle-hole spin-flip
excitations of the `bare' system, i.e  $\tilde{\Pi}^0(\vec{q},\omega)$,
cf. Eq.~(\ref{0897}). For both methods we show the excitations along
the $(100)$ and $(111)$ directions. The difference in these directions
mainly arises due to the orientation dependence of the particle-hole
dispersion which is significantly stronger along the diagonals.

One first important difference between HF and Gutzwiller approximation
concerns the difference in the magnetic band splitting 
$\lambda^- = E_F^\uparrow - E_F^\downarrow$. In HF theory this value
is just given by $\lambda^-(HF) = (U+J)m$ and thus for strongly
correlated systems produces a large gap ${\cal{O}}(U)$ between the
low energy magnon and a high energy Stoner continuum.
On the other hand, we find $\lambda^-(GA)$ is significantly reduced
with regard to its HF counterpart. For the present system
$\lambda^-(GA) \approx 1/4 \lambda^-(HF)$. Given the broadening
of the Stoner continuum with increasing transferred momentum,
the low energy magnon thus rapidly merges with the continuum
in the time-dependent GA as can be seen from Fig.\ \ref{fig2}.
As a consequence the excitation at $\omega \sim J$, corresponding
to a respective spin-flip in the two orbitals, is only
visible in HF+RPA  along the $(100)$ direction, whereas in
the time-dependent Gutzwiller approach it is already within the
continuum.
Note that the overestimation of the Stoner excitation energy within
HF+RPA is a longstanding problem in solid state theory as discussed in
Ref.\  \cite{stollhoff1990}.

At 
${\vec q}=0$ all the weight is contained in the zero frequency
Goldstone mode. The existence of this excitation provides an 
important consistency check of the Gutzwiller+RPA approach
similar to the analogous finding in HF+RPA. The positive
dispersion of the magnon further demonstrates that the underlying
Gutzwiller solution is a stable energy minimum which is not
destroyed by the fluctuations. 
The spin-wave stiffness, i.e. the quadratric coefficient
 of the magnon dispersion, is significantly larger in 
 HF+RPA than in time-dependent Gutzwiller theory. 
Note, however,  that to a certain extend this huge difference 
is caused by an instability of the ferromagnetic ground state 
with respect to an incommensurate phase which is found 
   for interaction parameters not much smaller
 than those used in Fig.\ \ref{fig2}.

  \section{Summary} \label{summary}
In this paper we have given a detailed derivation of the time-dependent
Gutzwiller approximation for multi-band Hubbard models. 
The basic assumptions which underlie the method can be summarised as
follows. First, it is assumed that the dynamics of the Slater-determinant
(upon which the Gutzwiller projector acts in the starting Ansatz) is
determined by the so-called Gutzwiller Hamiltonian, Eq.~(\ref{10.1220}),
which  leads to an equation of motion similar to standard RPA, 
Eq.\ (\ref{10.1230}). Second, the dynamics of the variational parameters
is determined from the assumption that at each instant of time
the energy is minimised. This leads to a linear relation between
variational parameter and density fluctuations, Eq.~(\ref{10.1170}).
Third, as in the standard HF+RPA approach it is assumed that the
external perturbation and thus the density fluctuations are small,
Eqs.~(\ref{10.930}), (\ref{10.940}).
We have seen that also in the multi-band case these assumptions lead
to a consistent theory in the sense, that an instability which is signalled
within the Gutzwiller+RPA corresponds to a (second-order) phase transition
which one would obtain from the bare variational Gutzwiller approximation.
We have further demonstrated that for ferromagnetic ground states 
the Gutzwiller+RPA leads to the appearance of the Goldstone mode as
expected for systems which break continuous spin symmetry.

The formalism as developed in its present form can now be straightforwardly
applied to the investigation of correlation functions in strongly
correlated multi-band systems as e.g. pnictides, manganites, cobaltates etc.
On the other hand, a natural application of the theory 
would also comprise the investigation of e.g. orbital quenches for
which the small amplitude assumption for the density matrices
has to be abandoned. For single-band Hubbard models such
a fully time-dependent formulation of the Gutzwiller approximation
has been recently presented by Schir\'{o} and Fabrizio 
\cite{schiro2010,schiro2011} where also the second assumption above
has been replaced by separate equations of motion for the
variational parameters. 
Future work should thus address the question whether their
approach reduces to the present theory in the small amplitude
limit and how it eventually can be extended to the multi-band case.

 \begin{appendix} 
\section{Second order expansion of determinants}\label{app1}
According to Eq.\  (\ref{miiprime}), section~\ref{locen},  
 expectation values $m^0_{I,I'}$ can be written as determinants
 of certain matrices $A$ with elements that are linear 
functions of the local density matrix $C^{0}_{\gamma,\gamma'}$. 
In the variational ground state, $C^{0}_{\gamma,\gamma'}$ is diagonal, 
 and, if we chose a proper order of the 
 orbitals $\gamma$,  to $0$-th  order the matrix $A$ is diagonal too, 
\begin{equation}
m^0_{I,I'}=|A|=|A^0|=\left|
\begin{array}{cccc}
A^0_{1,1}&0&\ldots&0\\
0&A^0_{2,2}&\ldots&0\\
0&\ldots&\ldots&\ldots\\
0&0&0&A^0_{N,N}
\end{array} \right|\;.
\end{equation}
For $I\neq I'$ at least one of the diagonal elements  $A^0_{i,i}$
 vanishes and we find 
\begin{equation}
m^0_{I,I'}= \delta_{I,I'}\prod_{i}A_{i,i}
=\delta_{I,I'}\prod_{\gamma \in I}n^0_{\gamma}
\prod_{\gamma \in (1\ldots N)\backslash I}(1-n^0_{\gamma})\;,
\end{equation}
as expected. In order to calculate the first and second derivative of $m^0_{I,I'}$ 
 we need to expand the determinant
\begin{equation}
|A|=|A^{0}+\delta a| 
\end{equation}
up to second order 
 with respect to the matrix elements $\delta a_{i,j}$. For this expansion
  one readily finds
\begin{equation}
|A|-|A^0|=|A^0|\sum_i\frac{\delta a_{i,i}}{A^0_{i,i}}
+|A^0|\sum_{i,j}
\frac{\delta a_{i,i} \delta a_{j,j}+\delta a_{i,j} \delta a_{j,i}}
{A^0_{i,i}A^0_{j,j} }
\;.
 \end{equation}
Note that for $A^0_{i,i}=0$ the right-hand side is defined by the
 corresponding limit $A^0_{i,i}\to 0$.

\section{Invariance of Second-Order Expansions}\label{jkio3}

\subsection{Equivalence of  the Lagrange-functional expansion}
\label{equi}
In this section, we show that the interaction kernel 
$\bar{K}^{\rho,\rho}_{Y,Y'}$ in~(\ref{hj74}) is identical to 
$K^{\rho,\rho}_{Y,Y'}$ in 
Eqs.\  (\ref{enexp})-(\ref{enexpb}).  To this end, we choose again some 
arbitrary  independent and dependent variational parameters 
$ \lambda^{\rm i}_{Z}$ and
$ \lambda^{\rm d}_{X}$, c.f. Eq.\  (\ref{2eys}).
By construction, the constraints~(\ref{const}) are automatically fulfilled
 as a function of $\vec{\lambda}^{\rm i}$ and $\vec{\rho}$,
 i.e., we have
 \begin{equation}\label{45ty}
g_{n}(\vec{\lambda}^{\rm d}(\vec{\lambda}^{\rm i},\vec{\rho}),\vec{\lambda}^{\rm i},\vec{\rho})=0\;.
\end{equation}
Consequently, all first or higher-order derivatives  of~(\ref{45ty}) 
with respect to  $\lambda^{\rm i}_{Z}$ and $\rho_Y$ vanish. For example, the 
 first-order derivatives lead to
\begin{eqnarray}\label{sd04}
 \frac{dg_n}{d\lambda^{\rm i}_{Z}} &=&\frac{\partial g_n}
{\partial \lambda^{\rm i}_{Z}}+
\sum_{X}\frac{\partial g_n}
{\partial \lambda^{\rm d}_{X}}
\frac{\partial \lambda^{\rm d}_{X}}
{\partial \lambda^{\rm i}_{Z}}=0\;,\\\label{sd04b}
 \frac{dg_n}{d\rho_{Y}} &=&\frac{\partial g_n}
{\partial \rho_{Y}}+
\sum_{X}\frac{\partial g_n}
{\partial \lambda^{\rm d}_{X}}
\frac{\partial \lambda^{\rm d}_{X}}
{\partial \rho_{Y}}=0\;.
\end{eqnarray} 
Using the matrices
\begin{equation}
G_{n,X}\equiv \frac{\partial g_n}
{\partial \lambda^{\rm d}_{X}} \;,\; \; \;
R_{X,Z}\equiv \frac{\partial \lambda^{\rm d}_{X}}
{\partial \lambda^{\rm i}_{Z}}\;,\; \; \;
Q_{X,Y}\equiv \frac{\partial \lambda^{\rm d}_{X}}
{\partial \rho_{Y}}\;,
\end{equation}
we can write Eqs.\ (\ref{sd04})-(\ref{sd04b}) as
\begin{eqnarray}\label{aus63}
\frac{\partial g_n}{\partial \lambda^{\rm i}_{Z}}&=&
-\big[\tilde{G} \tilde{R} \big]_{n,Z}\;,\\\label{aus63b}
\frac{\partial g_n}
{\partial \rho_{Y}}&=&
-\big[\tilde{G} \tilde{Q} \big]_{n,Y}\;.
\end{eqnarray}  

With the classification of dependent and independent variables we are in the 
 position to evaluate the antiadiabaticity conditions 
(\ref{lag2})-(\ref{lag3}). 
First, Eq.~(\ref{lag2}) leads to
\begin{equation}\label{lag2455}
\sum_{X}\frac{\partial g_n}{\partial\lambda^{\rm d}_X}\delta\lambda^{\rm d}_X 
+\sum_{Z}\frac{\partial g_n}{\partial\lambda^{\rm i}_Z}\delta\lambda^{\rm i}_Z 
+ \sum_Y \frac{\partial g_n}{\partial\rho_Y}\delta\rho_Y = 0
\end{equation}
which, together with Eqs.\ (\ref{aus63})-(\ref{aus63b}), yields
\begin{equation}\label{s36dd}
\tilde{G}\left[ \delta\vec{\lambda}^{\rm d}
- \tilde{R}  \delta\vec{\lambda}^{\rm i}
 - \tilde{Q} \delta\vec{\rho}
\right]=\vec{0}\;.
\end{equation}
Since the square matrix $\tilde{G}$ should be invertible, the bracket in 
(\ref{s36dd}) must vanish. Hence, we find the relation
  \begin{equation}\label{1ujrys}
\delta\vec{\lambda}^{\rm d}=\tilde{R} \delta\vec{\lambda}^{\rm i}
 + \tilde{Q} \delta\vec{\rho}
\end{equation}
 which determines the dependent-parameters fluctuations 
$\delta\vec{\lambda}^{\rm d}$ as a function of $\delta\vec{\lambda}^{\rm i}$
and $\delta\vec{\rho}$.

Applying the separation of dependent and independent parameter fluctuations 
to the second set of Eqs.\ (\ref{lag3}) yields
\begin{equation}\label{wkal}
\left(
\begin{array}{c}
\tilde{A}^{T} \\
\tilde{G}^{T}
\end{array}
\right)
\delta\vec{\Lambda}=
-\left(
\begin{array}{cc}
\tilde{L}^{\rm ii}&\tilde{L}^{\rm id} \\
\tilde{L}^{\rm di}&\tilde{L}^{\rm dd}
\end{array}
\right)
\left(
\begin{array}{c}
\delta\vec{\lambda}^{\rm i} \\
\delta\vec{\lambda}^{\rm d}
\end{array}
\right)-
\left(
\begin{array}{c}
\tilde{L}^{{\rm i}\rho} \\
\tilde{L}^{{\rm d}\rho}
\end{array}
\right)\delta\vec{\rho}\;.
\end{equation}
Here we introduced the six matrices
\begin{equation}\label{1ali}
{L}^{\rm ii}_{Z,Z'}\equiv \frac{\partial^2 L}
{\partial \lambda^{\rm i}_{Z} \partial  \lambda^{\rm i}_{Z'}}\;,\ldots,
{L}^{{\rm d}\rho}_{X,Y}\equiv\frac{\partial^2 L}
{\partial \lambda^{\rm d}_{X} \partial  \rho_{Y}}\;,
\end{equation}
of second derivatives. With~(\ref{1ujrys}) and 
the second `row' of Eqs.\  (\ref{wkal}) 
 one can write the 
 Lagrange-parameter fluctuations as a function of 
$\delta\vec{\lambda}^{\rm i}$ and $\delta\vec{\rho}$,
\begin{equation}\label{1poi}
\delta\vec{\Lambda}=-\big[\tilde{G}^{T}\big]^{-1}
\left[
(\tilde{L}^{\rm di}+\tilde{L}^{\rm dd}\tilde{R})
\delta\vec{\lambda}^{\rm i}
+(
\tilde{L}^{\rm dd}\tilde{Q}+\tilde{L}^{{\rm d}\rho}
)\delta\vec{\rho}
\right]\;.
\end{equation}
Inserting this expression into the first row 
of Eqs.\  (\ref{wkal}) and using 
\begin{equation}
\tilde{A}^{T}=-\tilde{R}^{T}\tilde{G}^{T}
\end{equation}
  we eventually find
\begin{eqnarray}\nonumber
\delta\vec{\lambda}^{\rm i}&=&
-\left[ 
\tilde{L}^{\rm ii}+\tilde{L}^{\rm id}\tilde{R}
+\tilde{R}^{T}+
\tilde{R}^{T}
\tilde{L}^{\rm dd}
\tilde{R}
\right]^{-1}\\\label{1qwvn}
&&\times\left[ 
\tilde{L}^{{\rm i}\rho}+
\tilde{L}^{\rm id}\tilde{Q}
+\tilde{R}^{T}
\tilde{L}^{{\rm d}\rho}+
\tilde{R}^{T}
\tilde{L}^{\rm dd}
\tilde{Q}
\right]\delta\vec{\rho}\;.
\end{eqnarray}

Equations~(\ref{1qwvn}),~(\ref{1poi}), and~(\ref{1ujrys}) now enable us
 to write all fluctuations  $\delta\vec{\lambda}^{\rm i}$,
 $\delta\vec{\lambda}^{\rm d}$, and $\delta\vec{\Lambda}$ as functions of 
the density fluctuations $\delta\vec{\rho}$. These relations can be inserted 
  into the second-order expansion of the Lagrange functional,
\begin{eqnarray}\nonumber
 &&\!\!\!2\delta{L}^{(2)}=
(\delta\vec{\rho})^{T}\tilde{L}^{\rho\rho}\delta\vec{\rho}
+(\delta\vec{\lambda}^{\rm i})^T\tilde{L}^{{\rm ii}}\delta\vec{\lambda}^{\rm i}
+(\delta\vec{\lambda}^{\rm d})^T\tilde{L}^{{\rm dd}}\delta\vec{\lambda}^{\rm d}\\\nonumber
&&+\Big[\big((\delta\vec{\rho})^{T}\tilde{L}^{\rho {\rm d}}\delta\vec{\lambda}^{\rm d}
+(\delta\vec{\rho})^{T}\tilde{L}^{\rho {\rm i}}\delta\vec{\lambda}^{\rm i}
+(\delta\vec{\lambda}^{\rm i})^T\tilde{L}^{{\rm id}}\delta\vec{\lambda}^{\rm d}
\big)\\\label{1gha}
&&\;\;\;\;+(\ldots)^T\Big] +2(\delta\vec{\Lambda})^T\tilde{G}
\left[ \delta\vec{\lambda}^{\rm d}
- \tilde{R}  \delta\vec{\lambda}^{\rm i}
 - \tilde{Q} \delta\vec{\rho}
\right]
\end{eqnarray}
in order to calculate $\bar{K}^{\rho\rho}_{Y,Y'}$ in Eq.\  (\ref{hj74}).
 However, to prove just the identity of $\bar{K}^{\rho\rho}_{Y,Y'}$ and  
$K^{\rho\rho}_{Y,Y'}$ in~(\ref{enexp}) it is 
sufficient to apply only 
 Eq.\  (\ref{1ujrys}) to the expansion~(\ref{1gha}). This leads to 
 \begin{eqnarray}\label{1eqnh}
2\delta{L}^{(2)}&=&
(\delta\vec{\rho})^{T}\big(
\tilde{L}^{\rho\rho}+\tilde{Q}^T\tilde{L}^{{\rm d}\rho}
+\tilde{L}^{\rho{\rm d}}\tilde{Q}+
\tilde{Q}^T\tilde{L}^{\rm dd}\tilde{Q}
\big)\delta\vec{\rho}\\\nonumber
&&+(\delta\vec{\lambda}^{\rm i})^T
\big(
\tilde{L}^{\rm ii}+\tilde{L}^{\rm id}\tilde{R}
+\tilde{R}^T\tilde{L}^{\rm di}+
\tilde{R}^T\tilde{L}^{\rm dd}\tilde{R}
\big)\delta\vec{\lambda}^{\rm i}\\\nonumber
&&+\Big[(\delta\vec{\rho})^{T}\big( 
\tilde{L}^{\rho{\rm i}}+\tilde{L}^{\rho{\rm i}}\tilde{R}
+\tilde{Q}^T\tilde{L}^{\rm di}
+\tilde{Q}^T\tilde{L}^{\rm dd}\tilde{R}
\big)\delta\vec{\lambda}^{\rm i}+\big(\ldots\bigl)^T\Big]\;.
 \end{eqnarray}

As we will show below, 
the matrices~(\ref{mat1})-(\ref{mat4}) which determine the second order
 expansion~(\ref{qqq1})  are the same 
 as the corresponding matrices in~(\ref{1eqnh}). Hence, we have
\begin{equation}
 \delta{E}^{(2)}=\delta{L}^{(2)}\;.
\end{equation}
Since the antiadiabaticity condition 
 \begin{equation}
\frac{\partial  \delta{E}^{(2)} }{\partial \delta\lambda^{\rm i}_{Z}}
=\frac{\partial  \delta{L}^{(2)} }{\partial \delta\lambda^{\rm i}_{Z}}=0
\end{equation}
for $\delta{E}^{(2)}$  reproduces Eq.~(\ref{1qwvn}), 
the  identity of  $\bar{K}^{\rho\rho}_{Y,Y'}$ and 
$K^{\rho\rho}_{Y,Y'}$ is then finally demonstrated.
 
It remains to be shown that the matrices~(\ref{mat1})-(\ref{mat4}) agree with 
those in~(\ref{1eqnh}).
To this end, we use the explicit form~(\ref{1uyr}) of the  
energy functional~(\ref{10.1030}) that appears in the definition of the 
 matrices~(\ref{mat1})-(\ref{mat4}).
As an example, we consider the matrix $\tilde{M}^{\rho\rho}$ 
and show that it is identical to the matrix in the first line of~(\ref{1eqnh}).
 With similar derivations one can prove the same for the other 
matrices~(\ref{mat2}),(\ref{mat4}) and their counterparts in~(\ref{1eqnh}).

 Using~(\ref{1uyr}) and~(\ref{mat1}) we find
 \begin{eqnarray}\nonumber
M^{\rho\rho}_{Y,Y'}&=&\big[\tilde{E}^{\rho\rho}+
\tilde{Q}^T\tilde{E}^{{\rm d}\rho}+
\tilde{E}^{\rho {\rm d}}  \tilde{Q} +
\tilde{Q}^T\tilde{E}^{\rm dd}\tilde{Q}\big]_{Y,Y'}
+2\sum_X\frac{\partial E}{\partial \lambda^{\rm d}_X}
\cdot \frac{\partial^2 \lambda^{\rm d}_X}{\partial \rho_Y \partial \rho_{Y'}}\;.
\\\label{1uye}
  \end{eqnarray}
Here, the matrices 
\begin{equation}
\tilde{E}^{\rho\rho}=\tilde{L}^{\rho\rho}-\sum_n\Lambda_n\tilde{g}_n^{\rho\rho}
\;,\ldots\;,
\tilde{E}^{\rm dd}=\tilde{L}^{\rm dd}-\sum_n\Lambda_n\tilde{g}_n^{\rm dd}\;,
 \end{equation}  
and  $\tilde{g}_n^{\rho\rho},\ldots,\tilde{g}_n^{\rm dd} $ are defined as 
in~(\ref{1ali}) only with $L$ replaced by $E$ or $g_n$ respectively.
Obviously, the matrix  in the first line of~(\ref{1eqnh}) is identical to 
 $\tilde{M}^{\rho\rho}$ if
\begin{equation}\label{1oart}
2\sum_X\frac{\partial E}{\partial \lambda^{\rm d}_X}
\cdot \frac{\partial^2 \lambda^{\rm d}_X}{\partial \rho_Y \partial \rho_{Y'}}
=-\sum_n\Lambda_n\big[
\tilde{g}_n^{\rho\rho}+\tilde{Q}^T\tilde{g}_n^{{\rm d}\rho}
+\tilde{g}_n^{\rho{\rm d}}\tilde{Q}+
\tilde{Q}^T\tilde{g}_n^{\rm dd}\tilde{Q}
\big]_{Y,Y'}\;.
\end{equation}
To prove~(\ref{1oart}), we
use the fact that the second (total) derivatives of~(\ref{45ty}) 
with respect to the densities $\rho_Y$ vanish
\begin{eqnarray}\nonumber
\frac{d g_n}{d\rho_Y d\rho_{Y'}}&=&\big[
\tilde{g}_n^{\rho\rho}+\tilde{Q}^T\tilde{g}_n^{{\rm d}\rho}
+\tilde{g}_n^{\rho{\rm d}}\tilde{Q}+
\tilde{Q}^T\tilde{g}_n^{\rm dd}\tilde{Q}
\big]_{Y,Y'}+2\sum_X\frac{\partial g_n}{\partial \lambda^{\rm d}_X}
\cdot \frac{\partial^2 \lambda^{\rm d}_X}{\partial \rho_Y \partial \rho_{Y'}}\\
&=&0\;.
\end{eqnarray}
Equation~(\ref{1oart}) is therefore fulfilled if
\begin{equation}
\sum_X\bigg(\frac{\partial E}{\partial \lambda^{\rm d}_X}+
\sum_n\Lambda_n\frac{\partial g_n}{\partial \lambda^{\rm d}_X}
\bigg)\frac{\partial^2 \lambda^{\rm d}_X}{\partial \rho_Y 
\partial \rho_{Y'}}=0\;.
\end{equation}
This equation, however, holds trivially, since~(\ref{2eiusy})
 leads to 
 \begin{equation}\label{1qazx}
 \frac{\partial L}{\partial \lambda_{Z}}=\frac{\partial E}{\partial \lambda_{Z}}+
\sum_n\Lambda_n\frac{\partial g_n}{\partial \lambda_{Z}}=0
\end{equation}
for all parameters $\lambda_{Z}$ and in particular for 
$\lambda_{Z}=\lambda^{\rm d}_{X}$ as it appears in~(\ref{1qazx}).

 \subsection{Linear transformations of the density matrix}\label{8uh}
In investigations of our translationally invariant lattice systems~(\ref{1.1}) 
 it turns out to be more convenient to work with fluctuations $\delta \vec{\mu}$
 which are linearly related to the density-matrix fluctuations, 
\begin{equation}\label{7dgsd}
\delta \vec{\rho}=\tilde{\Xi} \cdot \delta \vec{\mu}
\end{equation}
 c.f., Eqs.\  (\ref{4yut})-(\ref{4yutb}) and the resulting 
Green's functions~(\ref{8dgdd}). 
 The effective second-order functional~(\ref{enexp})-(\ref{enexpb})
 in terms of the 
fluctuations $\delta \vec{\mu}$ is then given as 
\begin{equation}\label{7dgs}
 \delta E^{(2)}(\delta \vec{\mu} ) =  \frac{1}{2} ( \delta \vec{\mu})^{\rm T}
 (\tilde{\Xi}^{T}\tilde{K}^{\rho \rho}  \tilde{\Xi}) \delta \vec{\mu} 
\end{equation}
with $\tilde{K}^{\rho \rho}$ as defined in~(\ref{enexpb}). 
 For numerical calculations it is important to show that one obtains 
 the same kernel 
\begin{equation}
\tilde{K}^{\mu \mu}\equiv \tilde{\Xi}^{T}\tilde{K}^{\rho \rho}  \tilde{\Xi}
\end{equation}
 as in~(\ref{7dgs}) if the transformation~(\ref{7dgsd}) and the 
  antiadiabaticity condition are applied in the reverse order:  
  If we apply~(\ref{7dgsd}) first to~(\ref{qqq1}), we obtain
\begin{equation}\label{qqq1gg}
\delta E^{(2)}=\frac{1}{2}\Big[
 (\delta \vec{\mu})^{T}\tilde{\Xi}^{T} \tilde{M}^{\rho\rho}\tilde{\Xi}
\delta \vec{\mu}
+2(\delta\vec{\lambda}^{\rm i})^{T}  \tilde{M}^{\lambda\rho}\tilde{\Xi}\delta \vec{\mu}+(\delta \vec{\lambda}^{\rm i})^{T} \tilde{M}^{\lambda\lambda}\delta 
\vec{\lambda}^{\rm i}\Big]\;.
\end{equation}
 The antiadiabaticity condition for $\delta\vec{ \mu}$ then reads 
\begin{equation} \label{} 
\delta \vec{\lambda}^{\rm i}=-\left[\tilde{M}^{\lambda\lambda}\right]^{-1}
\tilde{M}^{\lambda\rho}\tilde{\Xi}
\delta\vec{ \mu}\;.
\end{equation}
Inserted into~(\ref{qqq1gg}) this equation yields 
 \begin{eqnarray}
\delta E^{(2)}(\delta\vec{\mu})&=&E_0+\frac{1}{2} (\delta\vec{\mu})^{T}
\tilde{K}^{\mu\mu}\delta\vec{\mu}\;,\\\nonumber
\tilde{K}^{\mu\mu}&=&\tilde{\Xi}^{T}\tilde{M}^{\rho\rho}\tilde{\Xi}
-\tilde{\Xi}^{T}\tilde{M}^{\rho\lambda}
\left[\tilde{M}^{\lambda\lambda}\right]^{-1}
\tilde{M}^{\lambda\rho}\tilde{\Xi}
=\tilde{\Xi}^{T}\tilde{K}^{\rho \rho}  \tilde{\Xi}\;,
\end{eqnarray}  
as claimed above. 

\section{Explicit form of the second-order expansion}\label{seexp}
We calculate the second-order expansion of the Lagrange functional 
with respect to the variational parameters $\lambda_{i;\Gamma,\Gamma'}$
 and the density matrix~(\ref{10.780}). For the general consideration in 
 section~\ref{chap7.3.3} and \ref{jkio3} it was convenient to 
subsume  the  parameters $\lambda_{\Gamma,\Gamma'}$  and their 
conjugates $\lambda^*_{\Gamma,\Gamma'}$ in a set of  $n_{\rm p}$ 
parameters $\lambda_{Z}$,  c.f., Eq.\  (\ref{10.1030b}). Here in this 
appendix, where we aim to resolve the explicit structure of the second-order 
expansion, it is better to
 take the difference  between $\lambda_{\Gamma,\Gamma'}$ and 
 $\lambda_{\Gamma,\Gamma'}^*$ into account. 

The constraints~(\ref{1.10c})-(\ref{1.10d}),  
the local energy~(\ref{0agh}), and
  the renormalisation matrix~(\ref{qmat}) are all functions only 
 of $\lambda^*_{i;\Gamma,\Gamma'},\lambda_{i;\Gamma,\Gamma'}$ and of the 
 local density matrix $C^{0}_{i;\sigma,\sigma'}$. For simplicity we use the 
joint variables $A^{i}_{v},(A^{i}_{v})^*$ for all these local variables, i.e., 
it is either
\begin{equation}
A^{i}_{v}=A^{i}_{\sigma_1,\sigma_2}=\langle\hcd_{i,\sigma_2}
\hc_{i,\sigma_1}  \rangle\;\;\;\;{\rm or}\;\;\;\;A^{i}_{v}=A^{i}_{\Gamma,\Gamma'}
 =\lambda_{i;\Gamma,\Gamma'}\;.
\end{equation}
 With respect to the parameters $\lambda^*_{i;\Gamma,\Gamma'},
\lambda_{i;\Gamma,\Gamma'}$ the second derivatives 
 of~(\ref{1.10c})-(\ref{1.10d}),~(\ref{0agh}), 
and~(\ref{qmat}) are quadratic functions of the form 
$\sim (A^{i}_{v})^*A^{i}_{v'}$. Due to the 
 Hermiticity of the density matrix the same can be achieved with respect
 to the local density matrix. Then the only finite second derivatives 
 of the Lagrange functional
\begin{eqnarray}\label{3ued}
L&=&T+\sum_{i}E_{i,\rm loc}(\{(A_v^i)^*\},\{A_v^i\})
+\sum_{i,n}\Lambda_{i,n}g_{i,n}(\{(A_v^i)^*\},\{A_v^i\})\\\nonumber
&\equiv&T+L_{\rm loc}\\
T&=&\sum_{i\neq j}\sum_{\sigma_1,\sigma_2\atop \sigma'_1,\sigma'_2}
t^{\sigma_1,\sigma_2}_{i,j}
q_{i,\sigma_1}^{\sigma'_1}\left( q_{j,\sigma_2}^{\sigma'_2}\right)^{*}
\langle \hcd_{i\sigma'_1} \hc_{j\sigma'_2} \rangle
\end{eqnarray}
are
\begin{equation}
\frac{\partial^2 L}{\partial (A^{i}_{v})^*  \partial A^{i}_{v'} }\neq 0
\end{equation}
whereas
\begin{equation}
\frac{\partial^2 L}{\partial (A^{i}_{v})^* \partial (A^{i}_{v'})^* }=
\frac{\partial^2 L}{\partial A^{i}_{v} \partial A^{i}_{v'} }=0\;.
\end{equation}
 
The second-order expansion of the constraints
 and the local energy is straightforward since only 
 local fluctuations $\delta A^i_{v}$ couple,
\begin{equation}\label{4509}
\delta L^{(2)}_{\rm loc}=\sum_{q}\sum_{v,v'}
(\delta A^q_{v})^* 
K^{\rm loc}_{v,v'}\ 
\delta A^{q}_{v'}
\end{equation}
where we introduced
\begin{equation}\label{opu}
K^{\rm loc}_{v,v'}=\frac{\partial^2 L_{\rm loc}}{\partial (A^{i}_{v})^*  
\partial A^{i}_{v'} }
\end{equation}
and the Fourier transforms  of the local fluctuations
\begin{equation}\label{4eud}
\delta A^i_{v}=\frac{1}{\sqrt{L_s}}
\sum_q e^{-{\rm i}\vec{R}_i\cdot \vec{q}}\delta A^q_{v}\;.
\end{equation}
All derivatives in this section  (e.g., (\ref{opu}))  
have to be evaluated for the ground-state
 values of the variational parameters 
$\lambda_{i;\Gamma,\Gamma'}$, the density matrix $\tilde{\rho}$, and 
 the Lagrange parameters $\Lambda_{i,n}$.   
Note that the density-matrix fluctuations 
$\delta A^q_{\sigma_2,\sigma_1}$ can 
be written as
\begin{eqnarray}\nonumber
\delta A^q_{\sigma_2,\sigma_1}&=&\frac{1}{\sqrt{L_s}}
\sum_ie^{{\rm i}\vec{R}_i\cdot \vec{q}}
\delta\langle\hcd_{i,\sigma_1}
\hc_{i,\sigma_2}  \rangle=\frac{1}{\sqrt{L_s}}
\sum_k\delta\langle\hcd_{k,\sigma_1}
\hc_{k+q,\sigma_2}  \rangle\\\label{7yhu}
&&= \delta \langle \hat{A}^q_{\sigma_2,\sigma_1} \rangle
\end{eqnarray}
 where the operator $\hat{A}^q_{v}$ has been defined in~(\ref{4vbyt}).

In addition to (\ref{4509}), we need to take into account 
 the mixed terms $\sim \delta A^i_v \delta \Lambda_{i,n}$. 
 In real space, their contribution is given as 
\begin{equation}\label{qwdf}
\delta L^{(2)}_{\rm c}=\sum_{i,n,v}\Bigg(
\frac{\partial g_{i,n}}{\partial (A^i_{v})^*}
\delta (A^i_{v})^*+
\frac{\partial g_{i,n}}{\partial A^i_{v}}
\delta A^i_{v}
\Bigg)\delta \Lambda_{i,n}\;.
\end{equation}
If we introduce the Fourier transforms $\delta \Lambda^q_{n}$
 of the fluctuations $\delta \Lambda_{i,n}$, we can write
 (\ref{qwdf}) as
\begin{equation}
\delta L^{(2)}_{\rm c}=\sum_q\sum_{n,v}
(\delta A^q_v)^*K^{\rm c}_{v,n}\delta \Lambda^q_{n}+
(\delta \Lambda^q_{n})^*(K^{\rm c}_{v,n})^*\delta A^q_v\;.
\end{equation}
Here, we used that the constraints $g_{i,n}$ are assumed to be  real 
and lattice-site  independent such that 
\begin{equation}
K^{\rm c}_{v,n}\equiv\frac{\partial g_{i,n}}{\partial (A^i_{v})^*}
=\left(\frac{\partial g_{i,n}}{\partial A^i_{v}}\right)^*\;.
\end{equation}

 More involved than the calculation of~(\ref{4509}) is the 
 expansion of the kinetic energy. Here we find 
\begin{equation}\label{dghs1}
\delta T^{(2)}=\delta T^{(2)}_{\rm l} +\delta T^{(2)}_{\rm t}
\end{equation}
with
\begin{eqnarray}\nonumber
\delta T^{(2)}_{\rm l}&=&\sum_{i\neq j}
\sum_{\sigma_1,\sigma_2,\sigma'_1,\sigma'_2}t^{\sigma_1,\sigma_2}_{i,j}
\langle \hcd_{i\sigma'_1} \hc_{j\sigma'_2} \rangle
\sum_{v,v'}\Bigg[
\frac{\partial^2 q_{i,\sigma_1}^{\sigma'_1}}{\partial( A^i_{v})^*\partial A^i_{v'}}
\left( q_{j,\sigma_2}^{\sigma'_2}\right)^{*}(\delta A^i_{v})^*\delta A^i_{v'}
\\\nonumber
&&+\frac{1}{2}\Bigg(\frac{\partial q_{i,\sigma_1}^{\sigma'_1}}
{\partial (A^i_{v})^*}
\frac{\partial \left( q_{j,\sigma_2}^{\sigma'_2}\right)^{*}}{\partial A^j_{v'}}
(\delta A^i_{v})^*\delta A^j_{v'}
+\frac{\partial q_{i,\sigma_1}^{\sigma'_1}}
{\partial A^i_{v}}
\frac{\partial \left( q_{j,\sigma_2}^{\sigma'_2}\right)^{*}}
{\partial (A^j_{v'})^*}
\delta A^i_{v}(\delta A^j_{v'})^*\Bigg)
\Bigg]\\\label{3ed}
&&+{\rm c.c.}
\end{eqnarray}
and 
\begin{eqnarray}\label{3ef}
\delta T^{(2)}_{\rm t}&=&\sum_{i\neq j}
\sum_{\sigma_1,\sigma_2, \sigma'_1,\sigma'_2}t^{\sigma_1,\sigma_2}_{i,j}
\delta\langle \hcd_{i\sigma'_1} \hc_{j\sigma'_2} \rangle \\\nonumber
&&\!\!\!\!\!\times \sum_v
\Bigg[
\frac{\partial q_{i,\sigma_1}^{\sigma'_1}}{\partial (A^i_{v})^*}
\left(q_{j,\sigma_2}^{\sigma'_2}\right)^{*}
(\delta A^i_{v})^*
+q_{i,\sigma_1}^{\sigma'_1} \frac{\partial \left(q_{j,\sigma_2}^{\sigma'_2}\right)^{*}}{\partial (A^j_{v})^*}
(\delta A^j_{v})^*
\Bigg]+{\rm c.c.}\;.
\end{eqnarray}  
The fact that the complex conjugates give 
 the terms not explicitly shown in Eqs.\  (\ref{3ed})-(\ref{3ef}) 
follows from the relations 
\begin{eqnarray}
\left(\frac{\partial q_{\sigma}^{\sigma'}}{\partial A_{v}}\right)^*
&=&\frac{\partial\left( q_{\sigma}^{\sigma'}\right)^*}{\partial (A_{v})^*}\;,\\
\left(\frac{\partial^2 q_{\sigma}^{\sigma'}}{\partial (A_{v})^*
\partial A_{v'}}\right)^*
&=&\frac{\partial^2 \left( q_{\sigma}^{\sigma'}\right)^*}
{\partial (A_{v'})^*
\partial A_{v}}\;,\\
\left( t_{i,j}^{\sigma,\sigma'}\right)^*&=&t_{j,i}^{\sigma',\sigma}\;.
\end{eqnarray}  
For our translationally invariant ground state it is more convenient to 
write~(\ref{3ed})-(\ref{3ef})
in momentum space: With the Fourier transforms
of the local fluctuations the term~(\ref{3ed}) reads
\begin{equation}
\delta T^{(2)}_{\rm l}=\sum_{q}\sum_{v,v'}
(\delta A^q_{v})^* 
\big[K^{\rm l}_{q;v,v'}+(K^{\rm l}_{q;v',v})^*\big] 
\delta A^{q}_{v'}
\end{equation}
where 
\begin{eqnarray}\nonumber
K^{\rm l}_{q;v,v'}&\equiv&\sum_{\sigma_1,\sigma_2, \sigma'_1,\sigma'_2}\Bigg[
\frac{1}{2}E_{\sigma_1,\sigma_2, \sigma'_1,\sigma'_2}(\vec{q})
\Bigg(
\frac{\partial q_{\sigma_1}^{\sigma'_1}}{\partial (A_{v})^*}
\frac{\partial \left( q_{\sigma_2}^{\sigma'_2}\right)^{*}}{\partial A_{v'}}
+
\frac{\partial q_{\sigma_1}^{\sigma'_1}}{\partial A_{v'}}
\frac{\partial \left( q_{\sigma_2}^{\sigma'_2}\right)^{*}}{\partial (A_{v})^*}
\Bigg)
\\
&&\;\;\;\;\;\;\;\;\;+E_{\sigma_1,\sigma_2, \sigma'_1,\sigma'_2}
\frac{\partial^2 q_{\sigma_1}^{\sigma'_1}}{\partial (A_{v})^*\partial A_{v'}}
 \left( q_{\sigma_2}^{\sigma'_2}\right)^{*}\Bigg]\;.
\end{eqnarray}
Here we assumed that the renormalisation matrix 
is lattice-site independent and introduced the tensor 
  \begin{equation}\label{4tgh}
E_{\sigma_1,\sigma_2, \sigma'_1,\sigma'_2}(\vec{q})
=\frac{1}{L_{\rm s}}\sum_{k}\epsilon_{k+q}^{\sigma_1,\sigma_2}
\langle  \hcd_{k\sigma'_1} \hc_{k\sigma'_2}  \rangle
\end{equation}
with
\begin{equation}
\epsilon_{k}^{\sigma_1,\sigma_2}=\frac{1}{L_{\rm s}}\sum_{i\neq j} t^{\sigma_1,\sigma_2}_{i,j}e^{{\rm i}\vec{k}(\vec{R}_i-\vec{R}_j)}\;.
\end{equation}  
Note that for $\vec{q}=0$ the tensor~(\ref{4tgh}),
\begin{equation}
E_{\sigma_1,\sigma_2, \sigma'_1,\sigma'_2}
=E_{\sigma_1,\sigma_2, \sigma'_1,\sigma'_2}(0)
\end{equation}
has already been defined in~(\ref{0etsy}). For the evaluation of 
the second ('transitive') term 
(\ref{3ef}) we write the non-local density-matrix fluctuations as
\begin{equation}
\delta\langle \hcd_{i\sigma'_1} \hc_{j\sigma'_2} \rangle =\frac{1}{L_{\rm s}}\sum_{k,k'}
e^{{\rm i}(\vec{R}_i\cdot \vec{k}-\vec{R}_j\cdot \vec{k}')}
\delta \langle  \hcd_{k\sigma'_1} \hc_{k'\sigma'_2}  \rangle \;.
\end{equation}
 Together with~(\ref{4eud}) this yields
\begin{equation}\label{4jff}
\delta T^{(2)}_{\rm t}=\frac{1}{L_{\rm s}}\sum_{q,k}\sum_{v;\sigma'_1,\sigma_2'}
(\delta A^q_{v})^* \bar{K}^{\rm t}_{k,q;v,\sigma'_1,\sigma_2'} 
\delta \langle  \hcd_{k\sigma'_1} \hc_{k+q\sigma_2'}  \rangle+{\rm c.c}
\end{equation}
with
\begin{equation}\label{4jffb}
\bar{K}^{\rm t}_{k,q;v,\sigma'_1,\sigma_2'} =\sum_{\sigma_1,\sigma_2}\Big[
\frac{\partial q_{\sigma_1}^{\sigma'_1}}{\partial (A_{v})^*}
\left(q_{\sigma_2}^{\sigma_2'}\right)^{*}\epsilon^{\sigma_1,\sigma_2}_{k+q}
+q_{\sigma_1}^{\sigma'_1} \frac{\partial \left(q_{\sigma_2}^{\sigma_2'}\right)^{*}}
{\partial (A_{v})^*}\epsilon^{\sigma_1,\sigma_2}_{k}\Big]\;.
\end{equation}  

In principle, Eqs.\  (\ref{4jff})-(\ref{4jffb}) allow us to 
calculate all second-order 
couplings  of density-matrix and parameter fluctuations that arise
 from $\delta T^{(2)}_{\rm t}$. For numerical 
 calculations, however, these equations are not very useful due to the 
explicit $k$~dependence of~(\ref{4jffb}). It is much easier
 to introduce the two auxiliary fluctuations 
\begin{eqnarray}\label{4yut}
\delta B^{q}_{w}&\equiv&
\delta B^{q}_{\sigma_2,\sigma_1,\sigma_2',\sigma_1'}\equiv\frac{1}{\sqrt{L_s}}
\sum_k\epsilon^{\sigma_1,\sigma_2}_{k}
\delta \langle \hcd_{k\sigma'_1} \hc_{k+q\sigma'_2} \rangle\;,\\\label{4yutb}
\delta \bar{B}^{q}_{w}&\equiv&
\delta \bar{B}^{q}_{\sigma_2,\sigma_1,\sigma_2',\sigma_1'}\equiv\frac{1}{\sqrt{L_s}}
\sum_k\epsilon^{\sigma_1,\sigma_2}_{k+q}
\delta \langle \hcd_{k\sigma'_1} \hc_{k+q\sigma'_2}  \rangle\;,
\end{eqnarray}  
where $w\equiv(\sigma_2,\sigma_1,\sigma_2',\sigma_1')$ is an abbreviation
for quadruples of indices $\sigma$. With these definitions we can write 
(\ref{4jff}) as 
\begin{eqnarray}\label{4kfjf}
\delta T^{(2)}_{\rm t}&=&\sum_{q}\sum_{v,w}\Big[
(\delta A^q_{v})^*K^{{\rm t}(1)}_{vw}
\delta B^{q}_{w}+(\delta A^q_{v})^*K^{{\rm t}(2)}_{vw}
\delta \bar{B}^{q}_{w}\\\nonumber
&&+(\delta B^{q}_{w})^*(K^{{\rm t}(1)}_{vw})^*
\delta A^{q}_{v}+(\delta \bar{B}^{q}_{w})^*(K^{{\rm t}(2)}_{vw})^*
\delta A^{q}_{v}\Big]
\end{eqnarray}
where 
\begin{eqnarray}
K^{{\rm t}(1)}_{v(\sigma_2,\sigma_1,\sigma_2',\sigma_1')}&\equiv& 
q_{\sigma_1}^{\sigma'_1} \frac{\partial \left(q_{\sigma_2}^{\sigma'_2}\right)^{*}}
{\partial (A_{v})^*}\;,\\
K^{{\rm t}(2)}_{v(\sigma_2,\sigma_1,\sigma_2',\sigma_1')}&\equiv& 
\frac{\partial q_{\sigma_1}^{\sigma'_1}}{\partial (A_{v})^*}
\left(q_{\sigma_2}^{\sigma'_2}\right)^{*}\;.
\end{eqnarray}  
Note that we introduced  the {\sl two} different fluctuations 
(\ref{4yut}),(\ref{4yutb}) only because they allow us to 
 write the second-order expansion in a relatively simple form.
 In fact, these fluctuations  are not independent but 
related through
\begin{equation}
 \delta \bar{B}^{q}_{\sigma_1,\sigma_2,\sigma_1',\sigma_2'}
=\big(\delta B^{-q}_{\sigma_2,\sigma_1,\sigma_2',\sigma_1'}\big)^*\;.
\end{equation}

Altogether we end up with the following second-order expansion of the
 Lagrange functional 
\begin{equation}\label{4kfjfb}
\delta L^{(2)}=\frac{1}{L_{\rm s}}\sum_q
\left(
\begin{array}{cccc}
\delta \vec{A}^q&\delta \vec{B}^q&\delta \vec{\bar{B}}^q&
\delta \vec{\Lambda}^q
\end{array}
\right)^{*}
\tilde{K}^{q}
\left(
\begin{array}{c}
\delta \vec{A}^q\\
\delta \vec{B}^q\\
\delta \vec{\bar{B}}^q\\
\delta\vec{\Lambda}^q
\end{array}
\right)
\end{equation}
 where
\begin{equation}\label{7hdz}
\tilde{K}^{q}\equiv
\left(
\begin{array}{cccc}
\tilde{K}^{(A,A)}&\tilde{K}^{(A,B)}&\tilde{K}^{(A,\bar{B})}&\tilde{K}^{(A,\Lambda)}\\
\big(\tilde{K}^{(A,B)}\big)^{\dagger}&0&0&0\\
\big(\tilde{K}^{(A,\bar{B})}\big)^{\dagger}&0&0&0\\
\big(\tilde{K}^{(A,\Lambda)}\big)^{\dagger}&0&0&0
\end{array}
\right)
\end{equation} 
and
\begin{eqnarray}
&&\tilde{K}^{(A,A)}\equiv \tilde{K}^{\rm loc}+\tilde{K}_q^{\rm l}+(\tilde{K}_q^{\rm l})^{\dagger}\;,\\
&&\tilde{K}^{(A,B)}\equiv\tilde{K}^{{\rm t}(1)}\;,\;
\tilde{K}^{(A,\bar{B})}\equiv \tilde{K}^{{\rm t}(2)}\;,\\
&&\tilde{K}^{(A,\Lambda)}\equiv \tilde{K}^{\rm c}\;.
\end{eqnarray}
As described in section~\ref{lpe}, the antiadiabaticity condition  
leads to an  effective second-order functional only of the 
 density matrix. This condition can be evaluated  directly for  the 
second-order 
 expansion~(\ref{4kfjfb}) since the fluctuations $\delta \vec{A}^q$,
 $\delta \vec{B}^q$, $\delta \vec{\bar{B}}^q$ are some linear functions 
 of the density-matrix fluctuations $\delta\langle  \hcd_{k,\sigma_1}
 \hc_{k+q,\sigma_2}\rangle $, c.f., \ref{8uh}. To this end, we distinguish 
 the fluctuations of the local density matrix
 $\delta \vec{A}^{q}_{\rho}$ and of the variational parameters  
$\delta \vec{A}^{q}_{\lambda}$ as well as the corresponding blocks in the 
matrix~(\ref{7hdz}),
\begin{equation}
\tilde{K}^{(A,A)}=
\left(
\begin{array}{cc}
\tilde{K}^{(A,A)}_{(\lambda,\lambda)}&\tilde{K}^{(A,A)}_{(\lambda,\rho)}\\
\big(\tilde{K}^{(A,A)}_{(\lambda,\rho)}\big)^{\dagger}&\tilde{K}^{(A,A)}_{(\rho,\rho)}
\end{array}
\right)
\;,\;
\tilde{K}^{(A,B)}=
\left(
\begin{array}{c}
\tilde{K}^{(A,B}_{(\lambda)}\\
\tilde{K}^{(A,B}_{(\rho)}
\end{array}
\right)\;,\ldots
\end{equation}
The resulting functional is then given as
\begin{equation}\label{4kfjfbv}
\delta \bar{L}^{(2)}=\frac{1}{L_{\rm s}}\sum_q
\left(
\begin{array}{ccc}
\delta \vec{A}_{\rho}^q&\delta \vec{B}^q&\delta \vec{\bar{B}}^q
\end{array}
\right)^{*}
\tilde{V}^{q}
\left(
\begin{array}{c}
\delta \vec{A}_{\rho}^q\\
\delta \vec{B}^q\\
\delta \vec{\bar{B}}^q
\end{array}
\right)
\end{equation}
 with the new kernel 
\begin{eqnarray}\label{vq}
\tilde{V}^{q}&\equiv&\left(
\begin{array}{ccc}
\tilde{V}^{(A,A)}&\tilde{V}^{(A,B)} &\tilde{V}^{(A,\bar{B})}  \\
\tilde{V}^{(B,A)}&\tilde{V}^{(B,B)}&\tilde{V}^{(B,\bar{B})} \\
\tilde{V}^{(\bar{B},A)}&\tilde{V}^{(\bar{B},B)} &\tilde{V}^{(\bar{B},\bar{B})}
\end{array}
\right)\\
&=&\left(
\begin{array}{ccc}
\tilde{K}^{(A,A)}_{\rho,\rho}&\tilde{K}_{\rho}^{(A,B)}&\tilde{K}_{\rho}^{(A,\bar{B})}  \\
\big(\tilde{K}_{\rho}^{(A,B)}\big)^{\dagger}&0&0 \\
\big(\tilde{K}_{\rho}^{(A,\bar{B})}\big)^{\dagger}&0&0
\end{array}
\right)-\Delta\tilde{V}^{q}
\end{eqnarray}
where
\begin{eqnarray}\nonumber
\Delta\tilde{V}^{q}&\equiv&
\left(
\begin{array}{cc}
\tilde{K}^{(A,A)}_{\rho,\lambda}&\tilde{K}^{(A,\Lambda)}_{\rho}\\
\big(\tilde{K}^{(A,B)}_{\lambda}\big)^{\dagger}&0\\
\big(\tilde{K}^{(A,\bar{B})}_{\lambda}\big)^{\dagger}&0
\end{array}
\right)
\times
\left(
\begin{array}{cc}
\tilde{K}^{(A,A)}_{\lambda,\lambda}&\tilde{K}^{(A,\Lambda)}_{\lambda,\Lambda}\\
\big(\tilde{K}^{(A,\Lambda)}_{\lambda,\Lambda}\big)^{\dagger}&0
\end{array}
\right)^{-1}\\
&&\times\left(
\begin{array}{ccc}
\tilde{K}^{(A,A)}_{\lambda,\rho}&\tilde{K}^{(A,B)}_{\lambda}&
\tilde{K}^{(A,\bar{B})}_{\lambda}\\
\big(\tilde{K}^{(A,\Lambda)}_{\rho}\big)^{\dagger}&0&0
\end{array}
\right)
\end{eqnarray}
Note that $\tilde{V}^{q}$  
(unlike $\tilde{K}^{q}$) includes finite couplings also between the 
 fluctuations $\delta \vec{B}^q$, $\delta \vec{\bar{B}}^q$.
The calculation of $\tilde{V}^{q}$ (for fixed $\vec{q}$) only 
involves  the handling of finite-dimensional matrices. In contrast, 
 the evaluation 
 of the functional~(\ref{4jff}) (instead of~(\ref{4kfjf}))   
would have lead to significantly more complicated equations.

\section{Explicit form of the Gutzwiller-RPA equations}\label{app56}
In this appendix, we prove that the general Gutzwiller RPA Eqs.\  
(\ref{10.1250})  lead to the Green's function 
matrix~(\ref{0897}) if applied to our multi-band Hamiltonian~(\ref{1.1}). 
With the abbreviations
$\delta D^q_{\mu}$, $\hat{D}^q_{\mu}$ for the three fluctuations 
$\delta A^{q}_{v}$, $\delta B^{q}_{w}$,    $\delta \bar{B}^{q}_{w}$ and the
 corresponding operators $ \hat{A}^q_{v}$, $ \hat{B}^q_{w}$, 
$ \hat{\bar{B}}^q_{v}$, we have to show that the Green's function 
matrix
\begin{equation}
\Pi_{\mu,\mu'}(\vec{q},\omega)=\langle \langle 
\hat{D}^q_{\mu};(\hat{D}^q_{\mu'})^{\dagger}
\rangle \rangle_{\omega} \;,
\end{equation}
as given in~(\ref{0897}),  obeys the equation
\begin{equation}
\delta D^q_{\mu}=\sum_{\mu'}\langle \langle 
\hat{D}^q_{\mu};(\hat{D}^q_{\mu'})^{\dagger}
\rangle \rangle_{\omega} \delta f^q_{\mu'}\;.
\end{equation}
Using the explicit form~(\ref{0897}) of 
 $\tilde{\Pi}(\vec{q},\omega)$,  this equation can also be written as
\begin{equation}\label{10shd}
\sum_{\mu'}[1+\tilde{\Pi}^0(\vec{q},\omega)\tilde{V}^{q}]_{\mu,\mu'}\delta D^q_{\mu'}
=\sum_{\mu'}\Pi^0_{\mu,\mu'}(\vec{q},\omega)\delta f^q_{\mu'}
\end{equation}
Note that the excitation amplitudes $\delta f^q_{\mu}$ enter the problem 
 through the perturbation operator
\begin{eqnarray}\nonumber
\delta\hat{V}_f\equiv\sum_{\mu}\delta f^q_{\mu} (\hat{D}^q_{\mu})^{\dagger} 
&\equiv&\frac{1}{\sqrt{L_s}}\sum_k \sum_{\sigma_1,\sigma_2,\sigma'_1,\sigma'_2} \hcd_{k+q,\sigma'_1}
\hc_{k,\sigma'_2}
\Big(\delta f^{A;q}_{\sigma_1,\sigma_2}\delta_{\sigma_1,\sigma'_1}\delta_{\sigma_2,\sigma'_2}\\
&&
+\delta f^{B;q}_{\sigma_1,\sigma_2,\sigma'_1,\sigma'_2}\epsilon_k^{\sigma_1,\sigma_2}
+\delta f^{\bar{B};q}_{\sigma_1,\sigma_2,\sigma'_1,\sigma'_2}
\epsilon_{k+q}^{\sigma_1,\sigma_2}
\Big)
\end{eqnarray}
which is needed to define the general Green's functions~(\ref{8dgdd}).

Before we prove Eq.\  (\ref{10shd}), it is instructive
 to consider  
the case  $\tilde{V}^q=0$ in which the three fluctuations 
 $\delta A^{q}_{v}$, $\delta B^{q}_{w}$,    $\delta \bar{B}^{q}_{w}$ 
 are decoupled and we can set $f^{B;q}_w=f^{\bar{B};q}_w=0$.
We start this derivation in the eigenbasis of the Gutzwiller Hamiltonian  
(\ref{9stys}). It leads to the simplest form of the matrix $\tilde{E}$ 
in Eq.\  (\ref{10.1250}) which then reads
\begin{eqnarray}\label{10isd}
&&\Big(\omega-(E_{k+q,\alpha_1}-E_{k,\alpha_2})\Big)
\delta\langle\hhd_{k,\alpha_2} 
\hh_{k+q,\alpha_1} \rangle^{\rm hp/ph}\\\nonumber 
&&=\frac{1}{\sqrt{L_s}}(n^0_{k,\alpha_2}-n^0_{k+q,\alpha_1})
\delta f_{(k+q,\alpha_1),(k,\alpha_2)}\;.
\end{eqnarray}
Here the excitation amplitude is given as
\begin{equation}
\delta f_{(k+q,\alpha_1),(k,\alpha_2)}=\sum_{\sigma_1,\sigma_2}
\delta f^{A;q}_{\sigma_1,\sigma_2}
\big( u^{k+q}_{\sigma_1,\alpha_1}\big)^*u^k_{\sigma_2,\alpha_2}\;.
\end{equation}
Note that the factor $n^0_{k,\alpha_2}-n^0_{k+q,\alpha_1}  =\pm 1$
 in~(\ref{10isd})
 represents the 
 particle-hole and the hole-particle channels in  Eq.\  (\ref{10.1250}). 
For simplicity, we will drop the corresponding labels hp/ph in the following. 

With the transformations~(\ref{9oshss}),(\ref{9oshssb}), 
Eq.~(\ref{10isd}) leads to
\begin{eqnarray}\label{10try}
\delta A_{\sigma_1,\sigma_2}^{q}&=&\frac{1}{\sqrt{L_s}}\sum_k
\delta\langle\hcd_{k,\sigma_2} 
\hc_{k+q,\sigma_1}\rangle\\\nonumber
&=&\frac{1}{\sqrt{L_s}}\sum_k\sum_{\alpha_1,\alpha_2}
\big(u^{k}_{\sigma_2,\alpha_2}\big)^*u^{k+q}_{\sigma_1,\alpha_1}
\delta\langle\hhd_{k,\alpha_2} 
\hh_{k+q,\alpha_1}\rangle\\\nonumber
&=&\frac{1}{L_s}\sum_k\sum_{\alpha_1,\alpha_2 \atop \sigma'_1,\sigma'_2}
\frac{\big(u^{k}_{\sigma_2,\alpha_2}\big)^*u^{k+q}_{\sigma_1,\alpha_1}
\big( u^{k+q}_{\sigma'_1,\alpha_1}\big)^*u^k_{\sigma'_2,\alpha_2}}
{\omega-(E_{k+q,\alpha_1}-E_{k,\alpha_2})}(n^0_{k,\alpha_2}-n^0_{k+q,\alpha_1})
\delta f^{q}_{\sigma'_1,\sigma'_2}\;.
\end{eqnarray}
As expected, we therefore find
\begin{equation}
\delta A_{\sigma_1,\sigma_2}^{q}=\sum_{\sigma'_1,\sigma'_2}
\langle\langle\hat{A}^{q}_{\sigma_1,\sigma_2};(\hat{A}^{q}_{\sigma'_1,\sigma'_2})^{\dagger}
\rangle\rangle^0_\omega\,\delta f^{q}_{\sigma'_1,\sigma'_2}
\end{equation}
with the (`retarded') Green's function
\begin{eqnarray}\label{10afs}
&&\langle\langle\hat{A}^{q}_{\sigma_1,\sigma_2};(\hat{A}^{q}_{\sigma'_1,\sigma'_2})^{\dagger}
\rangle\rangle^0_\omega\\\nonumber
&&=
\frac{1}{L_s}\sum_k\sum_{\alpha_1,\alpha_2}
\frac{\big(u^{k}_{\sigma_2,\alpha_2}\big)^*u^{k+q}_{\sigma_1,\alpha_1}
\big( u^{k+q}_{\sigma'_1,\alpha_1}\big)^*u^k_{\sigma'_2,\alpha_2}}
{\omega-(E_{k+q,\alpha_1}-E_{k,\alpha_2})+{\rm i}\delta}
(n^0_{k,\alpha_2}-n^0_{k+q,\alpha_1})
\end{eqnarray}
 as introduced in~(\ref{9ghd}).

Now we consider the case of a finite interaction matrix  $\tilde{V}^{q}$. 
 Using our abbreviation  $\delta D_{\mu}$ for the amplitudes $\delta A_{v}$, $\delta B_{w}$, $\delta \bar{B}_{w}$ the Lagrange functional
 $\delta\bar{L}^{(2)}$ has the form
\begin{equation}
\delta\bar{L}^{(2)}= \sum_{q,\mu,\mu'}(\delta D^q_{\mu})^{*}V^{q}_{\mu,\mu'}
(\delta D^q_{\mu'})\;.
\end{equation}
 With this additional
  interaction term, Eq.\  (\ref{10isd}) reads
\begin{eqnarray}\label{10psd}
&&\big(\omega-(E_{k+q,\alpha_1}-E_{k,\alpha_2})\big)
\delta\langle\hhd_{k,\alpha_2} 
\hh_{k+q,\alpha_1} \rangle
+(n^0_{k,\alpha_2}-n^0_{k+q,\alpha_1})\\\nonumber
&&\times\sum_{k',\alpha_3,\alpha_4}U^{k',\alpha_3,\alpha_4}_{k,\alpha_1,\alpha_2}(q)
\delta\langle\hhd_{k',\alpha_4} 
\hh_{k'+q,\alpha_3} \rangle=\frac{1}{\sqrt{L_s}}
 (n^0_{k,\alpha_2}-n^0_{k+q,\alpha_1})
\delta f_{(k+q,\alpha_1),(k,\alpha_2)}
\end{eqnarray}
where
\begin{eqnarray}\label{10893}
U^{k',\alpha_3,\alpha_4}_{k,\alpha_1,\alpha_2}(q)
&=&\frac{\partial }{\partial \delta\langle\hhd_{k+q,\alpha_1} 
\hh_{k,\alpha_2} \rangle }
\frac{\partial }{\partial \delta\langle\hhd_{k',\alpha_4} 
\hh_{k'+q,\alpha_3} \rangle}
\delta\bar{L}^{(2)}\\\label{10893b}
&=&\sum_{\mu,\mu'}V^{q}_{\mu,\mu'}\frac{\partial (\delta D^q_{\mu})^{*}}
{\partial \delta\langle\hhd_{k+q,\alpha_1} 
\hh_{k,\alpha_2}\rangle}
\frac{\partial \delta D^q_{\mu'}}
{\partial \delta\langle\hhd_{k',\alpha_4} 
\hh_{k'+q,\alpha_3} \rangle }\;.
\end{eqnarray}
and 
\begin{eqnarray}
\delta f_{(k+q,\alpha_1),(k,\alpha_2)}&=&\sum_{\sigma_1,\sigma_2,\sigma'_1,\sigma'_2}
 \big(u^{k+q}_{\sigma'_1,\alpha_1}\big)^*u^k_{\sigma'_2,\alpha_2}
\Big(\delta f^{A;q}_{\sigma_1,\sigma_2}\delta_{\sigma_1,\sigma'_1}
\delta_{\sigma_2,\sigma'_2}\\\nonumber
&&
+\delta f^{B;q}_{\sigma_1,\sigma_2,\sigma'_1,\sigma'_2}\epsilon_k^{\sigma_1,\sigma_2}
+\delta f^{\bar{B};q}_{\sigma_1,\sigma_2,\sigma'_1,\sigma'_2}
\epsilon_{k+q}^{\sigma_1,\sigma_2}
\Big)\;.
\end{eqnarray}
The derivatives in~(\ref{10893b}) can be further evaluated using the 
transformation~(\ref{9oshss}),(\ref{9oshssb}), 
\begin{eqnarray}\label{10yjf}
\frac{(\partial \delta D^q_{\mu})^{*}}
{\partial \delta\langle\hhd_{k+q,\alpha_1} 
\hh_{k,\alpha_2}\rangle}&=&
\sum_{\sigma_1,\sigma_2}\frac{\partial (\delta D^q_{\mu})^{*}}
{\partial \delta\langle\hcd_{k+q,\sigma_1} 
\hc_{k,\sigma_2}\rangle}\big(u^{k+q}_{\sigma_1,\alpha_1}\big)^{*}
u^{k}_{\sigma_2,\alpha_2}\;,\\\label{10yjfb}
\frac{\partial \delta D^q_{\mu'}}
{\partial \delta\langle\hhd_{k',\alpha_4} 
\hh_{k'+q,\alpha_3} \rangle}&=&\sum_{\sigma_3,\sigma_4}
\frac{\partial \delta D^q_{\mu'}}
{\partial \delta\langle\hcd_{k',\sigma_4} 
\hc_{k'+q,\sigma_3} \rangle}
\big(u^{k'}_{\sigma_4,\alpha_4}\big)^{*}u^{k'+q}_{\sigma_3,\alpha_3}\;.
\end{eqnarray}
 Depending on the  particular fluctuations $\delta D^q_{\mu}$, 
 the remaining derivatives on the r.\ h.\ s.\ of 
equations~(\ref{10yjf}), (\ref{10yjfb}) are given as 
\begin{eqnarray}\label{10atsa}
\delta D^q_{\mu}=\delta A^q_{v}&:&
\frac{\partial \delta A^q_{\sigma_2,\sigma_1}}
{\partial \delta\langle\hcd_{k,\sigma} 
\hc_{k+q,\sigma'} \rangle}
=\frac{\partial \big(\delta A^q_{\sigma_1,\sigma_2}\big)^*}
{\partial \delta\langle
\hcd_{k+q,\sigma} 
\hc_{k,\sigma'} 
\rangle}
=\frac{\delta_{\sigma,\sigma_1}\delta_{\sigma',\sigma_2}}{\sqrt{L_s}}\;,
\\\label{10atsb}
\delta D^q_{\mu}=\delta B^q_{w}
&:&
\frac{\partial \delta B^q_{\sigma_2,\sigma_1,\sigma'_2,\sigma'_1} }
{\partial \delta\langle\hcd_{k,\sigma} 
\hc_{k+q,\sigma'} \rangle}
=\frac{\partial \big(  B^q_{\sigma_1,\sigma_2,\sigma'_1,\sigma'_2}  \big)^*}
{\partial \delta\langle
\hcd_{k+q,\sigma} 
\hc_{k,\sigma'} 
\rangle}
=\frac{\delta_{\sigma,\sigma'_1}\delta_{\sigma',\sigma'_2}}{\sqrt{L_s}}
\epsilon_k^{\sigma_1,\sigma_2}\;,\\\label{10ats}
\delta D^q_{\mu}=\delta \bar{B}^q_{w}
&:&
\frac{\partial \delta \bar{B}^q_{\sigma_2,\sigma_1,\sigma'_2,\sigma'_1} }
{\partial \delta\langle\hcd_{k,\sigma} 
\hc_{k+q,\sigma'} \rangle}
=\frac{\partial \big(  \bar{B}^q_{\sigma_1,\sigma_2,\sigma'_1,\sigma'_2}  \big)^*}
{\partial \delta\langle
\hcd_{k+q,\sigma} 
\hc_{k,\sigma'} 
\rangle}
=\frac{\delta_{\sigma,\sigma'_1}\delta_{\sigma',\sigma'_2}}{\sqrt{L_s}}
\epsilon_{k+q}^{\sigma_1,\sigma_2}\;.
\end{eqnarray}
With Eqs.\ (\ref{10893})-(\ref{10ats}) we are now in the position to 
evaluate~(\ref{10psd}). To this end, we proceed as in~(\ref{10try}),
\begin{eqnarray}\label{10tryb}
\delta A_{\sigma_1,\sigma_2}^{q}=\frac{1}{\sqrt{L_s}}
\sum_k\sum_{\alpha_1,\alpha_2}
\big(u^{k}_{\sigma_2,\alpha_2}\big)^*u^{k+q}_{\sigma_1,\alpha_1}
\delta\langle\hhd_{k,\alpha_2} 
\hh_{k+q,\alpha_1}\rangle\\\nonumber
=-\sum_{\mu,\mu'}V_{\mu,\mu'}^q
\Bigg\{
\Bigg[\frac{1}{\sqrt{L_s}}\sum_k
\sum_{\alpha_1,\alpha_2 \atop \sigma'_1,\sigma'_2}
\frac{\big(u^{k}_{\sigma_2,\alpha_2}\big)^*u^{k+q}_{\sigma_1,\alpha_1}
\big(u^{k+q}_{\sigma'_1,\alpha_1}\big)^*u^{k}_{\sigma'_2,\alpha_2}
}{\omega-(E_{k+q,\alpha_1}-E_{k,\alpha_2})}
(n_{k,\alpha_2}^0-n_{k+q,\alpha_1}^0)\\\nonumber
\times\frac{\partial (\delta D^q_{\mu})^{*}}
{\partial \delta\langle\hcd_{k+q,\sigma'_1} 
\hc_{k,\sigma'_2}\rangle}\Bigg]
\times \sum_{k'}\sum_{\sigma_3,\sigma_4} \frac{\partial \delta D^q_{\mu'}}
{\partial \delta\langle\hcd_{k',\sigma_4} 
\hc_{k'+q,\sigma_3} \rangle} \delta\langle\hcd_{k',\sigma_4} 
\hc_{k'+q,\sigma_3} \rangle  \Bigg\}\\\nonumber
+\sum_{\mu}
\langle\langle\hat{A}^{q}_{\sigma_1,\sigma_2};(\hat{D}^{q}_{\mu})^{\dagger}
\rangle\rangle^0_\omega\,\delta f^{q}_{\mu}\;.
\end{eqnarray}
The sums over $\mu$ and $\mu'$ lead to nine contributions which can all 
 be evaluated using Eqs.\  (\ref{10atsa})-(\ref{10ats}). As a result we
  find
\begin{eqnarray}\label{1098434}
\delta A_{v}^{q}+\sum_{\mu,\mu'}
\langle\langle\hat{A}^{q}_{v};(\hat{D}^{q}_{\mu})^{\dagger}
\rangle\rangle^0_{\omega}V^q_{\mu,\mu'}
\delta D_{\mu'}^{q}=\sum_{\mu}
\langle\langle\hat{A}^{q}_{v};(\hat{D}^{q}_{\mu})^{\dagger}
\rangle\rangle^0_\omega\,\delta f^{q}_{\mu}\;.
\end{eqnarray}
 where the `non-interacting' 
Green's functions 
$\langle\langle\hat{A}^{q}_{v};(\hat{D}^{q}_{\mu})^{\dagger}
\rangle\rangle^0_{\omega}$ in~(\ref{1098434}) are given as in 
(\ref{10afs}), apart from additional factors 
$\epsilon^{\sigma_3,\sigma_4}_{k}$ or $\epsilon^{\sigma_3,\sigma_4}_{k+q}$:
\begin{eqnarray}
&&\left(\begin{array}{c}
\langle\langle\hat{A}^{q}_{\sigma_1,\sigma_2};
(\hat{B}^{q}_{\sigma_3,\sigma_4,\sigma'_3,\sigma'_4})^{\dagger}\rangle\rangle^0_\omega\\
\langle\langle\hat{A}^{q}_{\sigma_1,\sigma_2};(\hat{\bar{B}}^{q}_{\sigma_3,\sigma_4,\sigma'_3,\sigma'_4})^{\dagger}

\rangle\rangle^0_\omega
\end{array}\right)
\\\nonumber
&&=
\frac{1}{L_s}\sum_k\sum_{\alpha_1,\alpha_2}
\frac{\big(u^{k}_{\sigma_2,\alpha_2}\big)^*u^{k+q}_{\sigma_1,\alpha_1}
\big( u^{k+q}_{\sigma'_3,\alpha_1}\big)^*u^k_{\sigma'_4,\alpha_2}}
{\omega-(E_{k+q,\alpha_1}-E_{k,\alpha_2})+{\rm i}\delta}
\left(\begin{array}{c}
\epsilon^{\sigma_3,\sigma_4}_{k}\\
\epsilon^{\sigma_3,\sigma_4}_{k+q}
\end{array}\right)(n^0_{k,\alpha_2}-n^0_{k+q,\alpha_1})\;.
\end{eqnarray}
With~(\ref{1098434}), we have proven the `first' set of Eqs.\  
 (\ref{10shd}), i.e., those with $\mu=v=(\sigma,\sigma')$. 
If we replace $\delta A_{\sigma_1,\sigma_2}$ in the first line of 
Eq.\  (\ref{10tryb}) by
\begin{equation}
\delta B_{\sigma_1,\sigma_2,\sigma'_1,\sigma'_2}=\frac{1}{\sqrt{L_s}}
\sum_k\sum_{\alpha_1,\alpha_2}
\big(u^{k}_{\sigma'_2,\alpha_2}\big)^*u^{k+q}_{\sigma'_1,\alpha_1}
\delta\langle\hhd_{k,\alpha_2} 
\hh_{k+q,\alpha_1}\rangle \epsilon_k^{\sigma_2,\sigma_1}
\end{equation}
 or by
\begin{equation}
\delta \bar{B}_{\sigma_1,\sigma_2,\sigma'_1,\sigma'_2}=\frac{1}{\sqrt{L_s}}
\sum_k\sum_{\alpha_1,\alpha_2}
\big(u^{k}_{\sigma'_2,\alpha_2}\big)^*u^{k+q}_{\sigma'_1,\alpha_1}
\delta\langle\hhd_{k,\alpha_2} 
\hh_{k+q,\alpha_1}\rangle \epsilon_{k+q}^{\sigma_2,\sigma_1}\;
\end{equation}
 the remaining Eqs.\ (\ref{10shd})  are derived in the very same way 
 as~(\ref{1098434}). 
This closes our proof of Eq.\  (\ref{0897}).
\end{appendix}
 \bibliographystyle{unsrt}
\bibliography{bib3}

\begin{thebibliography}{10}
\bibitem{note1}
Note that, in local quantities, we frequently drop lattice site indices.
\end{thebibliography}
\end{document}